\documentclass[prd,aps,10pt,notitlepage,floatfix,twocolumn]{revtex4-1}

\usepackage{amssymb}
\usepackage{amsmath}    
\usepackage{graphicx}   
\usepackage{verbatim}   
\usepackage{color}      
\usepackage{hyperref}   
\usepackage{float}
\usepackage{multirow}

\begin{document}

\title{CMB power spectrum of Nambu-Goto cosmic strings}
\author{Andrei Lazanu}
\email{A.Lazanu@damtp.cam.ac.uk}
\affiliation{Centre for Theoretical Cosmology, Department of Applied Mathematics and Theoretical Physics, Wilberforce Road, Cambridge CB3 0WA, United Kingdom}
\author{E. P. S. Shellard}
\email{E.P.S.Shellard@damtp.cam.ac.uk}
\affiliation{Centre for Theoretical Cosmology, Department of Applied Mathematics and Theoretical Physics, Wilberforce Road, Cambridge CB3 0WA, United Kingdom}\author{Martin Landriau}
\email{landriau@astro.as.utexas.edu}
\affiliation{The University of Texas at Austin, McDonald Observatory, 2515 Speedway, Stop C1402, Austin, Texas 78712-1206, USA}
\date{\today}

\begin{abstract}
We improve predictions of the cosmic microwave background (CMB) power spectrum induced by cosmic strings by using source terms obtained from Nambu-Goto network simulations in an expanding universe. We use three high-resolution cosmic string simulations that cover the entire period from recombination until late-time $\Lambda$ domination to calculate unequal time correlators (UETCs) for scalar, vector and tensor components of the cosmic-string energy-momentum tensor. We calculate the CMB angular power spectrum from strings in two ways. First, to aid comparison with previous work, we fit our simulated UETCs to those obtained from different parameter combinations from the unconnected segment model and then calculate the CMB power spectra using these parameters to represent the string network. Second and more accurately, we decompose the UETCs into their corresponding eigenvalues and eigenvectors and input them directly into an Einstein-Boltzmann solver to calculate the power spectrum for each of the three simulation time periods. We combine the three simulations together, using each of them in its relevant redshift range and we obtain overall power spectra in temperature and polarisation channels. Finally, we use the power spectra obtained with the latest Planck and BICEP2 likelihoods to obtain constraints on the cosmic string tension.
\end{abstract}

\maketitle
\tableofcontents
\section{Introduction}

Topological defects appear naturally during phase transitions in the early Universe. The field develops a spontaneous symmetry breaking phase, where topological defects may form: textures, monopoles, strings, and domain walls (for a review see Ref. \cite{shbook}). Of these topological defects, cosmic strings are one dimensional and may have been created at the end of inflation. Their large energy per unit length ($\mu$) is expected to give rise to observable effects, such as gravitational lensing and gravitational waves. They induce temperature linelike discontinuities, thus giving a characteristic signature in the cosmic microwave background (CMB) power spectrum \cite{kaiser, Gott:1984ef}. 

The CMB is a powerful method for distinguishing between early Universe models. Results from the Planck Collaboration \cite{planckres} provide strong constraints on cosmic strings, instead giving robust support for a nearly scale-invariant inflationary model with the standard six parameters. At present, however, cosmic string constraints are determined not from direct Nambu-Goto string simulations but from either a phenomenological string model, the unconnected segment model (USM) \cite{Pogosian}, or from field theory simulations of the Abelian-Higgs model of increasingly, but still with limited resolution \cite{hindmarsh2006}. The resulting CMB constraints are different, so there is good motivation for determining the Nambu-Goto results directly, which can also improve the calibration of the USM model.

A different approach for detecting cosmic strings has also been investigated, which is based on detecting non-Gaussian signatures generated by cosmic strings through CMB maps using higher-order correlation functions such as the bispectrum \cite{PhysRevD.78.043535, PhysRevD.86.023513, PhysRevD.82.063527}. These methods have yielded weaker constraints on the cosmic strings tension so far.

In addition, pulsar timings have been used to constrain the gravitational wave background, which in turn places stringent constraints on cosmic strings \cite{PhysRevD.85.122003, PhysRevD.89.023512}. These methods provide an independent bound on the cosmic string tension to the CMB one, and in the future they can be significantly improved by new constraints on gravitational waves.

In this paper, after a general review of cosmic strings in the literature, we make an estimate of the CMB power spectrum induced by Nambu-Goto cosmic strings. The main idea is to determine unequal time correlators (UETCs) at high resolution and precision, relevant for the Planck satellite. We use three simulations, covering the entire period from before the radiation to matter transition to late-time $\Lambda$ domination. These UETCs are then fitted with analytic ones characterised in terms of three parameters using the phenomenological USM \cite{copeland}. The parameters for which the analytic model best fits the simulations are used as parameters in the CMBACT code \cite{cmbact}, which determines the power spectrum of the cosmic strings. The UETCs are diagonalised and the eigenvectors are used directly as sources for the CMB fluctuations, thus obtaining a very accurate angular power spectrum. These UETCs are combined for the different epochs and the overall power spectra in the temperature and polarisation channels are obtained. Finally, the power spectrum is used to estimate the allowed cosmic string contribution in the power spectrum ($G\mu/c^2$ for the string tension and $f_{10}$ for the string fractional power) using Markov chain Monte Carlo parameter estimation (COSMOMC) with the latest CMB likelihoods.

\section{Review of cosmic strings and other topological defects}

As the size of the Universe has dramatically increased during inflation, the only defects that may be observable today must have been formed at the end of inflation or after its end \cite{shbook}. The existence of cosmic topological defects is related to the spontaneous symmetry breaking in the evolution of the Universe, during the cooling-down phase. They have been studied in analogy with condensed matter physics \cite{Kirzhnits} and particularly solid state physics \cite{weinberg1979prd}. From all these topological defects, strings are the most studied \cite{shbook}. 

Cosmic strings were once considered to be the primary source of anisotropies in the CMB \cite{shbook}. However, after the release of the Boomerang data it has been shown that the characteristics of the power spectrum they produce does not match the one observed in the CMB using the COBE mission and Boomerang  \cite{2000astro.ph..5115C}, WMAP and Planck probes. The cosmic string temperature power spectrum is smooth and has a unique peak, and hence it does not match the observed CMB power spectrum. In the meantime, a good agreement in the power spectrum has been obtained from the inflationary scenarios, effectively ruling out topological defects as the primary source of anisotropies \cite{planckstr}. However, cosmic strings can still be present. Current observational data allow a maximum of 3\% of the observed power to be due to cosmic strings \cite{planckstr}. Initially it was expected that the amplitude of the string tension was in the region of  $10^{16}$ GeV, which is the Grand Unification Theory scale, which corresponds to $G\mu/c^2 \sim 10^{-6}$. Such high energy is impossible to probe with terrestrial experiments, and identifying the existence of cosmic strings at these energies would offer a very interesting connection with particle physics. It would be possible to test particle collision patterns at very high energies and to identify signatures of extra dimensions from string theory \cite{planckstr, battye}.

More recent studies in string theory have shown that their tension could in fact be as low as the electroweak scale \cite{polchinski}. In this case, the allowed limit for the string tension would be $10^{-11}<G\mu/c^2<10^{-6}$ \cite{Copeland2004}. The more recent work is based on superstring theories and new methods of string compactification with large extra dimensions and/or large warp factors. These ideas are presented in detail in Refs. \cite{Arkani-Hamed, Randall}. Another option, which relies on supersymmetry, is presented in Ref. \cite{Jeannerot:2003qv}.

Very recently, various cosmic string models \cite{2014arXiv1403.4924L, 2014arXiv1403.6105M, 1475-7516-2015-02-024} have been discussed in trying to explain the BB polarisation obtained by the BICEP2 experiment \cite{bicep}.

String networks are formed of long strings and finite loops. When long strings intersect, there are two possibilities: they either pass through one another as if there ware no collision, or they disconnect and reconnect again in a different way. Loops can be formed in the latter case. When a string self-intersects, the reconnection probability is one for classical cosmic strings \cite{1126-6708-2005-10-013}. They then collapse inward and decay. During the decay process, their energy is converted into gravitational waves. 
 
There are two approaches for studying the evolution of cosmic strings: the Abelian-Higgs field theory model and the Nambu-Goto effective action.

\subsection*{Abelian-Higgs model} 
The Abelian-Higgs model is the relativistic extension of the Ginsburg-Landau theory and has the action \cite{shbook}:
\begin{eqnarray}
S=\int d^4y \sqrt{-g}(\partial_\mu+ieA_\mu)\bar\phi(\partial^\mu-ieA^\mu)\phi- \nonumber \\
\frac{1}{4}F_{\mu\nu}F^{\mu\nu}-\frac{1}{4}\lambda(|\phi|^2-\eta^2)^2
\label{actionAH}
\end{eqnarray}
where $\phi$ is a complex scalar field, $\lambda$ and $e$ are coupling constants, and $A_\mu$ is a four-dimensional U(1) gauge field satisfying $F_{\mu\nu}=\partial_\mu A_\nu-\partial_\nu A_\mu$. Using $D_\mu=\partial_\mu-ieA_\mu$, the equations of motion become:
\begin{equation}
D_\mu D^\mu \phi=-\frac{\lambda \phi}{2}(|\phi|^2-\eta^2)
\end{equation}
\begin{equation}
D_\nu F^{\nu\mu}=2e\Im(\phi^\star D^{\mu}\phi)
\end{equation}

The action described in Eq. (\ref{actionAH}) has vortex-type solutions \cite{abrikosov, nielsen}, which are static and cylindrically symmetric:
\begin{equation}
\phi_s(\textbf{r})=e^{in\theta}f(r)
\end{equation}
\begin{equation}
A_{sa}(\textbf{r})=\epsilon_{ab}x_b\frac{n}{er^2}\alpha(r)
\end{equation}
with $a,b=1,2$ and $\epsilon$ being an antisymmetric tensor. Fixing suitable boundary conditions, the large $r$ asymptotic solutions to these equations can be obtained in terms of modified Bessel functions of the second kind:
\begin{eqnarray}
\alpha(r)=1-rK_1\left(\sqrt{2}er\right) \\
f(r)=1-K_0\left(\sqrt{\lambda}r\right)
\label{fr}
\end{eqnarray}

In the case of a curved string, one can express any point near the string world sheet in terms of tangent vectors to the world sheet and normal vectors:
\begin{equation}
y^{\mu}(\xi)=x^{\mu}(\zeta)+\rho^An_A^\mu(\zeta) 
\end{equation}
where $n_A^\mu$ are the normal vectors, $x_{,a}^\mu$ are the tangent vectors, $y^\mu$ is a point near the world sheet and $\xi^\mu=(\zeta^a,\rho^A)$.

The approximate solution is thus:
\begin{equation}
\phi\left(y(\xi)\right)=\phi_s(r)
\end{equation}
\begin{equation}
A^\mu\left(y(\xi)\right)=n_B^\mu(\zeta)A_{sB}(r)
\end{equation}
 
When reexpressing the action in terms of these new coordinates, one needs to calculate the Jacobian of the transformation from the $y$ to $\xi$ coordinates. This is given by the square root of the modulus of the determinant of the world sheet metric $M_{\alpha\beta}$, which can be expressed as:
\begin{equation}
M_{\alpha\beta}=\text{diag}\left(\gamma_{ab},-\delta_{AB}+O(r/R)\right)
\end{equation}
where
\begin{equation}
\gamma_{ab}=g_{\mu\nu}x^{\mu}_{,a}x^{\nu}_{,b}
\label{gmunu}
\end{equation}
The integration over the normal coordinates $\rho^A$ can be performed, yielding just the constant $\mu$. The asymptotic solutions (\ref{fr}) decaying exponentially, the correction is reduced to $O(\delta/R)$. Hence, if one considers the string curvature small with respect to the string length, the Nambu-Goto action is obtained as the first-order approximation \cite{shbook}. It will be discussed in the next subsection.

\subsection*{Nambu-Goto model}
\label{Nambu-Goto-model}
A one-dimensional reduction of the Abelian-Higgs action gives rise to the Nambu-Goto action, described below (see Refs. \cite{shbook, kibble}). Hence, the Nambu-Goto strings have just one dimension (0 width) and live in a two-dimensional space-time parametrised by $X^{\mu}=X^{\mu}(\zeta^a)$ with $a = 0,1$ . The physical motivation for using this approximation is that higher-order corrections are small when strings are considered to be long enough compared to their width \cite{turok1988}. Nambu-Goto strings can be derived as solutions of the Nambu-Goto action:
\begin{equation}
S=-\mu\int\sqrt{-\gamma}d^2\zeta
\label{nambuS}
\end{equation}
where $\gamma_{ab}=g_{\mu\nu}\partial_aX^{\mu}\partial_bX^{\nu}$ is the two-dimensional world sheet metric and $\gamma=\det(\gamma_{\mu\nu})$ [same as Eq. (\ref{gmunu}), with $x^{\mu} \to X^{\mu}$].

($\zeta^0$, $\zeta^1$) is an arbitrary parametrisation of the string world sheet, with one of the parameters timelike and the other spacelike. Hence, in an expanding universe, one may choose to take $\zeta^0:=\tau$ (conformal time) and $\zeta^1:=\sigma$ (the spacelike parameter of the string).

Simulations usually start with Vachaspati-Vilenkin initial conditions \cite{vilenkin}. When two strings segments meet, they split and then reconnect the other way (intercommutation). In this process, loops are being formed and they decay and radiate energy. In the time evolution of the cosmic string network, the strings are expected to reach a scaling solution, i.e. the number of cosmic strings crossing each horizon volume is fixed \cite{bouchet1990}. This energy loss mechanism in fact makes cosmic strings cosmologically viable (otherwise cosmic strings would eventually dominate the Universe) \cite{kibble76} and also the initial conditions considered for the simulations less important. 
In Fourier space, the energy-momentum tensor arising from action  (\ref{nambuS}) can be expressed as:
\begin{eqnarray}
\label{emtensor}
\Theta_{\mu\nu}({\bf k},\tau)= \int d^3{\bf x}
e^{i{\bf k}.{\bf x}} \Theta_{\mu\nu}({\bf x},\tau)= \nonumber \\
= \mu\int d \sigma e^{i{\bf k}.{\bf X(\sigma,\tau)}} 
\left( \epsilon \dot{\bf X}^\mu\dot{\bf X}^\nu - \epsilon^{-1}  
{\bf X}'^\mu {\bf X}'^\nu \right)\,
\end{eqnarray}
where the prime denotes differentiation with respect to $\sigma$ and dot denotes differentiation with respect to $\tau$ and $\epsilon=\sqrt{{\bf X}'^2/(1-\dot{\bf{X}}^2)}$ represents the energy density along the string.
For the Nambu-Goto strings, a good phenomenological model is given by the velocity-dependent one-scale (VOS) model \cite{shellard96, shellard96_2, shellard2000}. This model assumes that the string population is formed by long strings (denoted by $\infty$) and small loops (denoted by $l$). The long strings are characterised by the correlation length $L$ and by the root-mean-square velocity $v$: 
\begin{equation}
v^2=\frac{\int{\dot{X}^2\epsilon d\sigma}}{\int{\epsilon d\sigma}}
\end{equation}
The averaged energy density of the long strings is: 
\begin{equation}
\rho_{\infty}=\frac{\mu}{L^2} 
\end{equation}
and the parameter $\tilde{c}$ is a constant which expresses the loop production rate and is defined by the following formula:
\begin{equation}
\frac{d\rho_{\infty}}{dt}=\tilde{c}v_\infty \frac{\rho_\infty}{L}
\end{equation}

The evolution equations for the correlation length $L$ and for the velocity of long strings $v_\infty$ can be derived from the microscopic equations of motion and Newton's second law:
\begin{equation}
2\frac{dL}{dt}=2HL(1+v_\infty^2)+\tilde{c}v_\infty
\label{eom1}
\end{equation}
\begin{equation}
\frac{dv_\infty}{dt}=(1-v_\infty)\left(\frac{k}{L}-2Hv_\infty\right)
\label{eom2}
\end{equation}
where $k$ is a parameter which characterises the small scale structure of the string network and which expresses the loop production rate \cite{shellard96}:
\begin{equation}
k=\frac{\langle (1-\dot{\textbf{x}})(\dot{\textbf{x}} \cdot \hat{\textbf{u}})\rangle}{v(1-v^2)}
\end{equation}
where $\hat{\textbf{u}}$ is a unit vector parallel to the curvature radius one. For the relativistic regimes considered in the case of cosmic strings, a suitable asymptotic ansatz is:
\begin{equation}
k_{\text{rel}}=\frac{2\sqrt{2}}{\pi}\cdot \frac{1-8v^6}{1+8v^6}
\end{equation}
while in the nonrelativistic limit a consistent asymptotic limit is found \cite{shellard2000}:
\begin{equation}
k_{\text{non-rel}}=\frac{2\sqrt{2}}{\pi}
\end{equation}

Numerical simulations have fixed $\tilde{c}=0.23$ regardless of epoch.
Scale-invariant solutions, which are characterised by $v_\infty=\text{constant}$ and $L \propto t$ exist only when the scale factor is evolving as a power law.

\subsection*{Phenomenological unconnected segment model}
\label{phenousm}
For Nambu-Goto strings, the USM model has been devised, as described in Refs. \cite{Pogosian, albrecht,  robinson2, copeland}. In the this model, the cosmic string network is described by a Brownian network which is formed from a set of independent, uncorrelated straight segments with random velocities. All segments are produced early in the evolution of the Universe, and then, at each epoch, part of the strings decay such that scaling is preserved throughout the history of the Universe. Each segment has comoving length equal to the correlation length, and its position is randomly chosen, in such a way such that the equations of motion (\ref{eom1}) and (\ref{eom2}) are satisfied for each particular string segment. Hence, the magnitude of the velocity is determined by these equations, but its orientation is arbitrary and is taken from a flat distribution.

As the model is made from straight segments, the small scale structure of the strings is not taken into account. This has been adjusted phenomenologically, by adding a new ``wiggliness'' parameter  $\alpha$ \cite{carter}, which, however, describes only the macroscopic evolution of the strings. This modifies the energy momentum tensor (\ref{emtensor}):
\begin{equation}
\label{emtensorwiggly}
\Theta_{\mu\nu}({\bf k},\tau)= 
\mu\int d \sigma e^{i{\bf k}.{\bf X(\sigma,\tau)}} 
\left( \epsilon \alpha \dot{\bf X}^\mu\dot{\bf X}^\nu - \frac{1}{\epsilon \alpha}
{\bf X}'^\mu {\bf X}'^\nu \right)\,
\end{equation}

The string segment decay is realised through a function $T^{\text{off}}$ that is a smooth approximation to the Heaviside function, such that after a certain time the particular string segment disappears and similarly for the appearance of the segment through a similar function $T^{\text{on}}$. The total stress-energy tensor is calculated as the sum of the individual components for the segments:
\begin{eqnarray} 
\Theta_{\mu\nu} ( {\bf k} ,\tau ) &=&
\sum_{m} \Theta^m_{\mu\nu}  ( {\bf k} ,\tau ) T^{\rm off}
\left(\tau,\tau_m^{\rm on}
\right)  T^{\rm on} \left(\tau,\tau_m^{\rm off}
\right)\,
\label{emtsum}
\end{eqnarray} 

The energy-momentum tensor of one segment is of the form of Eq. (\ref{emtensorwiggly}): 
\begin{eqnarray}
\Theta_{\mu\nu}(\textbf{k},\tau) = \nonumber \\
= \mu \int_{-{\it l}/2}^{{\it l}/2} d \sigma e^{i\textbf{k}\cdot\textbf{X}}
(\epsilon\alpha \dot{X}^\mu \dot{X}^\nu - \frac{1}{\epsilon\alpha}
X^{\prime\mu} X^{\prime\nu})
\end{eqnarray}
where $l$ is the comoving correlation length $l=L/a$. The number of string segments at $N$ at each particular time satisfies
\begin{equation}
N(\tau) \propto \frac{1}{\tau^3}
\end{equation}
and hence scaling is preserved \cite{robinson2}. However, in this case in order to have one string segment today, one would need at least $10^{12}$ initial string segments, which is not possible numerically. The problem was overcome by considering only one of the segments decaying at each particular time and multiplying it by a suitable weighting function, chosen such that scaling is preserved. An equation for the evolution of the wiggliness parameter $\alpha$ is used \cite{bouchet1990}:
\begin{equation}
\alpha(\tau)=1+\frac{0.9}{H\tau} 
\end{equation}
such that it satisfies the expected behaviour in the radiation, matter and cosmological constants eras.

As the equations describing the matter perturbations and the power spectra do not depend on the direction of the wave-vector $\textbf{k}$, this can be taken to be along the $k_3=k_z$ axis. Thus, the energy-momentum tensor components become:
\begin{equation}
\Theta_{00} = \frac{\mu \alpha} {\sqrt{1-v^2}}
\frac{\sin(k\hat{X}_3^\prime{\it l}/2)}{k\hat{X}_3^\prime/2}
\cos(k \cdot X_0 + k\hat{\dot{X}}_3 v\tau)
\label{th00usm}
\end{equation}
\begin{equation}
\Theta_{ij} = \left[ v^2 \hat{\dot{X}}_i \hat{\dot{X}}_j -
\frac{(1-v^2)}{\alpha^2} \hat{X}^\prime_i \hat{X}^\prime_j \right]
\Theta_{00}
\label{thijusm}
\end{equation}
while $\Theta_{0i}$ can be expressed using the conservation of the stress-energy tensor $\Theta_{\mu\nu}$. With this choice of the wave vector, the components required for the Boltzmann integrator CMBACT \cite{cmbact}, which in turn is based on CMBFAST \cite{seljak} are:
\begin{eqnarray}
\label{thSusm}
\Theta^S = (2\Theta_{33}-\Theta_{11}-\Theta_{22})/2 \\
\Theta^V = \Theta_1^V=\Theta_{13} \\
\label{thTusm}
\Theta^T = \Theta_{12}^T=\Theta_{12} \\
\Theta = \Theta_{ii}  \\
\Theta^D = \Theta_{03}
\end{eqnarray}
These are the anisotropic scalar, the vector component, the tensor component, the tracem and the velocity field.

This model has been used to mimic the behaviour of Abelian-Higgs strings, by tuning its parameters. The results are in good agreement with the field theory simulations \cite{battye}.

\subsection*{CMB comparison for Abelian-Higgs and Nambu-Goto simulations}
As described in the previous subsections, field theory simulations have a much lower dynamical range than Nambu-Goto simulations. They are, however, able to resolve scales of sizes comparable to the string width, and the decay products appear naturally out of the simulation. In the case of Nambu-Goto simulations, loops are clearly visible, but in field theory simulations, energy moves directly into massive modes of the fields because of the limited dynamical range. A comparison between the two types of simulations appears in Fig. 1 of Ref. \cite{hindmarsh2011}. 

This can be illustrated by the different shapes and amplitudes of the temperature power spectra determined from these two models, as it can be seen in Fig. 3 of Ref. \cite{planckstr}. These plots were created with the standard parameters from the code CMBACT \cite{cmbact} for the USM (Nambu) and AH mimic and with field theory simulations for the Abelian-Higgs cosmic strings.

The difference may be due to the fact that the USMs are not able to model the velocity correlations between the strings, but also to the fact that the field theory simulations rely on extrapolation over many orders of magnitude \cite{hindmarsh2006}. Even though extensive simulations have been performed for the Abelian-Higgs model, the Nambu-Goto strings have mostly been described using the simplified USM model.
 
In this paper we are using the Allen and Shellard code \cite{Allen} to generate Nambu-Goto string networks with Vachaspati-Vilenkin initial conditions and evolve them in time in different epochs of the Universe (as described later). The code outputs the string parameters for all the points from the string network at each time. Another code is used to read in all the parameters for all points at a particular time step, evaluate the local energy-momentum tensor using the real-space version of Eq. (\ref{emtensor}) and then interpolate it on a three dimensional grid of chosen size. The outcome of this is an energy-momentum tensor for the whole network at a specific time evaluated on a 3D grid. This is Fast Fourier Transformed, and it is then decomposed into scalar, vector, and tensor parts (SVT decomposition) in order to determine the components required \cite{landriau2003}.

The first code treats each time step separately. It reads the coordinates of each point and the data required to calculate the energy-momentum tensor at that particular place according to Eq. (\ref{emtensor}). This energy-momentum tensor is interpolated on a given grid, user-specified according to the resolution required, using a triangular cloud-in-cell interpolation method. This method interpolates each of the given points onto the 27 closest neighbours on the three-dimensional grid (weighted appropriately according to the distance to each point and ensuring energy conservation in this process), and the results are added up. Thus, the full stress-energy tensor is created on the grid at that particular time in real space. Then the full 3D matrix is converted to Fourier space using a Fast Fourier Transform routine. The new grid, now in Fourier space, is smoothed out by multiplying it with a Gaussian and then the energy-momentum tensor is split into scalar, vector, and tensor parts. For the scalar parts, we have chosen to output the $\Theta_{00}$ (energy density) and $\Theta^S$ (anisotropic scalar) components, but other choices can be made according to what one needs; for the vector parts, we have output two of the vector components and similarly for tensors.

\section{Cosmic string simulations}

\begin{figure*}[!htb]
\begin{center}$
\begin{array}{ccc}
\includegraphics[width=1.61in]{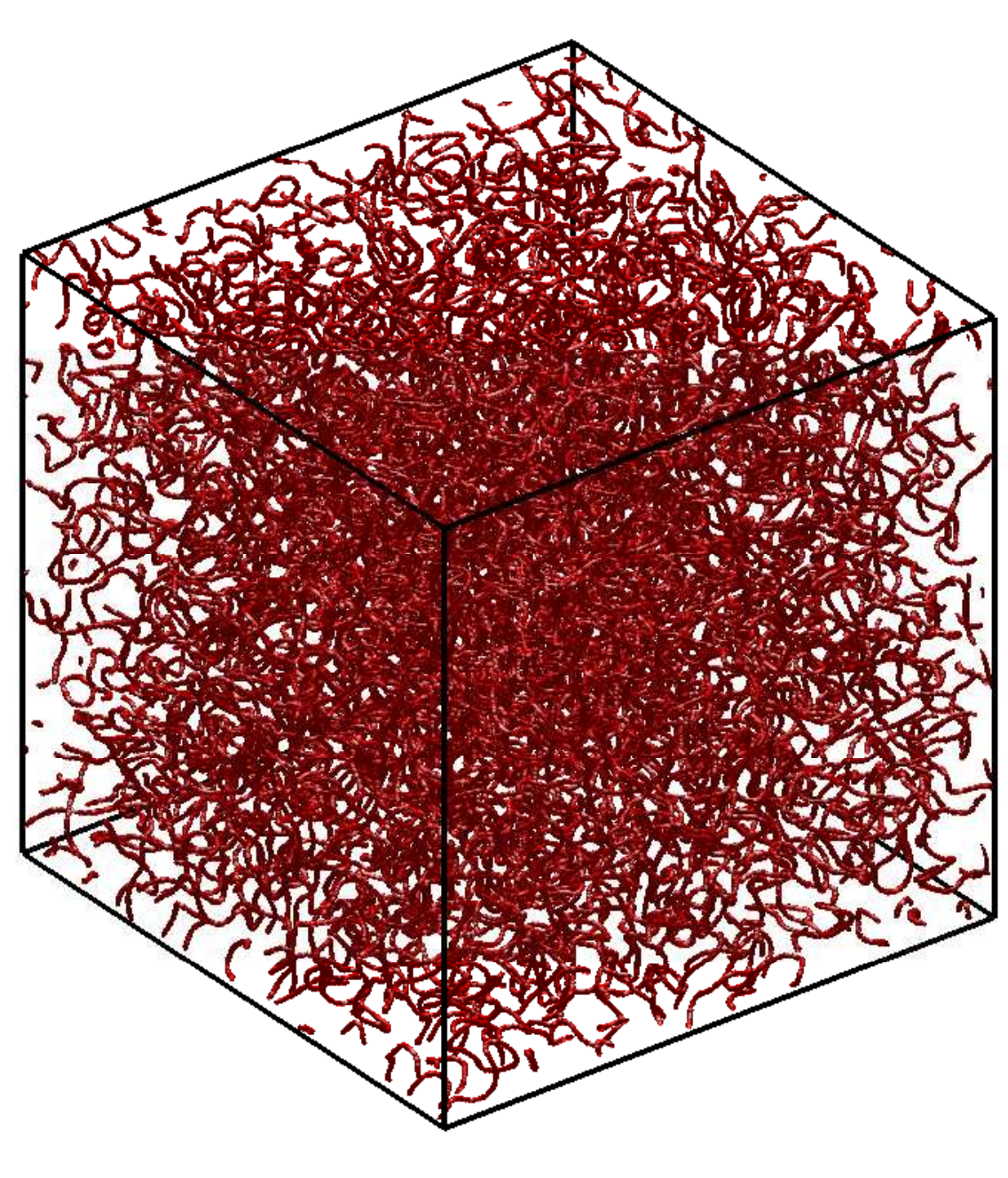} &
\includegraphics[width=1.61in]{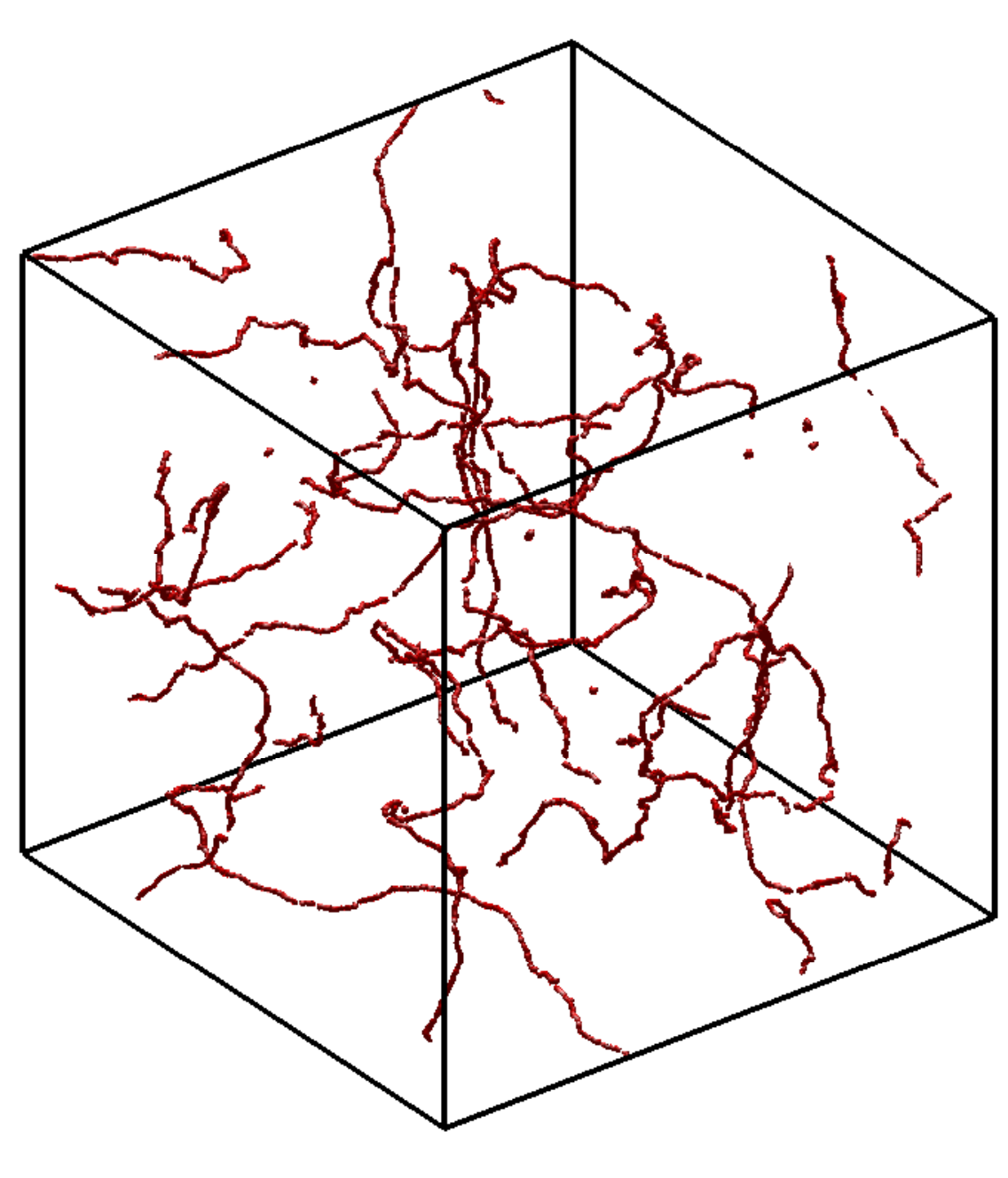} &
\includegraphics[width=1.61in]{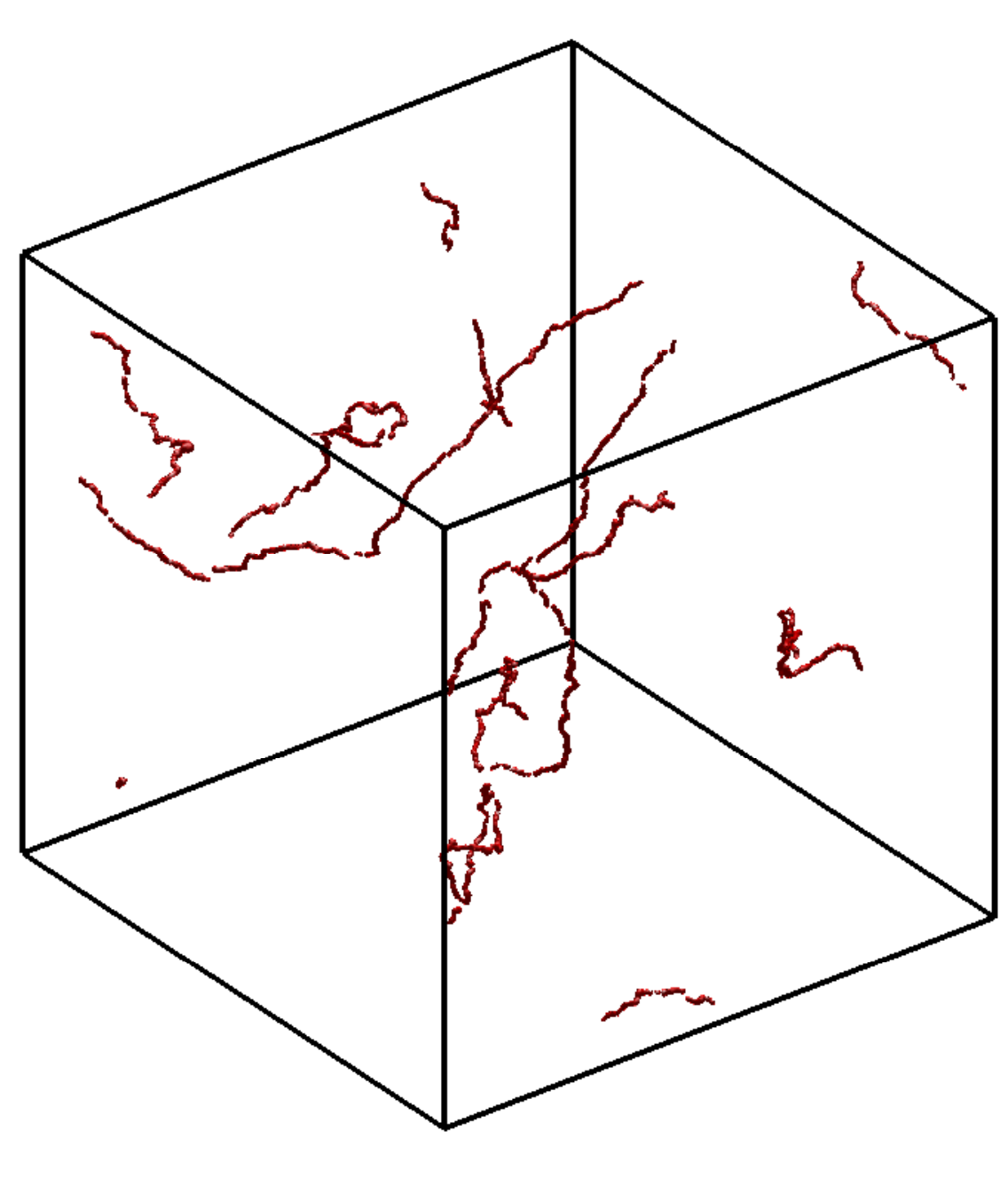} 
\end{array}$
\caption{Evolution of the string network in the simulation covering the matter era (redshift range 945 to 37.2)}
\label{strn_network2}
\end{center}
\end{figure*}

To obtain an accurate prediction for the cosmic string power spectrum, we have used three simulations, covering in total a redshift range from 5900 to 0 as follows. The first simulation (Simulation 1) starts deep into the radiation era, goes through radiation-matter transition, and ends in the matter era, corresponding to redshifts from 5900 to 700. The second simulation lies entirely in the matter era, with redshifts from 860 to 37. The third simulation starts in the matter era (redshift 48) and goes into the cosmological constant future, to z=0. All three simulations have Vachaspati-Vilenkin initial conditions \cite{shbook} and evolve in time. All three simulations had earlier initial times, but we have removed around 1.5\% of the time steps of each of them in order to remove the excessive correlations in the initial conditions. The important quantity in this context is the dynamical range of the simulations. After removing these initial time steps, we decrease the dynamical range of each of the simulations by roughly 15\%. In Fig. \ref{strn_network2} the time evolution of the string network simulation covering the matter epoch is shown by plotting the energy component for the strings at three time steps corresponding the first, middle and last time used in the calculation of the UETCs. The density of strings is decreasing with the expansion of the Universe. The simulations are the ones described in Ref. \cite{PhysRevD.83.043516}, and the same cosmological parameters are used.

The three simulations cover the entire cosmological history of the Universe which is of interest when determining the CMB power spectrum.  One can see that the network is initially very dense (Fig. \ref{strn_network2}) in each of the simulations, and Vachaspati-Vilenkin initial conditions are used.

Large loops are kept in the simulation and contribute to the total energy-momentum tensor of the network. In a physical context, small loops decay into gravitational radiation. Those that are smaller than the resolution of the simulation are not resolved and hence could be treated as point mass sources. Their effect on the overall string network is negligible in linear theory and therefore are neglected in practice because it accelerates the network simulation to remove very small nonintersecting loops. These tiny loops were also found to have a small effect in Refs. \cite{PhysRevLett.77.3061, doi:10.1142/S0218271802001299}. By ignoring these small loops, we obtain a conservative bound on cosmic strings. An alternative simulation technique has been developed in Refs. \cite{PhysRevD.74.063527, PhysRevD.75.063521} where the evolution of these small loops can be more efficiently continued during network evolution.

\section{Unequal-time correlator approach}
\label{UETCapp}
Cosmic strings are active sources. This means that unlike primordial perturbations, which are seeded at the end of inflation but primarily act after last scattering, cosmic strings continuously seed perturbations throughout the history of the Universe \cite{turok-causality}. The presence of these cosmic strings induces modifications to the usual perturbation equations, in the sense that a term corresponding to the energy-momentum tensor of cosmic strings must be added and then the Einstein-Boltzmann hierarchy must be solved. For example, in the scalar case, we present a simplified discussion showing the modification of the equations of interest in the synchronous gauge:
\begin{equation}
\ddot{\delta}_C+\frac{\dot{a}}{a}\dot{\delta}_C=4\pi G \sum_N (1+3c_N^2)\rho_Na^2\delta_N+S
\label{scalar1}
\end{equation}
\begin{equation}
\ddot{\delta}_R+\frac{\dot{a}}{a}(1-3c_S^2)\dot{\delta}_R=c_S^2\nabla^2\delta_R+\frac{4}{3}\ddot{\delta}_C+\frac{4}{3}\frac{\dot{a}}{a}(1-3c_S^2)\dot{\delta}_C
\label{scalar2}
\end{equation}
where $\delta_C$ and $\delta_R$ are the cold dark matter (CDM) and radiation overdensities, $c_S$ is the sound speed, $N$ represents the CDM, photon-baryon fluid and the species of neutrinos, while $S$ (the source term) can be expressed in terms of the stress-energy tensor as:
\begin{equation}
S=4\pi G (\Theta_{00}+\Theta_{ii})
\end{equation}

Uniform energy density and space curvature must be taken as initial conditions in the Boltzmann equations. In this case, to first order in perturbation theory, by integrating the full Boltzmann equations with these initial conditions one can obtain the string multipoles $C_l^{\text{string}}$. In this case, as the active sources are uncorrelated with the primordial fluctuations, the total angular power spectrum can be expressed as:
\begin{equation}
C_l=C_l^{\text{inflationary}}+C_l^{\text{string}} 
\end{equation}

To do the integration, there are two methods: (1a) ignore the full Boltzmann hierarchy and use Green's functions (e.g. Refs. \cite{ue-li-pen2, PhysRevD.78.043535}) or (1b) use a first-order equivalent to Greens's functions and treat the full Boltzmann hierarchy (Ref. \cite{landriau2003}); and (2) use UETCs. Indeed, Eqs. (\ref{scalar1})-(\ref{scalar2}) are linear and their homogeneus part only depends on the magnitude of the wave vector, which makes it possible to use the UETC approach.

To calculate the CMB power spectrum \cite{durrer} from active sources, one has to solve an equation of the form:
\begin{equation}
DX = S_T 
\end{equation}
where \textit{D} is a differential operator and $S_T$ is the active source.
The power spectrum is then a quadratic quantity which has the general form
\begin{equation}
\left\langle {{X_i}\left( {{\tau _0},\textbf{k}} \right)X_j^ * \left( {{\tau _0},\textbf{k}'} \right)} \right\rangle .
\end{equation}
This can be expressed in terms of Green's functions as follows:
\begin{eqnarray}
\left\langle {{X_i}\left( {{\tau _0},\textbf{k}} \right)X_j^*\left( {{\tau _0},\textbf{k}'} \right)} \right\rangle \int\limits_{{\tau _{in}}}^{{\tau _0}} {d\tau {G_{jm}}} \left( {\tau ,k} \right) \times \nonumber \\
\times \int\limits_{{\tau _{in}}}^{{\tau _0}} {d\tau G_{lm}^ * } \left( {\tau ',k'} \right){S_m}\left( {\tau ,k} \right)S_n^ * \left( {\tau ',k'} \right)
\end{eqnarray}

Hence, to calculate the influence of strings on the CMB power spectrum, only the following quantity is needed:
\begin{equation}
\left\langle {{S_m}\left( {\tau ,k} \right)S_n^ * \left( {\tau ',k'} \right)} \right\rangle 
\end{equation}

In particular, the string energy-momentum tensor UETC can be written as:
\begin{equation}
\langle\Theta_{\mu\nu}(\textbf{k},\tau)\Theta_{\rho\sigma}(-\textbf{k},\tau')\rangle=X_{\mu\nu,\rho\sigma}(k,\tau,\tau')
\end{equation}

Using scaling, one can reexpress this correlation function as \cite{ue-li-pen2}:
\begin{equation}
X_{\mu\nu,\rho\sigma}(k,\tau,\tau')=\frac{c_{\mu\nu,\rho\sigma}(k\tau,k\tau')}{\sqrt{\tau\tau'}}
\end{equation}

This new UETC matrix $c_{\mu\nu,\rho\sigma}(k\tau,k\tau')$ is obtained as the expectations value of a squared quantity and hence is positive definite \cite{turok-causality}. It is thus diagonalisable and can be expressed in terms of its eigenvalues and eigenvectors \cite{contaldi, ue-li-pen}:
\begin{equation}
c_{\mu\nu,\rho\sigma}(k\tau,k\tau')=\sum_i \lambda_i v_{\mu\nu}^{(i)}(k\tau)v_{\rho\sigma}^{(i)T}(k\tau')
\end{equation}
where $v_i$ are the a set of orthonormal eigenvectors of the matrix $c$. 

The eigenmodes are coherent \cite{contaldi} and hence each of them can be fed individually into a Boltzmann equation solver and then the total angular power spectrum can be expressed as:
\begin{equation}
C_l^{\text{string}}=\sum_i \lambda_i C_l^{(i)}
\label{sumcls}
\end{equation}

As the unequal time correlators have been multiplied by $\sqrt{\tau\tau'}$, the source terms in the Boltzmann equation are substituted as:
\begin{equation}
\Theta(k\tau) \to \frac{v^{(i)}(k\tau)}{\sqrt{\tau}}
\label{theta2v}
\end{equation}

To calculate the power spectrum of the cosmic strings, one has to modify the sources of the Einstein equations by adding the contribution from the strings as sources \cite{landriau2003}. The Einstein equation is:
\begin{equation}
G_{\mu\nu}+\Lambda g_{\mu\nu}=8\pi G T_{\mu\nu}
\label{einstein}
\end{equation}

For an expanding universe, the metric can be expressed as:
\begin{equation}
g_{\mu\nu}=a^2 (\eta_{\mu\nu}+h_{\mu\nu})
\label{pert}
\end{equation}
where $h_{\mu\nu}$ is a perturbation to $\eta$. In the synchronous gauge, $h_{00}=h_{0i}=0$.

A general tensor expressed in Fourier space can split into its scalar, vector, and tensor parts as:
\begin{eqnarray}
T_{ij}(\textbf{k})=\frac{1}{3} T \delta_{ij}+\left( \hat{k_i} \hat{k_j} - \frac{1}{3}\delta_{ij} \right)T^S+ \nonumber \\
+ \left( \hat{k_j}T_i^V+\hat{k_i}T_j^V\right)+T_{ij}^T
\label{SVT-g}
\end{eqnarray}
where the vector and tensor parts are transverse and the tensor part is traceless. Both the metric perturbation $h_{\mu\nu}$ and the energy-momentum tensor can be split according to Eq. (\ref{SVT-g}). To find the equations satisfied by the components of the metric perturbations, one has to consider the first-order perturbations to both the metric and the energy-momentum tensor and then use Eq. (\ref{einstein}). 

The metric perturbation tensor is split according to Eq. (\ref{SVT-g}), while for the stress-energy tensor one needs to consider the usual matter perturbations (as in Ref. \cite{ma}) and the perturbations given by the cosmic strings:
\begin{eqnarray}
\delta T_0^0 = - \delta \rho + \Theta_0^0 \\
\delta T_i^0 = \left(\rho+P \right)v_i +\Theta_i^0 \\
\delta T_j^i = \delta P \delta_j^i + p\Sigma_j^i + \Theta_j^i
\end{eqnarray}

Using the Einstein equation (\ref{einstein}) and its conservation $G_{\mu\nu}^{;\nu}=0$, one obtains the evolution equations for the metric perturbations:
\begin{eqnarray}
\label{modif2}
k\bar{\eta}'=4\pi G a^2 \sum_i \left( \rho_i+p_i\right)v_i-\frac{4\pi G}{k} \Theta^D\\
\ddot{h}^S+2\frac{a'}{a} \dot{h}^S-2k^2 \eta=16 \pi G \left(a^2 p \Sigma^S+\Theta^S \right) \\
\ddot{h}^V+2\frac{a'}{a} \dot{h}^V=16 \pi G \left(a^2 p \Sigma^V+\Theta^V \right) \\
\ddot{h}^T+2\frac{a'}{a} \dot{h}^T+k^2 h^T=16 \pi G \left(a^2 p \Sigma^T+\Theta^T \right)
\label{modif3}
\end{eqnarray}
where $\bar{\eta}=\frac{h-h^S}{6}$ and $\Theta^D$ satisfies the equation
\begin{equation}
\dot{\Theta}^D=\Theta^D \left(-2\frac{\dot{a}}{a}-\frac{k^2 a}{3\dot{a}} \right)-\frac{k^2}{3} \left(2\Theta^S-\Theta_{00}-\frac{a\dot{\Theta}_{00}}{\dot{a}} \right)
\label{modif}
\end{equation}

Equations (\ref{modif2})-(\ref{modif3}) have been implemented into a Boltzmann solver (CMBFAST), by modifying the relevant equations to accommodate the cosmic string sources. The energy-momentum tensor of the cosmic strings needed to be substituted with the relevant eigenvector, as described in Eq. (\ref{theta2v}).

For the scalar part of the power spectrum, one requires the components $\Theta_{00}$ and $\Theta^S$. In this situation, it is not possible to diagonalise each of the UETC matrices corresponding to $\langle \Theta_{00}\Theta_{00} \rangle$ and $\langle \Theta^S\Theta^S \rangle$ separately because the cross-correlator $\langle \Theta_{00}\Theta^S \rangle$ is nonzero. One has to build the block matrix
\begin{equation}
\begin{pmatrix}
\langle \Theta_{00}\Theta_{00} \rangle & \langle \Theta_{00}\Theta^S \rangle \\ 
\langle \Theta^S\Theta_{00} \rangle & \langle \Theta^S\Theta^S \rangle 
\end{pmatrix}
\end{equation}
and to diagonalize it. The first half of each of the eigenvectors would correspond to $\Theta_{00}$, and the second half would correspond to $\Theta^S$. The eigenvalues are common to both.

In the case of vectors and tensors, the situation is different. The two vector modes  $\Theta^{V1}$ and $\Theta^{V2}$ evolve independently, but their autocorrelators are the same:
\begin{equation}
\langle \Theta^V\Theta^V \rangle:=\langle \Theta^{V1}\Theta^{V1} \rangle=\langle \Theta^{V2}\Theta^{V2} \rangle
\end{equation}
and their cross-correlators vanish $\langle \Theta^{V1}\Theta^{V2} \rangle=0$, due to statistical isotropy. The same is true for the two tensor modes. Furthermore, the correlators between a vector and a tensor mode also vanish. We will discuss the results that we obtained using this method in Sec. \ref{eigendecomp}.

We have used the decomposition section of the Landriau and Shellard code \cite{landriau2003} to calculate the energy-momentum components of the UETCs. The energy-momentum tensor of the string network has been interpolated on a 3D grid in Fourier space, and it has been decomposed into scalar, vector and tensor parts. The relevant UETCs described in the previous paragraphs were then calculated.

\section{Evolution of the UETCs and resolution effects}

\begin{figure*}[!htb]
\begin{center}$
\begin{array}{ccc}
\includegraphics[width=2.25in]{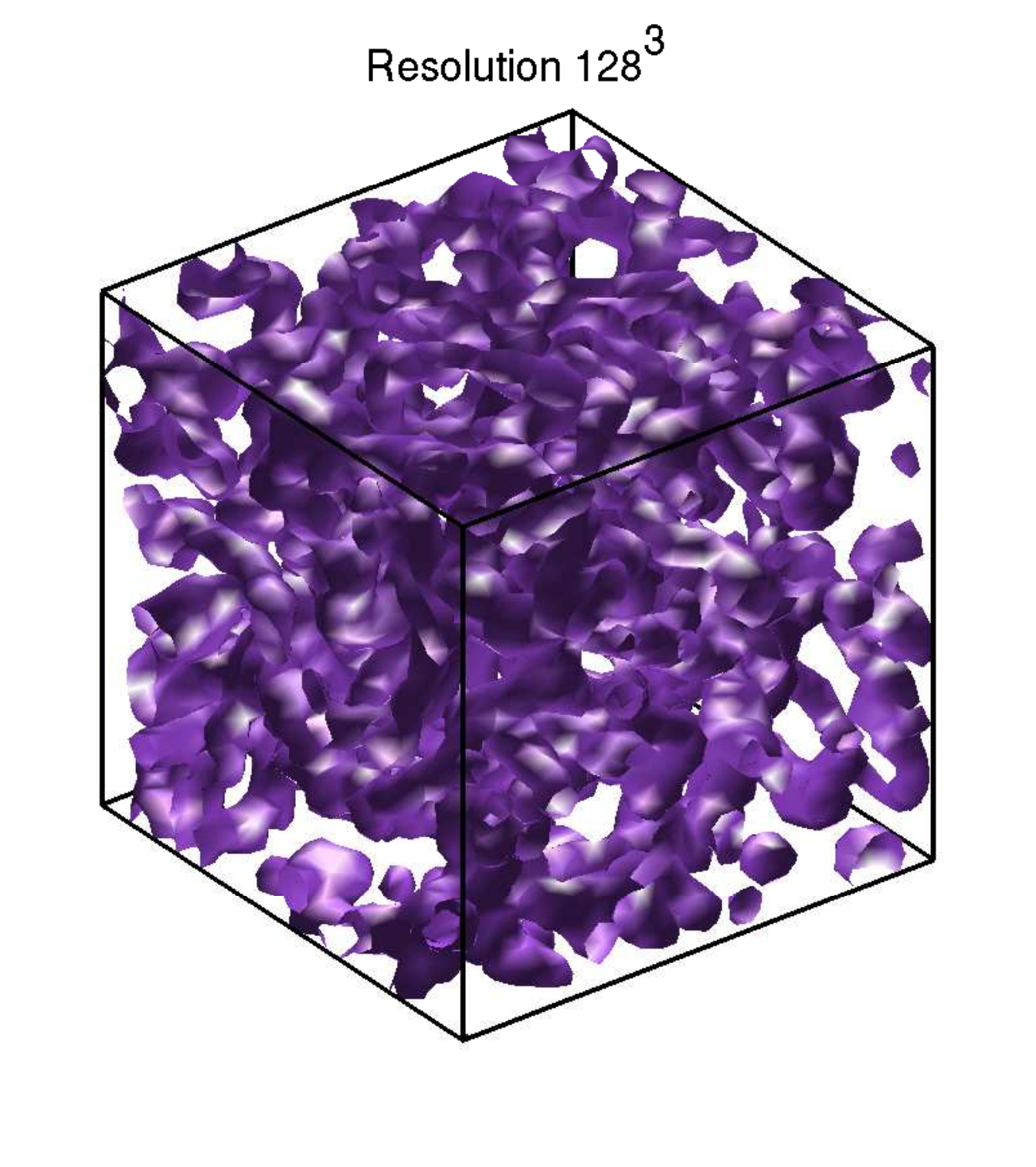} &
\includegraphics[width=2.25in]{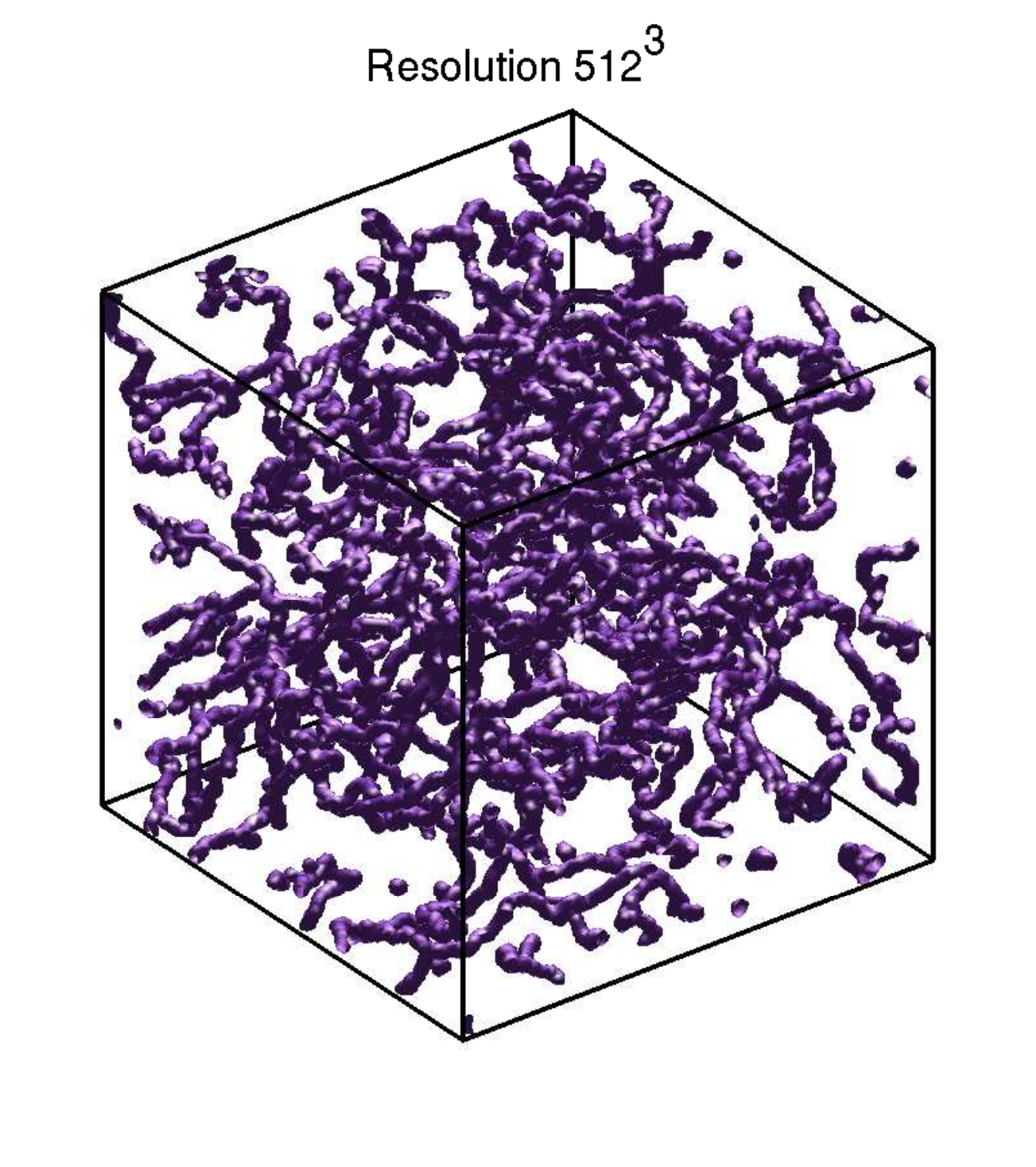} &
\includegraphics[width=2.25in]{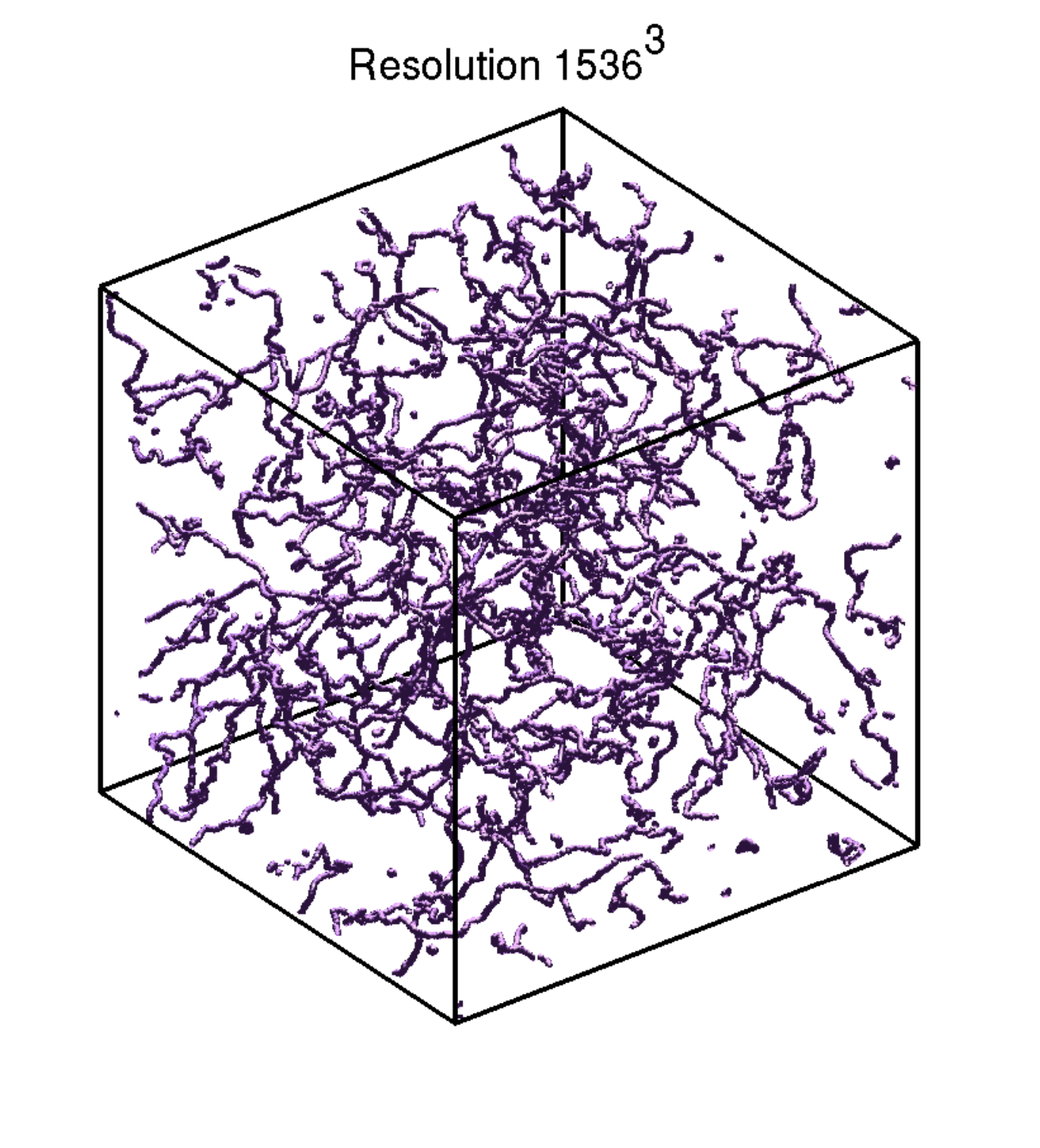}
\end{array}$
\caption{Energy density component of the string network in real space evaluated at time 384 out of 1536 for the simulation in the radiation era for resolutions of $128^3$, $512^3$, and $1536^3$.}
\label{00-144}
\end{center}
\end{figure*}

The most important aspect when calculating the UETCs is to make sure that the resolution considered is high enough so that it can capture all the physical scales of relevance for sourcing the main CMB signal. A first step in order to achieve this was to analyse the energy density of the string network in real space at a given time for a range of grid resolutions. Boxes of $128^3$, $256^3$, $512^3$, $768^3$, $1024^3$, and $1536^3$, respectively, points have been chosen. In Fig. \ref{00-144} the energy density of the string network for time 384 out of 1536 for the simulation in the radiation era has been plotted for the resolutions of $128^3$, $512^3$, and $1536^3$ .

For the lowest resolution, important information is smoothed out, and the strings do not have a threadlike appearance. As the resolution is increased, the strings become thinner as one would expect with better grid sampling. However, one cannot increase the resolution indefinitely because, after getting in the vicinity of the resolution of the simulation itself, the network would appearas made up of disconnected bulbs. The effect of resolution on a string network is especially apparent at earlier times (as shown in Fig. \ref{00-144}), when the string density is much higher. However, when one is interested in ray tracing through the simulation, e.g.  to compute CMB maps, the difference in resolution does not affect the results at early times because of the very high string density but will cause the late-time features to have increasing levels of sharpness; however, as we shall now show, adequate resolution is critical for the accurate computation of UETCs.

Even though in recent years the computational capacity has radically increased, it is still challenging to go to very high resolutions in simulations. Increasing by a factor of 2 the linear grid resolution increases each file size by a factor of 8 and the time required by a similar amount. Due to these time and disk space considerations, we chose to use a grid size of $1024^3$ for the simulations. The huge grid size limits, however, our possibility of using a very high time resolution as well, and for each of the simulations, we use around 100 time steps. We have checked that the time sampling does not modify the UETCs noticeably. To ensure the symmetry of the UETCs, we are using the same sampling for $\tau_1$ and $\tau_2$ for the computations.

\begin{figure*}[!htb]
\begin{center}$
\begin{array}{cc}
\includegraphics[width=2.75in]{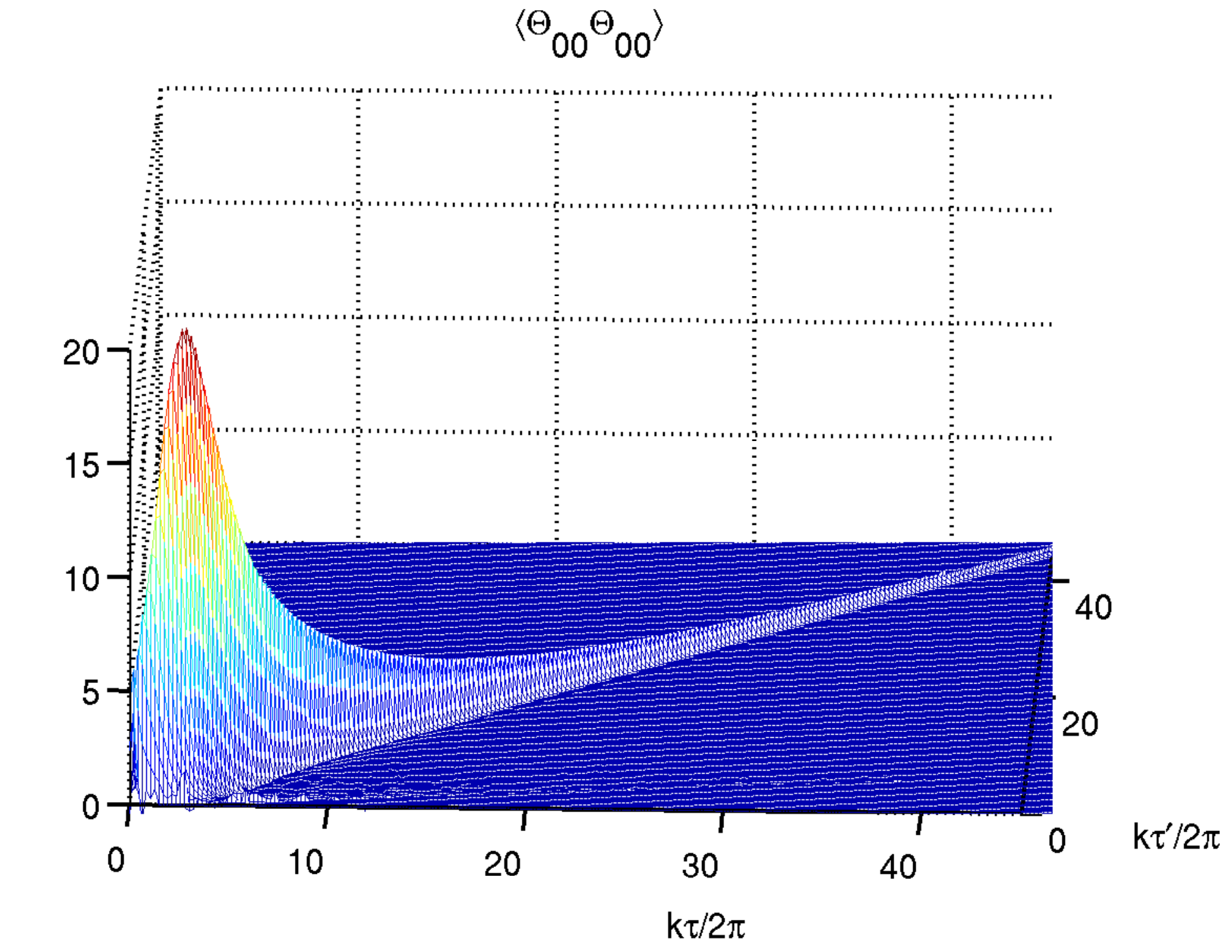} &
\includegraphics[width=2.75in]{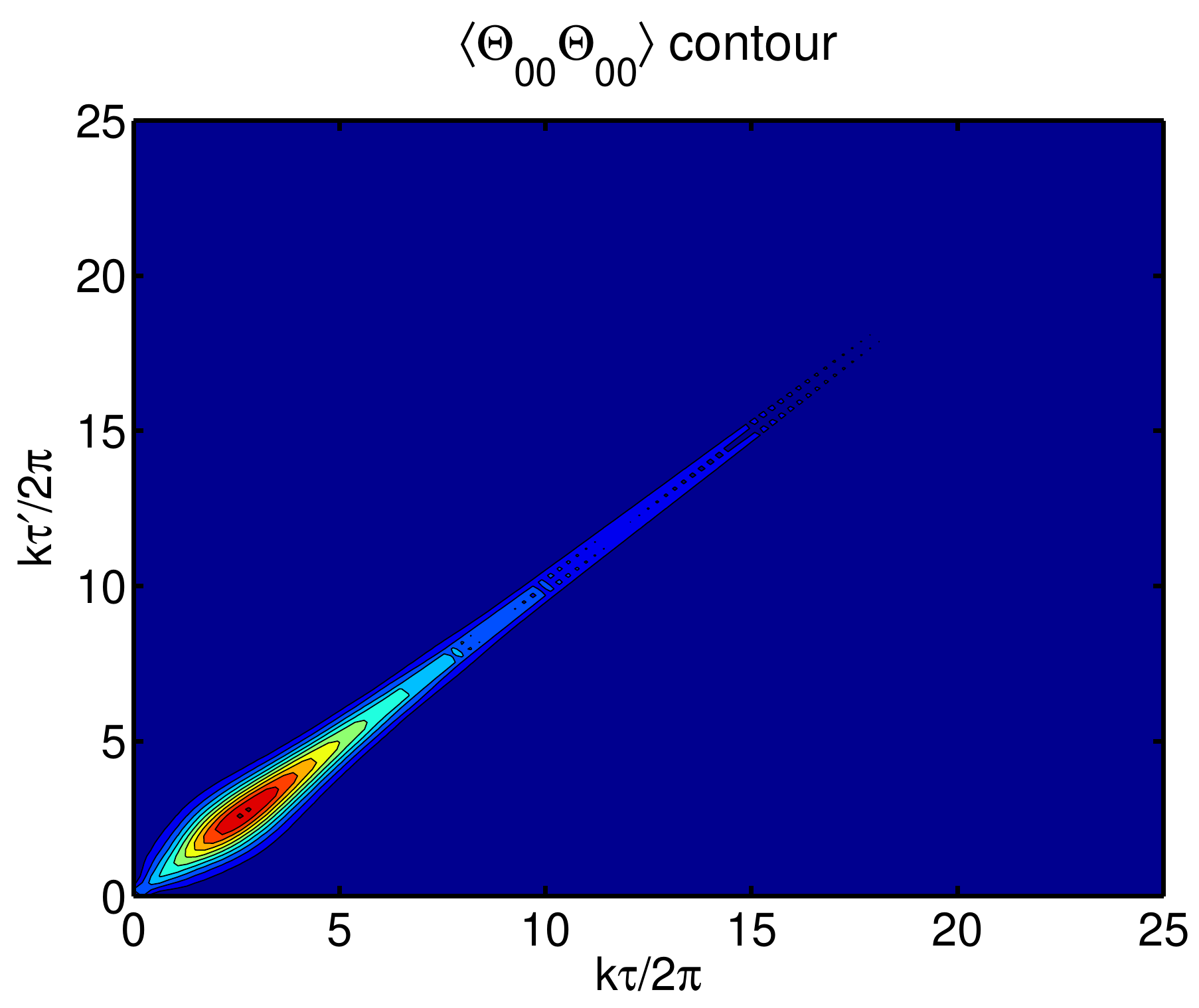} \\
\includegraphics[width=2.75in]{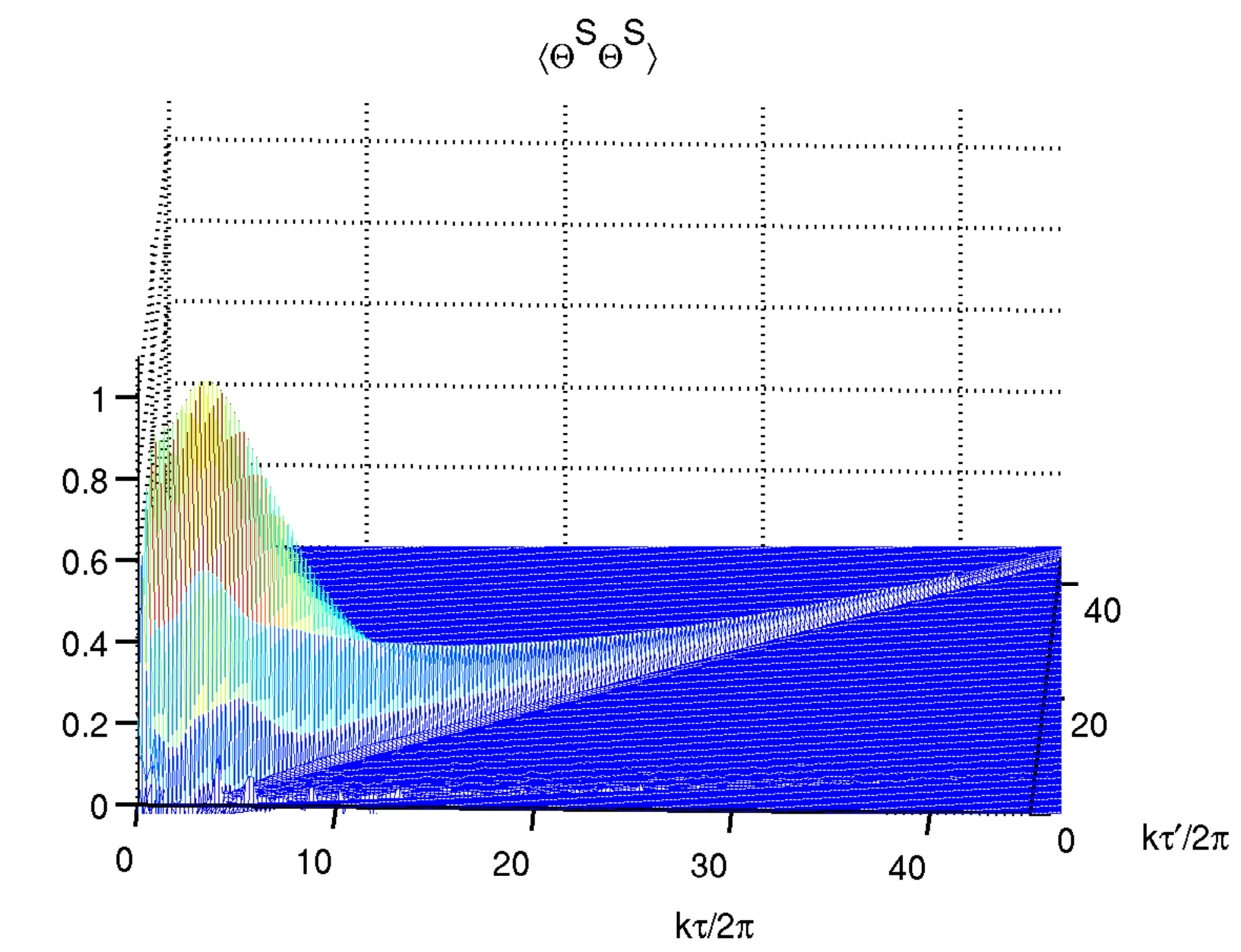} &
\includegraphics[width=2.75in]{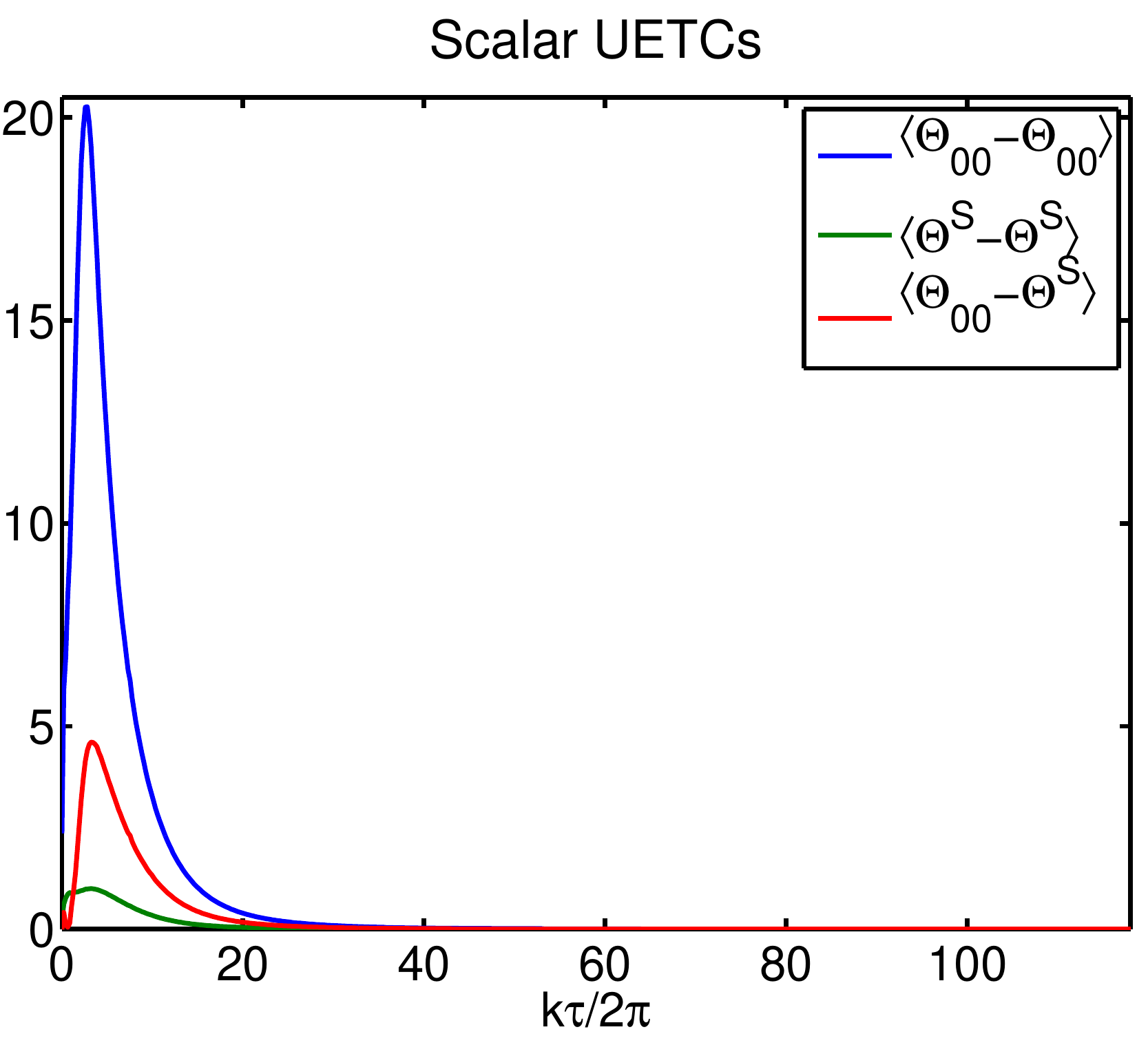} \\
\includegraphics[width=2.75in]{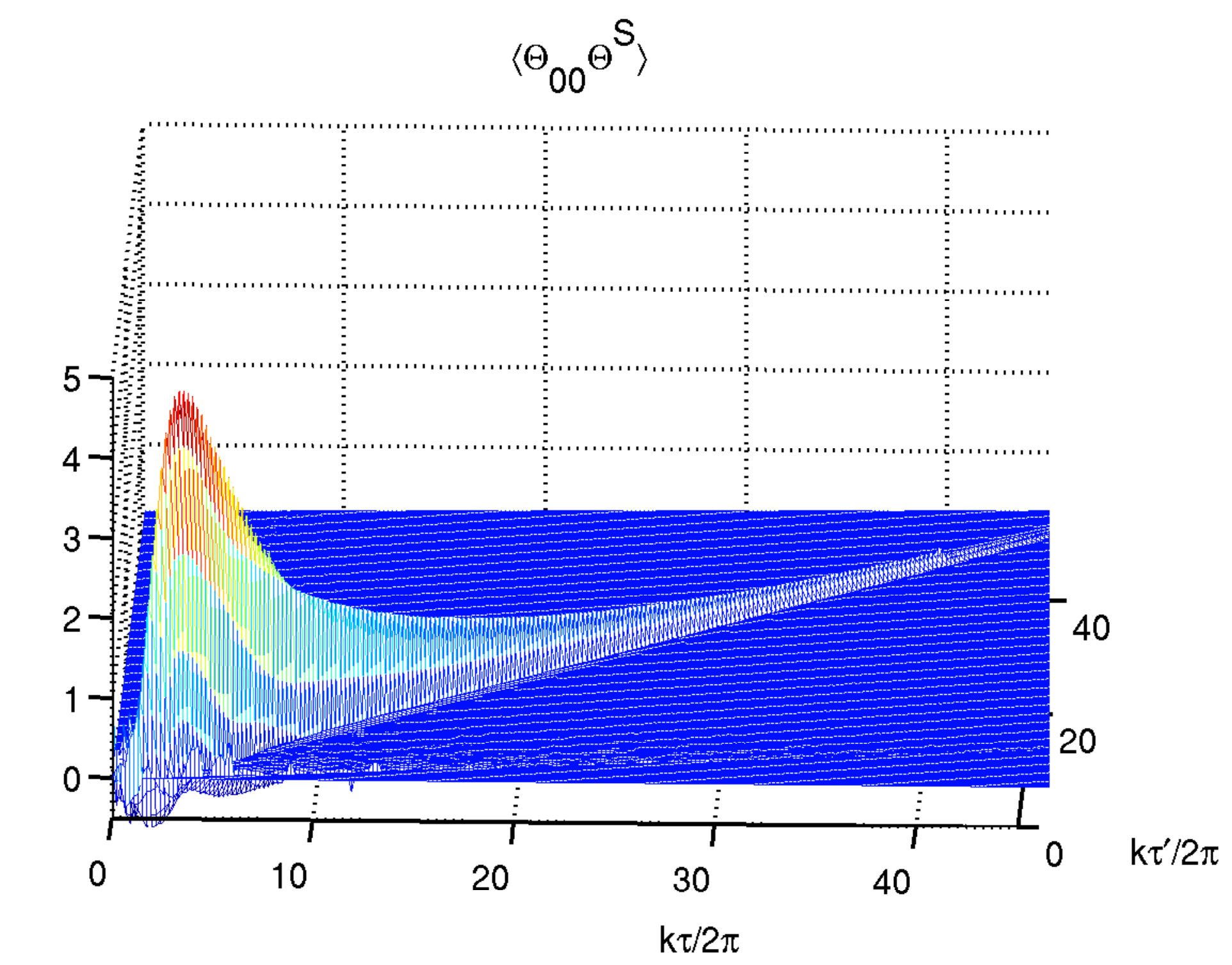} &
\includegraphics[width=2.75in]{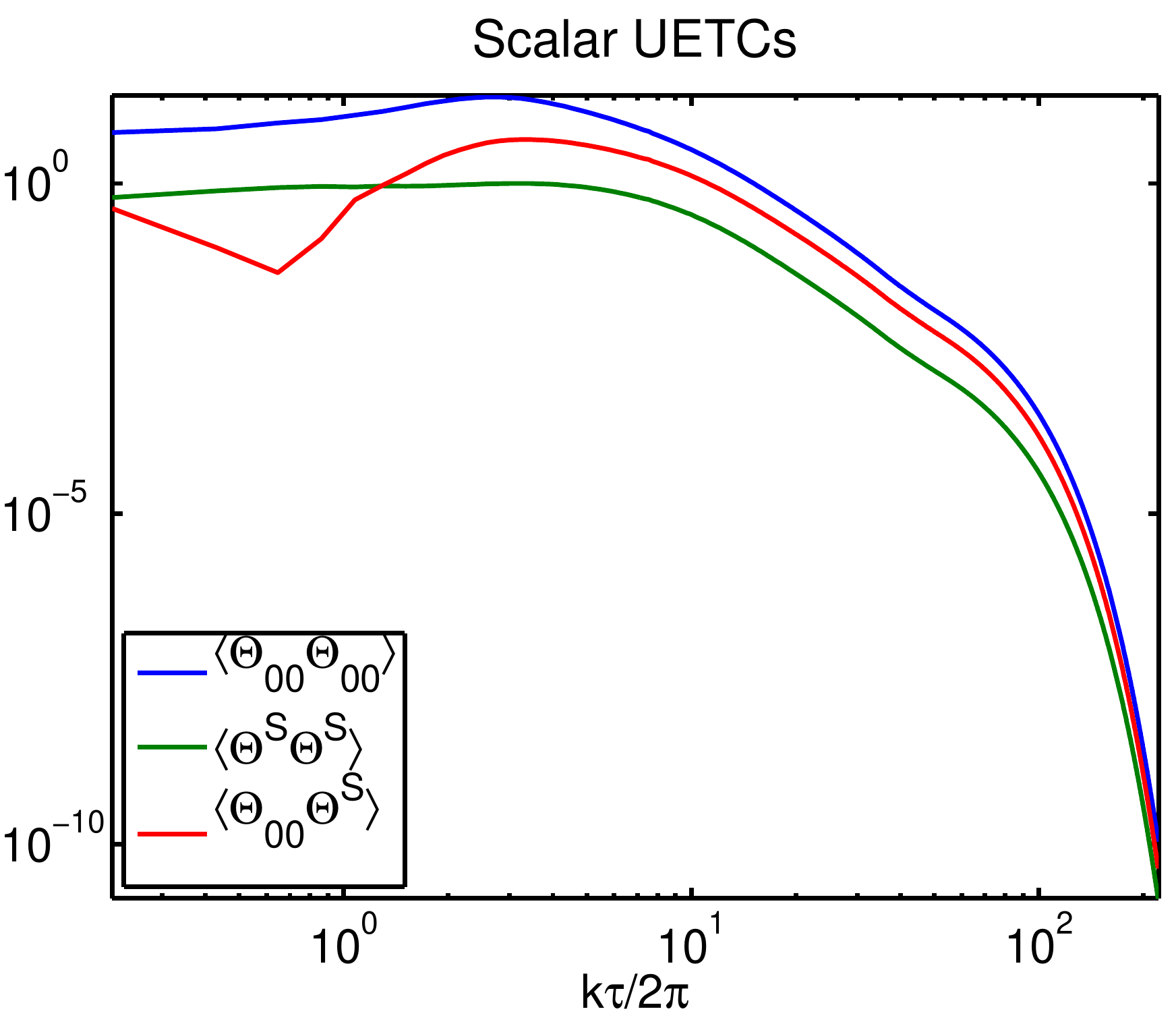} 
\end{array}$
\caption{Scalar UETCs obtained from a grid resolution of 1024: the figures of the left represent oblique 3D views of the three scalar UETCs ($\langle\Theta_{00}\Theta_{00}\rangle$ - top, $\langle\Theta^S\Theta^S\rangle$ - middle and $\langle\Theta_{00}\Theta^S\rangle$ - bottom), the top right plot represents a contour plot of the 00-00 UETC in linear scale and two bottom right plots represent the three scalar UETCs in linear and logarithmic scales}
\label{UETC-scalars}
\end{center}
\end{figure*}

\begin{figure*}[!htb]
\begin{center}$
\begin{array}{cc}
\includegraphics[width=2.55in]{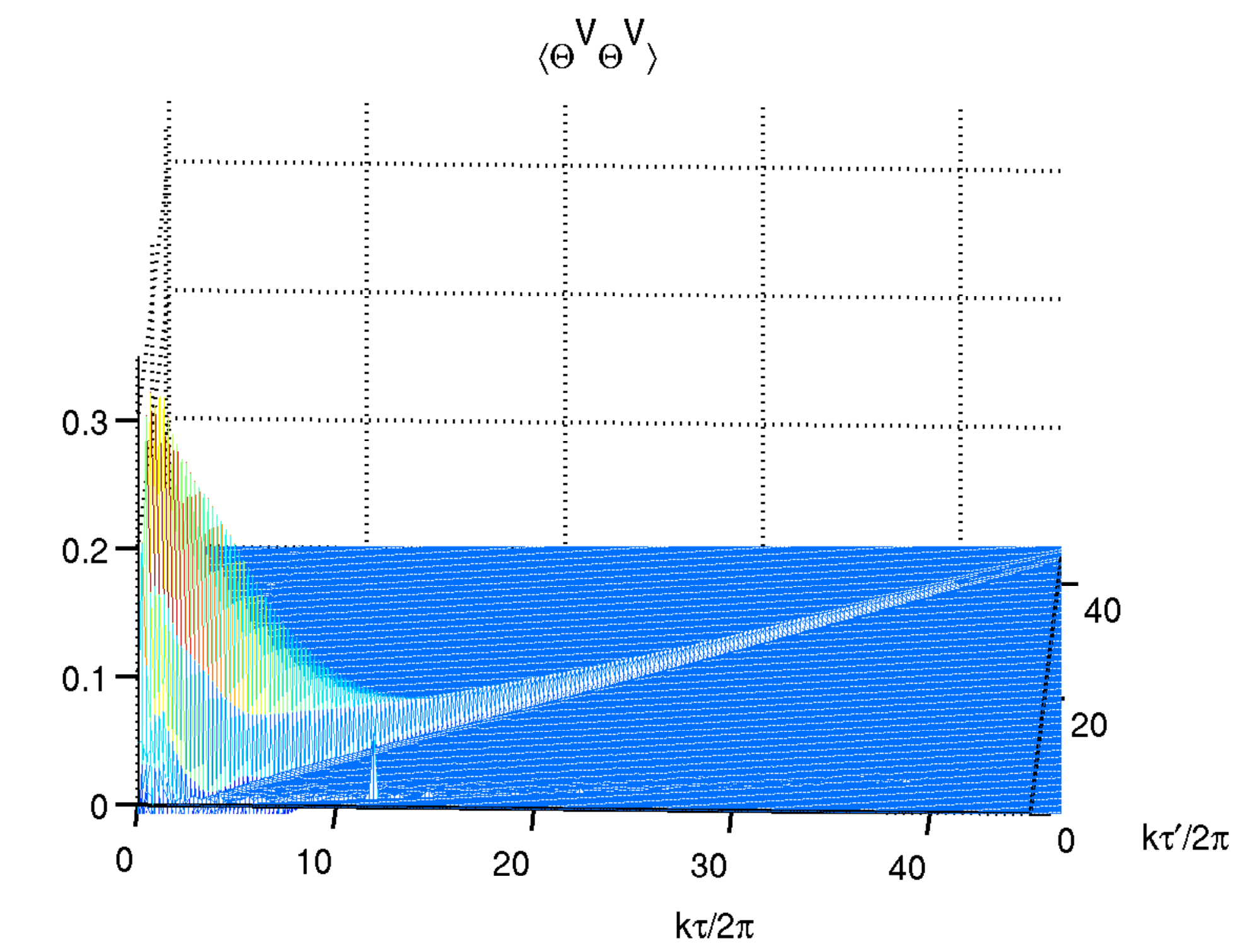} &
\includegraphics[width=2.55in]{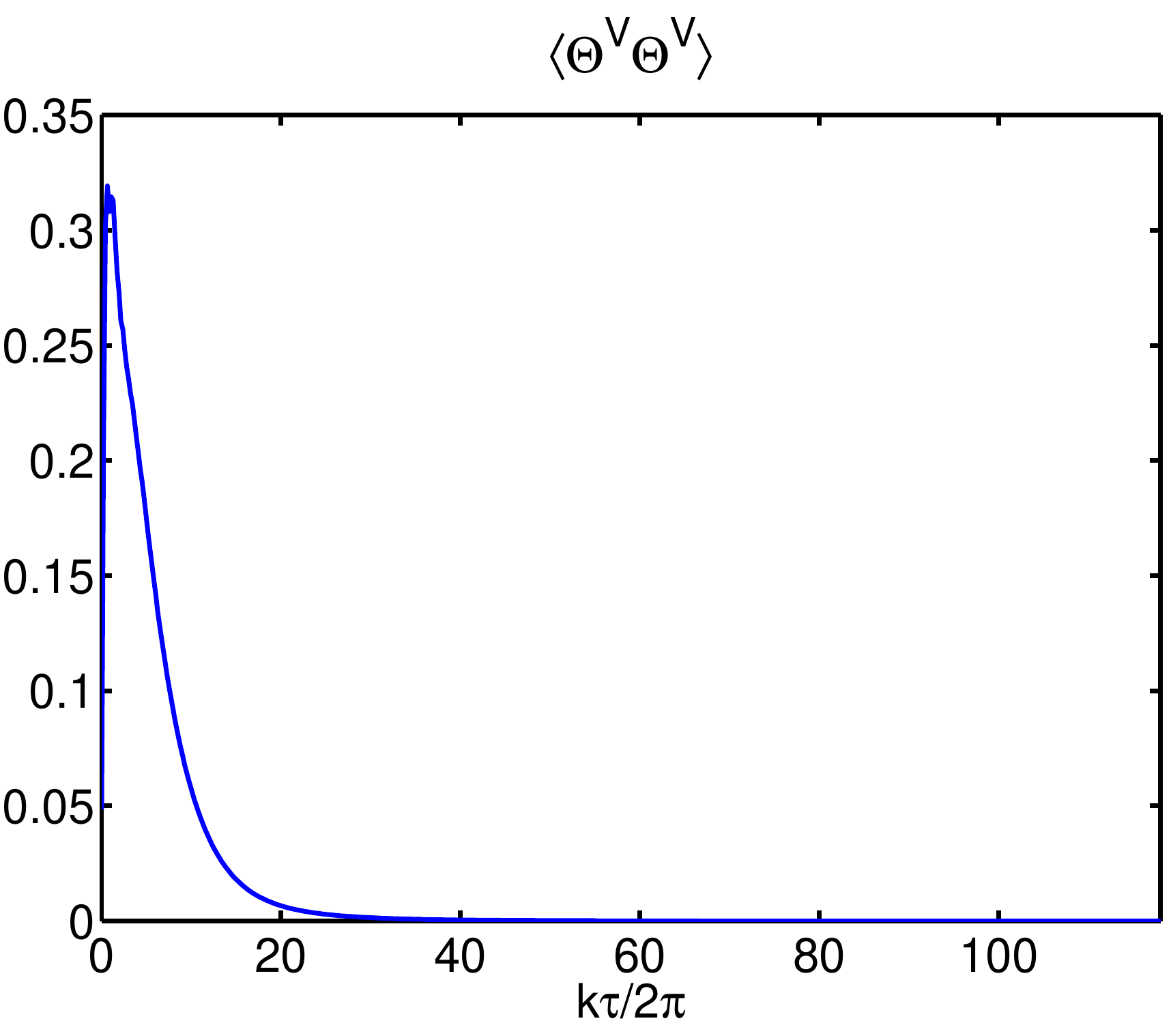} \\
\includegraphics[width=2.55in]{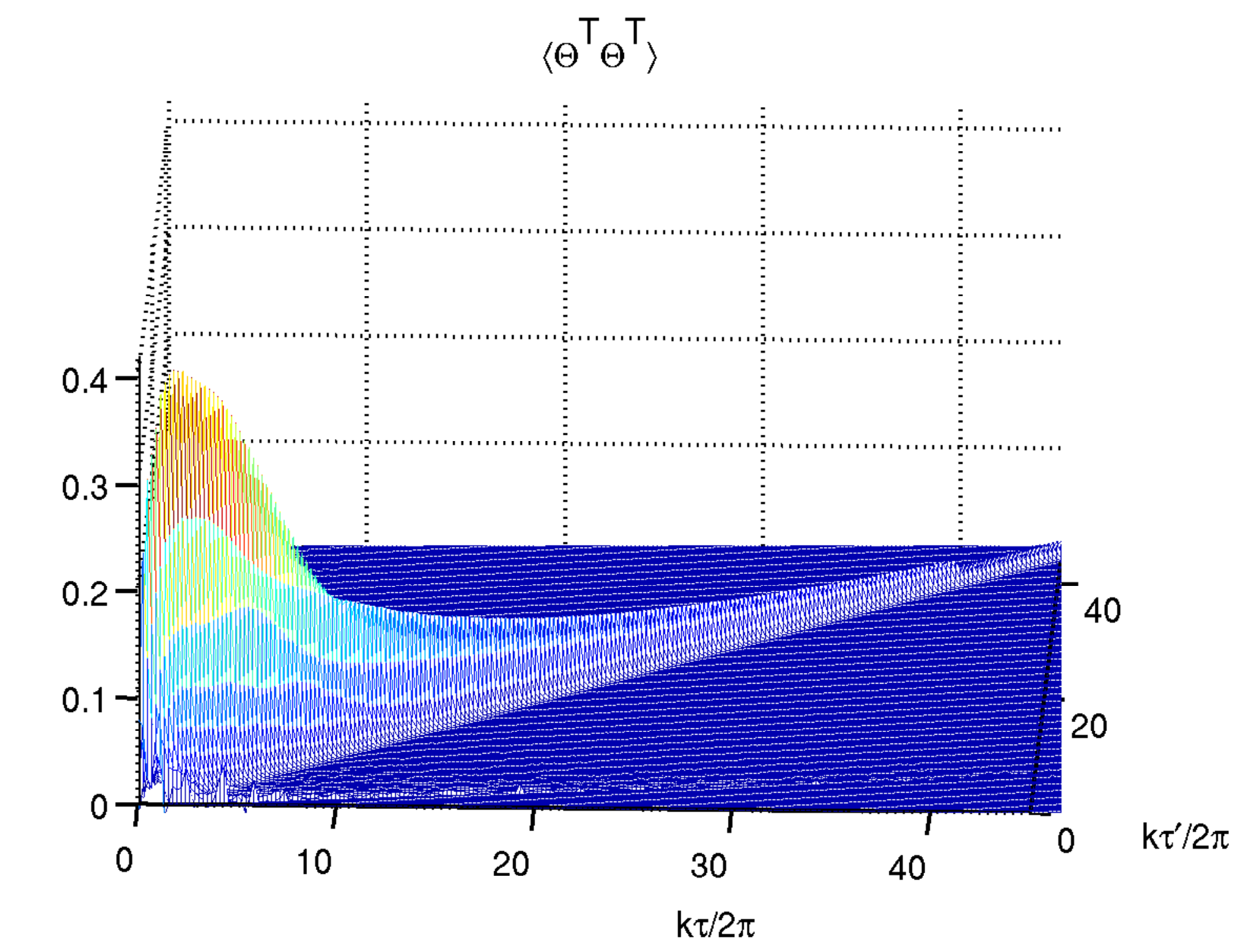} &
\includegraphics[width=2.55in]{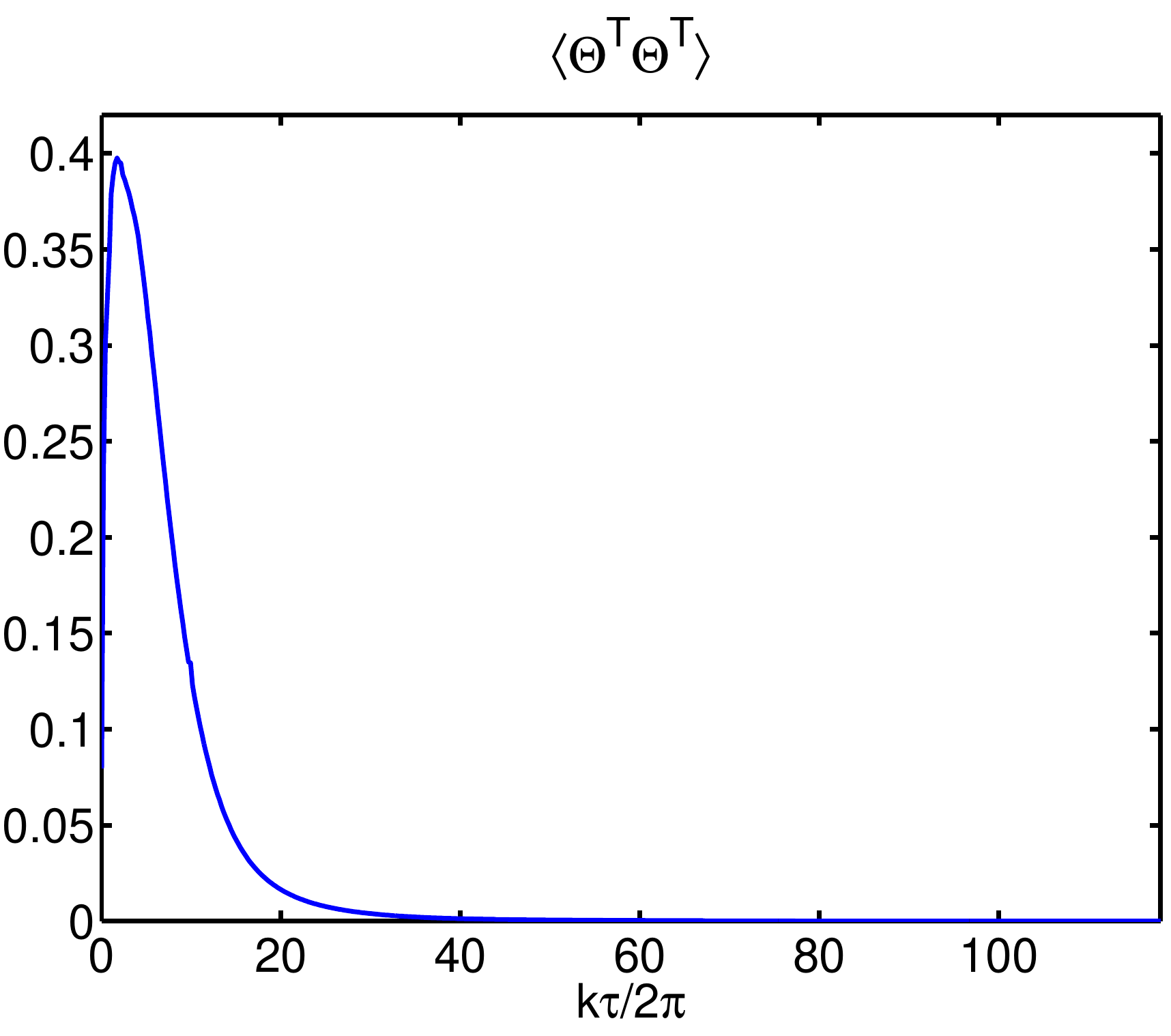} 
\end{array}$
\caption{Vector and tensor UETC components obtained from a grid resolution of 1024: oblique 3D views (left) and diagonal sections in linear scale (right)}
\label{UETCs}
\end{center}
\end{figure*}

An alternative approach is being developed \cite{landriau-later}, which uses a lower spatial resolution but a greater time resolution. To obtain the full UETCs at this resolution a total CPU time of approximately 20000 h is required using 200 Intel Xeon processors with a clock speed of 2.6GHz. We have performed all the calculations on the COSMOS supercomputer.
Typical UETCs obtained at resolution of $1024^3$ are plotted in Figs. \ref{UETC-scalars} and \ref{UETCs} from the simulation covering the matter era.

\begin{figure}[!htb]
\begin{center}$
\begin{array}{c}
\includegraphics[width=2.5in]{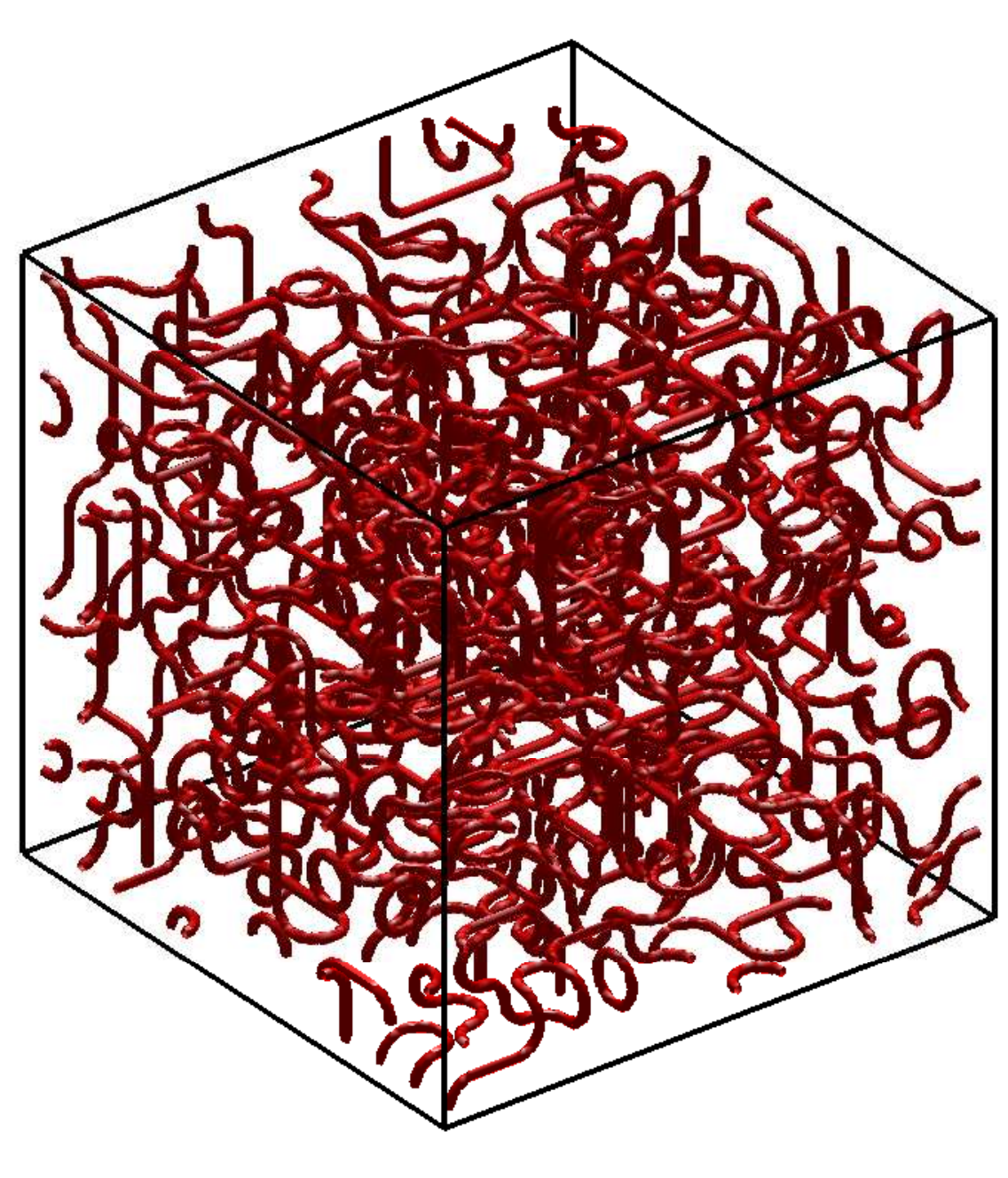} \\ 
\includegraphics[width=2.5in]{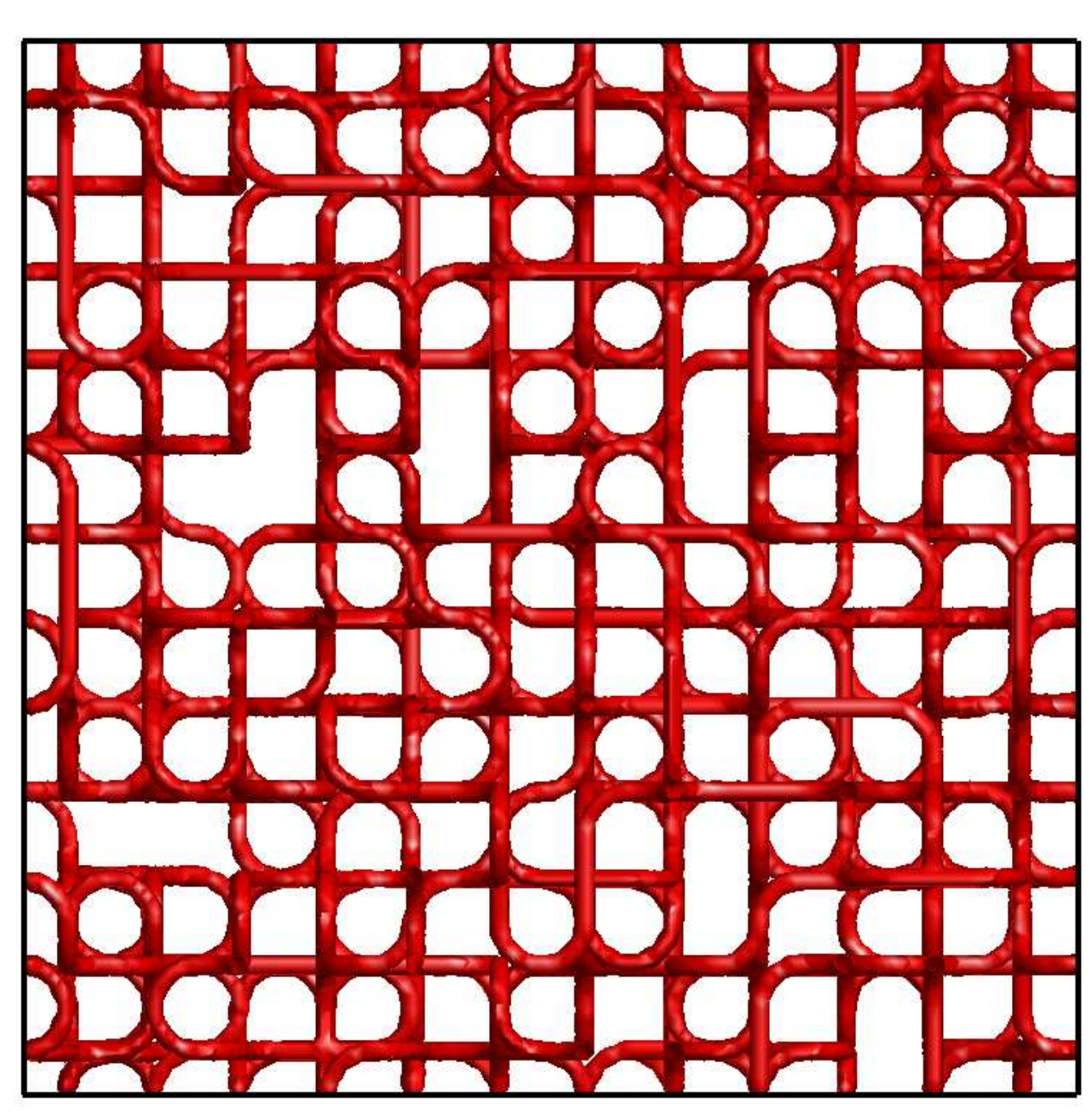} 
\end{array}$
\caption{Network correlation in the initial conditions from the simulation covering the matter era: left - oblique 3D view, right - front view}
\label{initcond}
\end{center}
\end{figure}

At resolutions greater than or equal to $512^3$, spurious peaks appear in the UETCs if the first 1\% of the  time steps of the simulation is considered. This is due to the appearance of loops over the length scale of the resolution size, \textit{i.e.} excessive correlation in the Vachaspati-Vilenkin initial conditions. In Fig. \ref{initcond}, we have represented the initial appearance of the string network, both as a 3D view and a projection of the energy density to illustrate the correlation between the segments forming the string network.  To get accurate predictions for the UETCs, the first time steps should be discarded, as they represent only the effect of the initial conditions and not of the physics involved.

\begin{figure*}[!htb]
\begin{center}$
\begin{array}{cc}
\includegraphics[width=2.75in]{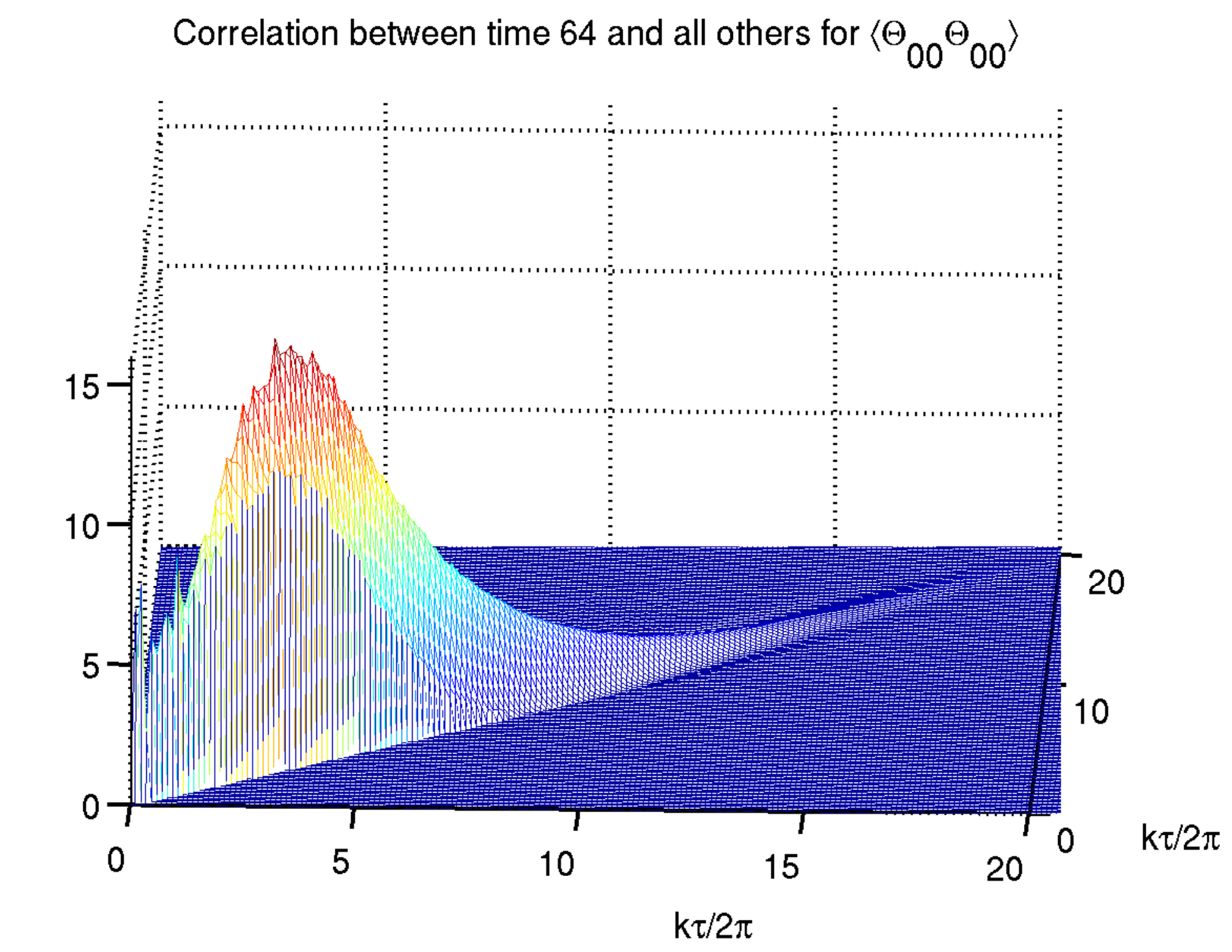} &
\includegraphics[width=2.75in]{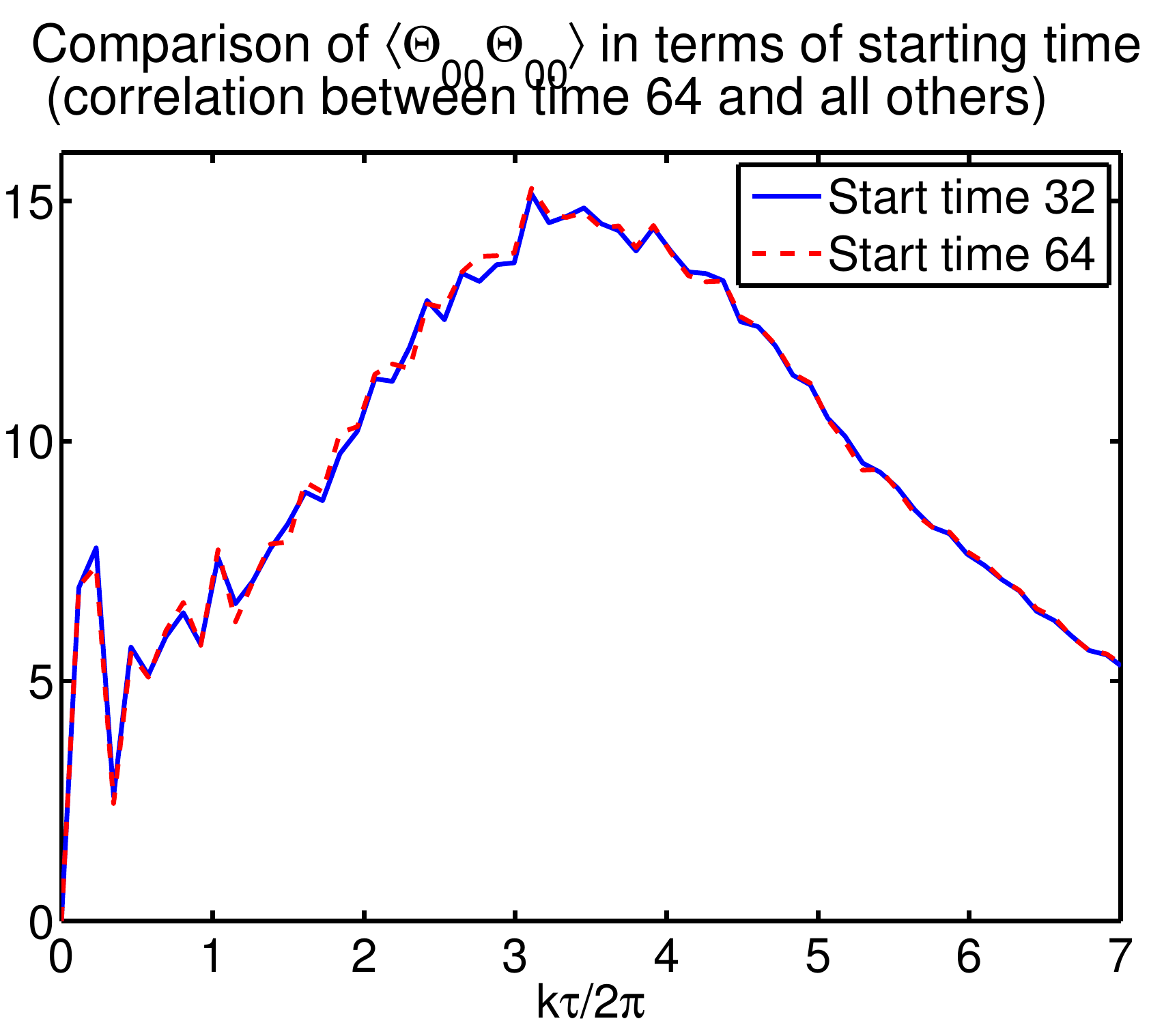} \\
\includegraphics[width=2.75in]{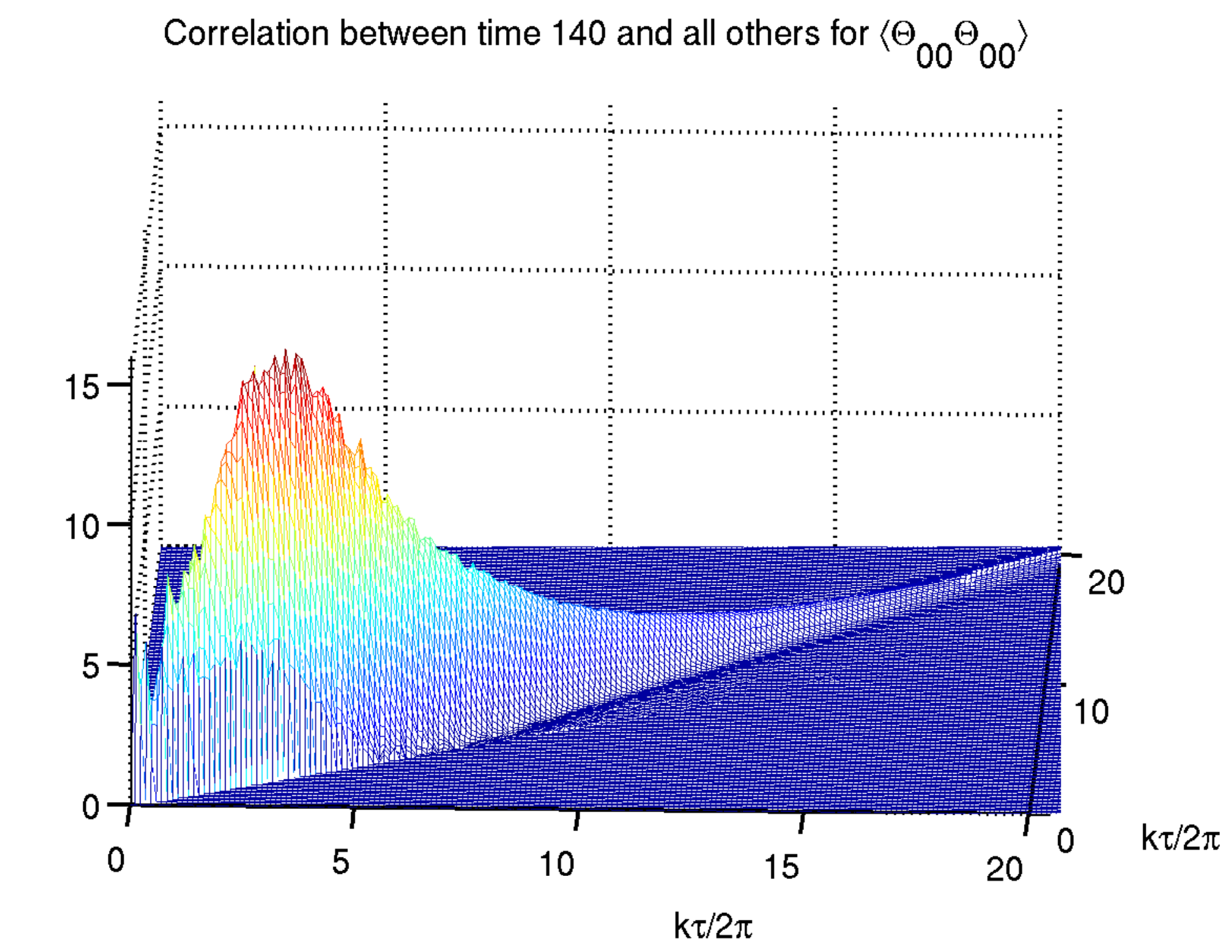} &
\includegraphics[width=2.75in]{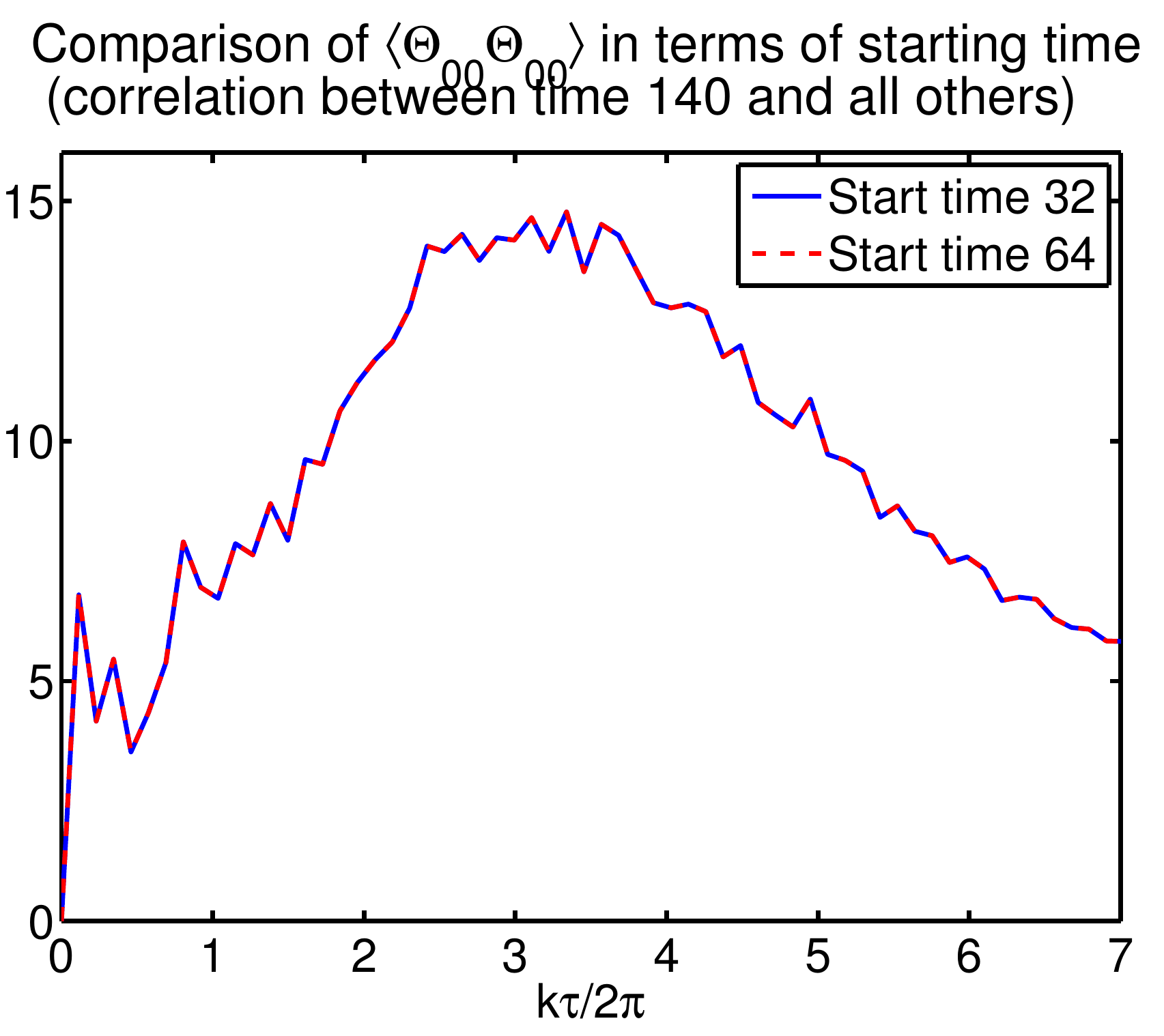} \\
\includegraphics[width=2.75in]{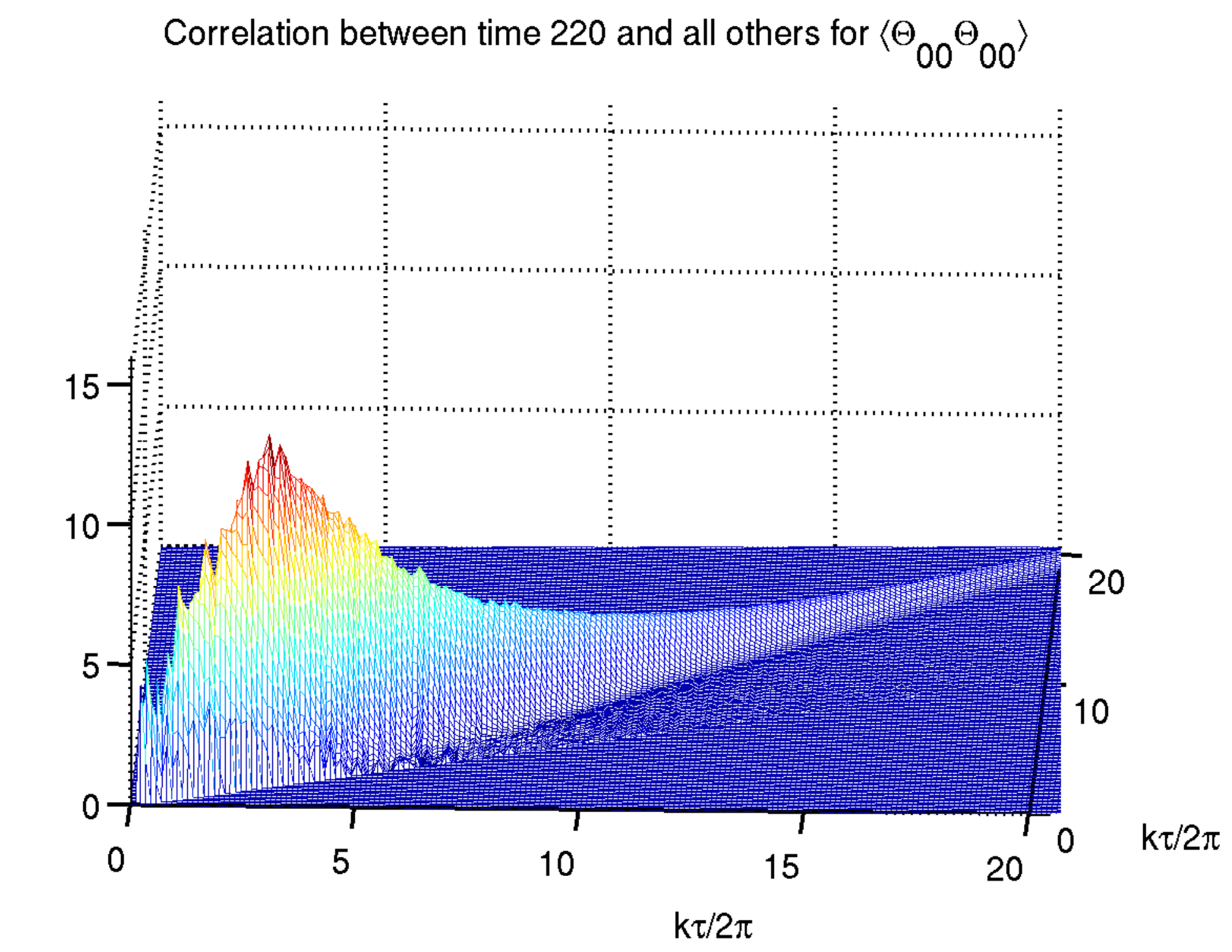} &
\includegraphics[width=2.75in]{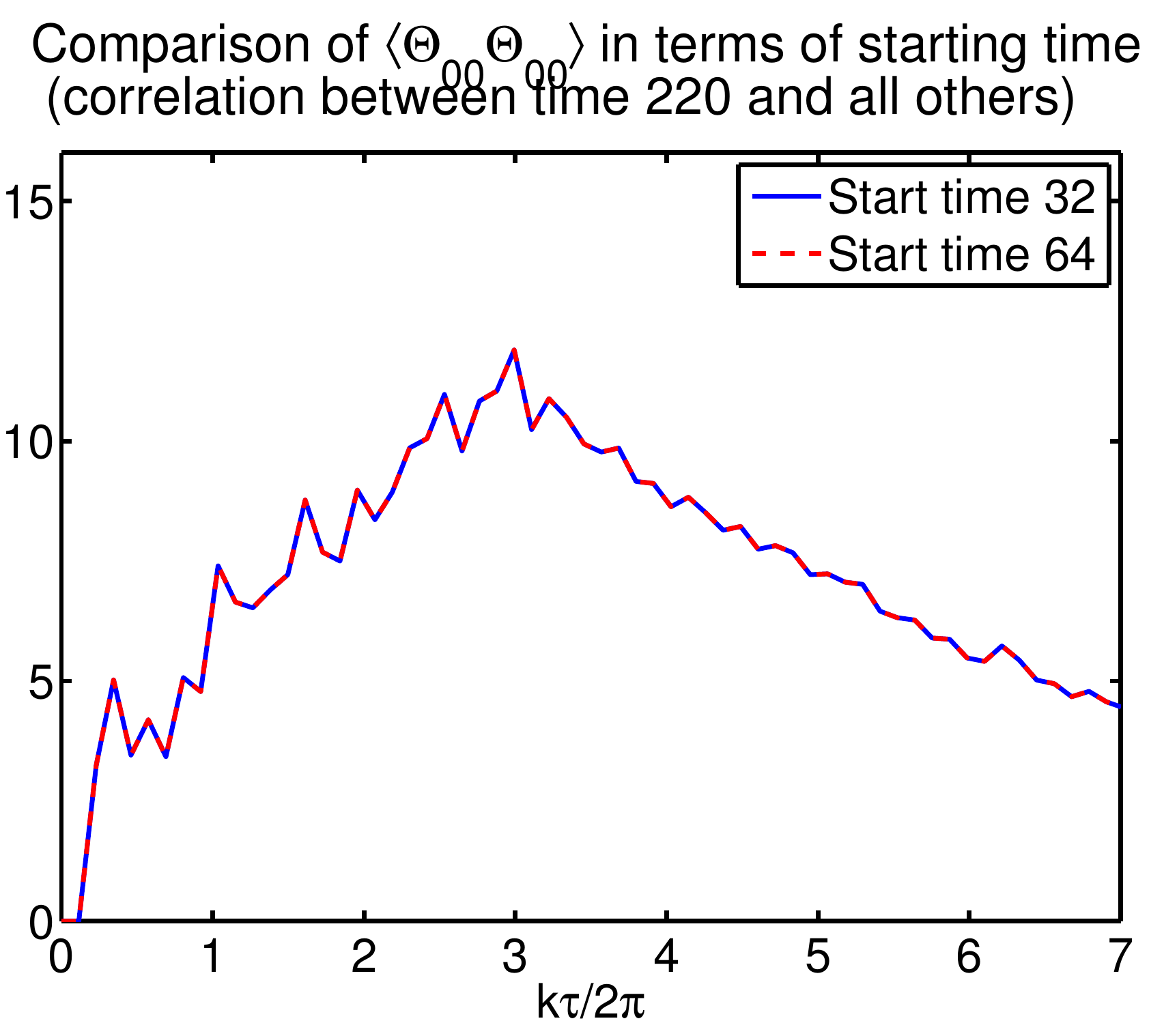}
\end{array}$
\caption{$\langle\Theta_{00}\Theta_{00}\rangle$ UETCs exhibiting scale invariance. In the bottom two rhs plots, the two plotted curves are indistinguishable. Scaling can be observed between the figures despite the correlation time used.}
\label{UETC-scaling}
\end{center}
\end{figure*}

Another important feature that needs to be checked is the scale invariance of the UETCs. This can be checked by verifying whether the shape of the UETC depends on which part of the simulation is used (after discarding the initial conditions). The UETCs in Figs. \ref{UETC-scalars}-\ref{UETCs} are scale-invariant. They are almost independent of the starting time of the simulation. We have illustrated this behaviour by plotting the $\langle\Theta_{00}\Theta_{00}\rangle$ UETC between three times (64, 140, and 220) and all times between 32 and 223 and 64 and 223 (Fig. \ref{UETC-scaling}). The plots have been zoomed in around the peak in order to show the scale invariance. In the case of the 3D plot, the differences in terms of starting time are imperceptible, and hence only the one with the starting time 32 is represented. When correlating components of the energy-momentum tensor from early times with all the corresponding components from a certain time until the end of the simulation, there appears to be a small difference in the UETC corresponding to that starting point. If we choose, however, a later time to correlate with all the others, the difference becomes imperceptible. This is due to the fact that for earlier times there is more information in the string network due to the higher string density. As the Universe expands, the strings become less dense in the Universe. This can be seen in the fact that the correlators in Fig. \ref{UETC-scaling} have slightly lower amplitudes from top to bottom as the time used for correlations increases. Nevertheless, scale invariance is a good approximation just throughout each of the simulations; the string network is not scale-invariant throughout the history of the Universe, as the UETCs are not identical in the three simulations.

\subsection*{Resolution convergence} 

\begin{figure}[!htb]
\begin{center}$
\begin{array}{c}
\includegraphics[width=2.75in]{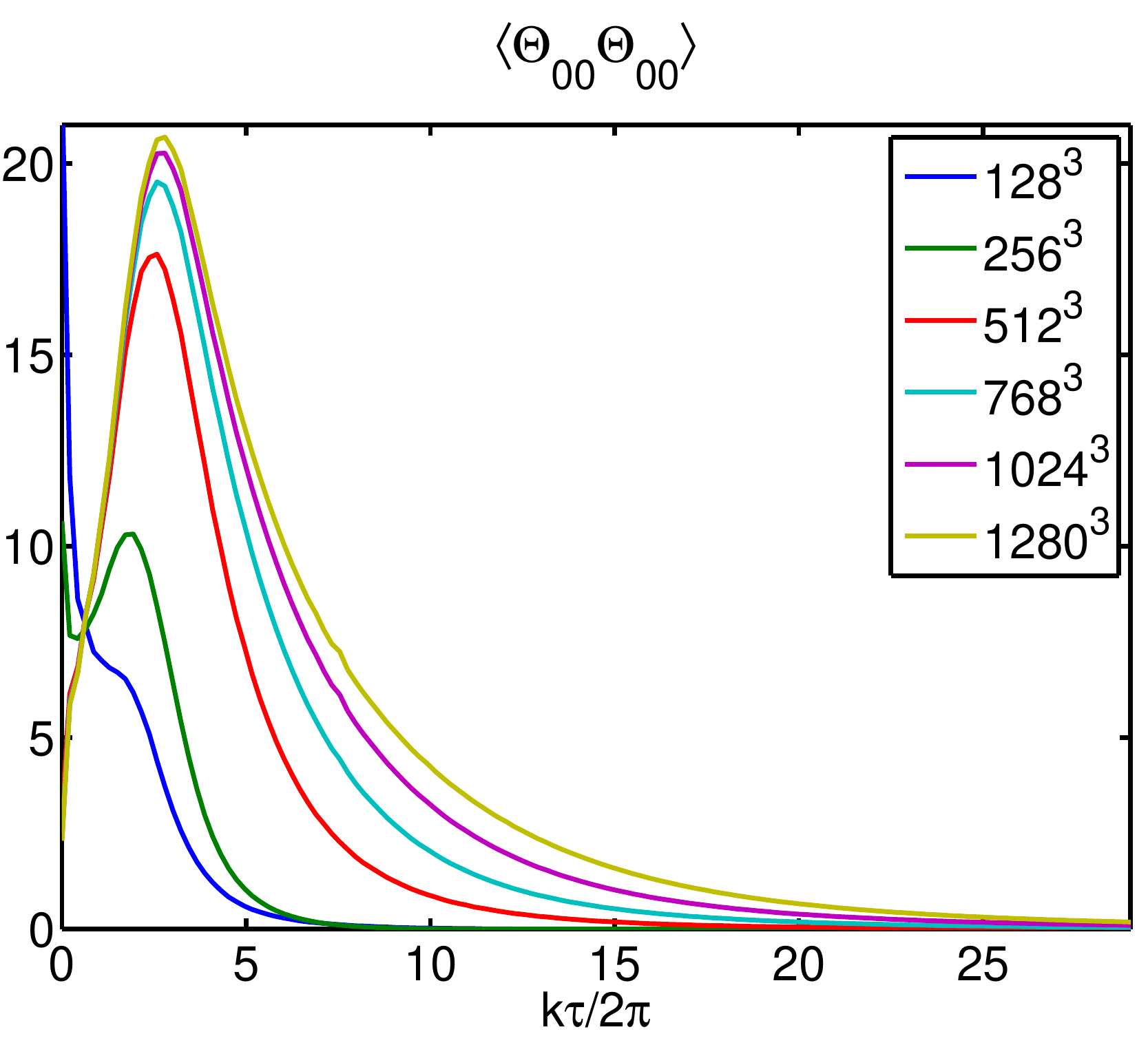} \\
\includegraphics[width=2.75in]{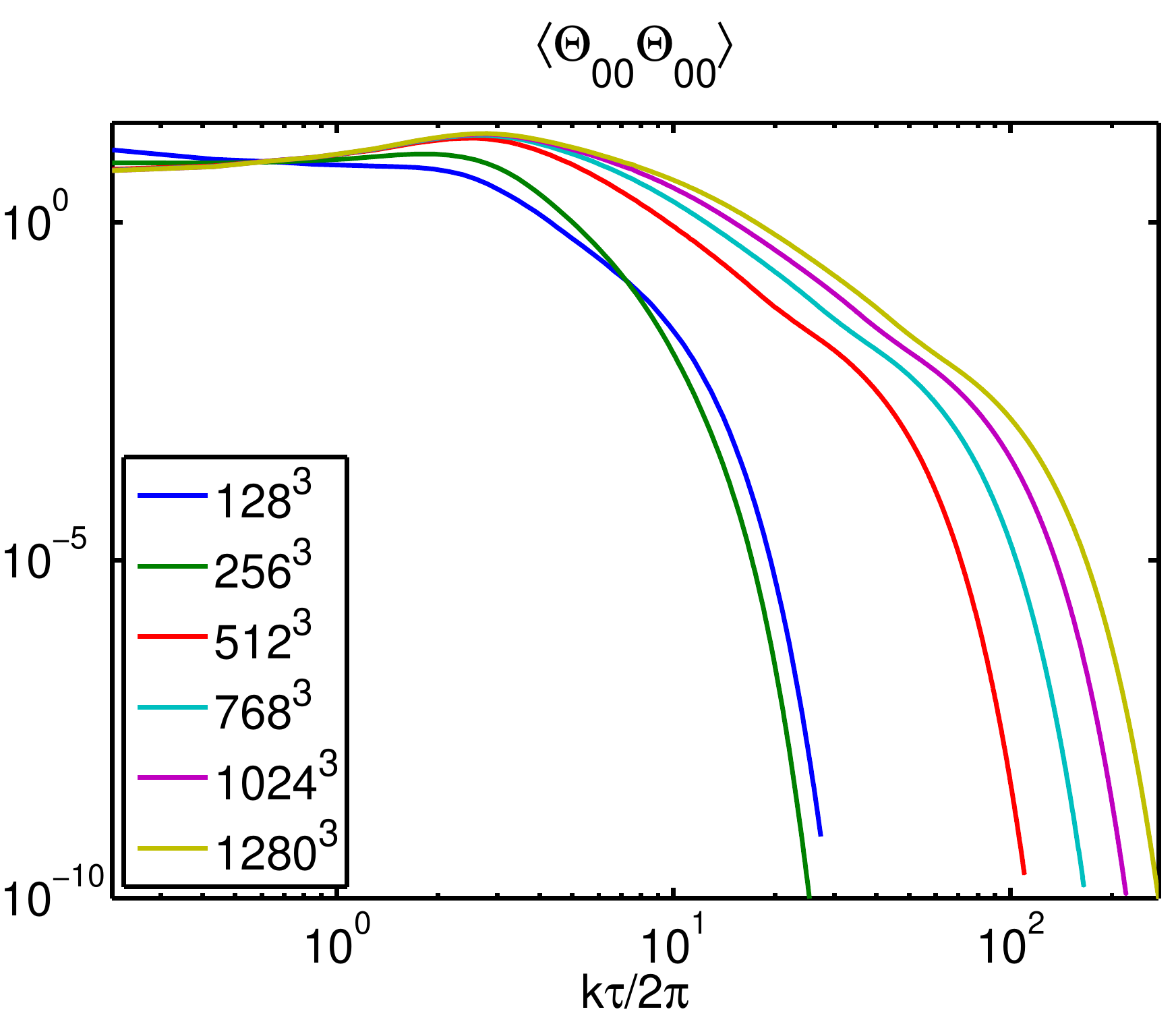}  
\end{array}$
\caption{$\langle\Theta_{00}\Theta_{00}\rangle$ equal time correlators at different resolutions}
\label{resolution}
\end{center}
\end{figure}

We have studied the convergence of both the shape and the amplitude of the UETCs in terms of resolution of the grid. To illustrate this, we have chosen the simulation in the matter era. In Fig. \ref{resolution}, we have plotted the equal time correlator (diagonal component of the UETC) of the energy density for the various resolutions considered, from $128^3$ until $1280^3$. The peak is still increasing as the resolution is increased, but one can observe that relative differences from consecutive resolutions are getting smaller. However, technical constraints do not allow us yet to increase the resolution further and get the results in a reasonable amount of time. Currently the full simulation at a resolution of 1280 takes around 40000 CPU hours on Intel Xeon processors with a clock speed of 2.6GHz on the COSMOS supercomputer.

\begin{figure}[!htb]
\begin{center}$
\begin{array}{c}
\includegraphics[width=2.75in]{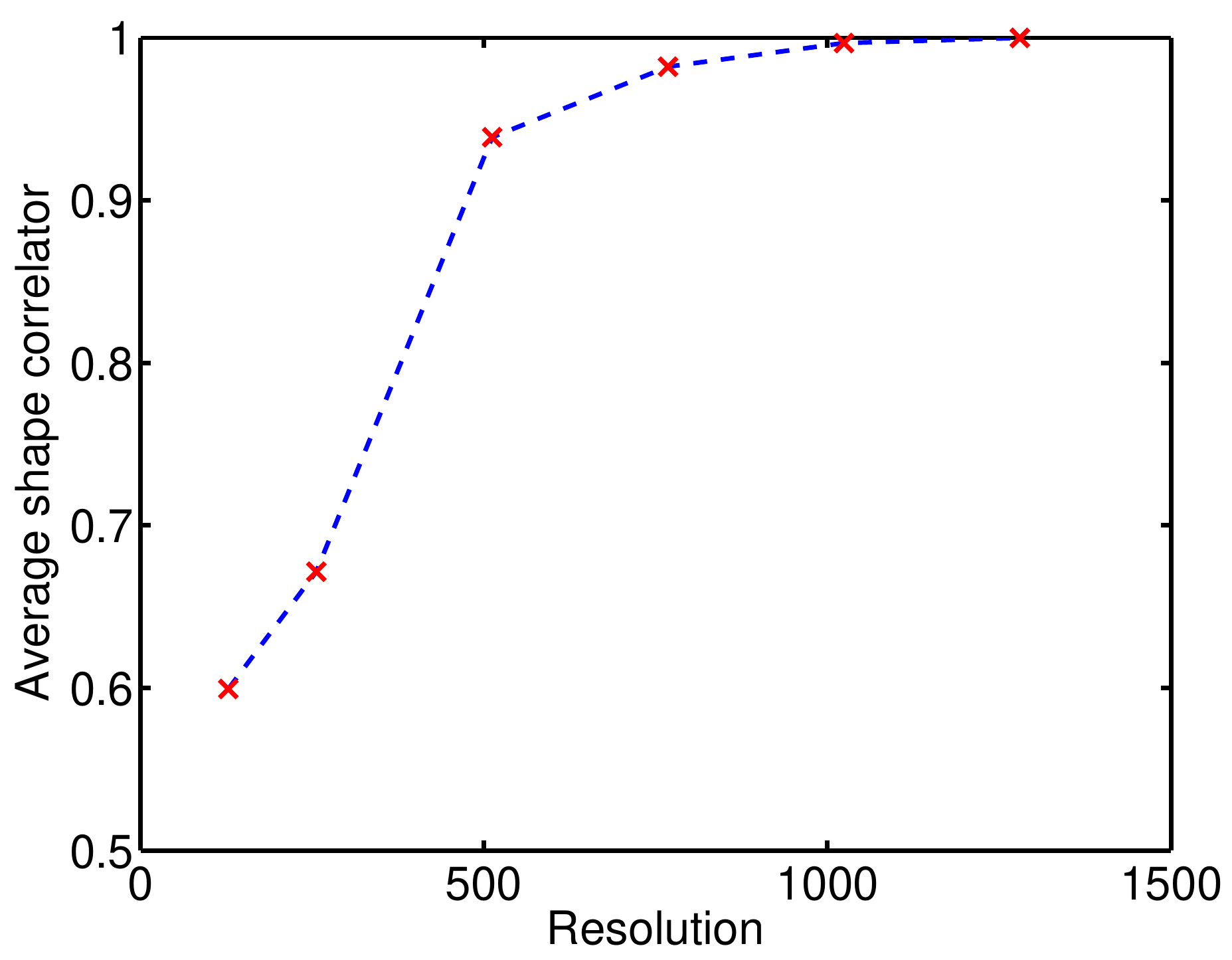} \\
\includegraphics[width=2.75in]{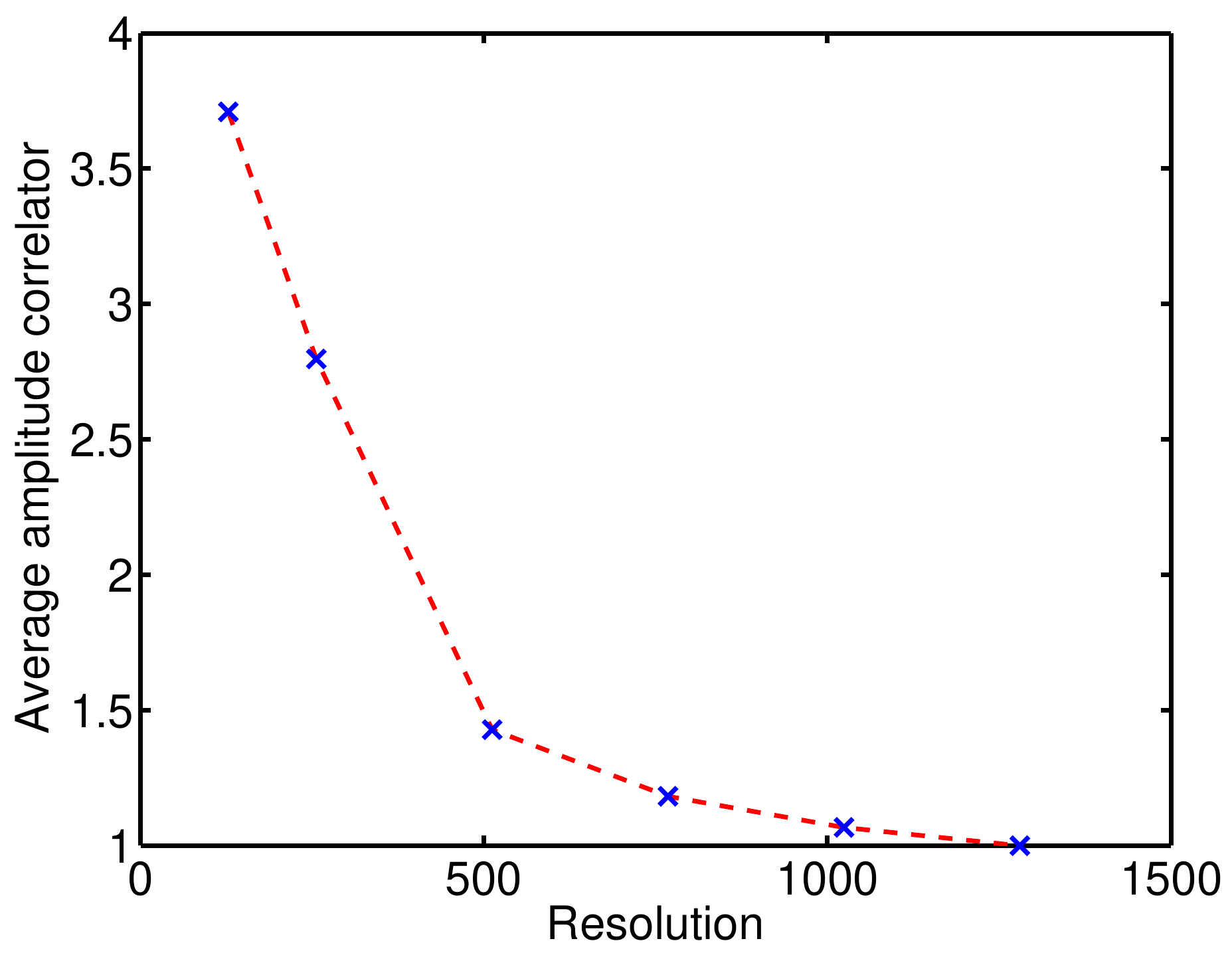}  
\end{array}$
\caption{Evolution of the averaged shape and amplitude correlators}
\label{shape_amp}
\end{center}
\end{figure}

From Fig. \ref{resolution} it can be seen that the two lowest resolutions do not give accurate results. This was expected since the string network is not properly resolved at this resolution (see Fig. \ref{00-144}). The behaviour of the other correlators that were calculated is similar and has not been plotted. 
We have used the UETCs obtained at resolutions of $128^3$, $256^3$, $512^3$, $768^3$, $1024^3$ and $1280^3$ and we have determined the correlations between them in terms of the \textit{shape} and \textit{amplitude correlators} defined by the two formulae:
\begin{equation}
\label{shapec}
s_{A,B}^{(c)}= \frac{\sum_i \sum_j U^{A}(i,j)U^{B}(i,j)} {\sqrt{\sum_{i,j} (U^{A}(i,j))^2}{\sqrt{\sum_{i,j} (U^{B}(i,j))^2}}}
\end{equation}

\begin{equation}
\label{amplitudec}
r_{A,B}^{(c)}=\frac{\sqrt{\sum_{i,j} (U^{A}(i,j))^2}}{\sqrt{\sum_{i,j} (U^{B}(i,j))^2}}
\end{equation}
where $(c)$ is taken to be $\langle\Theta_{00}\Theta_{00}\rangle$, $\langle\Theta^S\Theta^S\rangle$, $\langle\Theta_{00}\Theta^S\rangle$, $\langle\Theta^V\Theta^V\rangle$, and $\langle\Theta^T\Theta^T\rangle$, respectively. These represent measures of the goodness of fit between the different simulations considered in terms of their shapes and amplitudes respectively. We have taken \textit{A} to be the simulation at a resolution of $1280^3$, and for \textit{B} we took in turn each of the simulations from resolutions of $128^3$, $256^3$, $512^3$, $768^3$, and $1024^3$ respectively. The results obtained are shown in Table \ref{table5}, and the convergence trend is displayed in Fig. \ref{shape_amp}.

\begin{table}[!htb]
\centering
\caption{Shape and amplitude correlators for the UETCs at different simulation resolutions}
\begin{tabular}{|c|c|c|c|c|c|c|}
\hline
UETC & Correlator & $128^3$ & $256^3$ & $512^3$ & $768^3$ & $1024^3$ \\ \hline
\multirow{2}{*}{$\langle\Theta_{00}\Theta_{00}\rangle$} & Shape & 0.6048&  0.7783 & 0.9735 & 0.9939 & 0.9991        \\ \cline{2-7}
 & Amplitude & 3.0939 & 2.4376 & 1.2889 & 1.1123 & 1.0384 \\ \hline
\multirow{2}{*}{$\langle\Theta^S\Theta^S\rangle$} & Shape &0.5628 & 0.6026 & 0.9192 & 0.9750 & 0.9952\\ \cline{2-7}
 & Amplitude & 3.7040 & 2.6621 & 1.5079 & 1.2220 & 1.0831  \\ \hline
\multirow{2}{*}{$\langle\Theta_{00}\Theta^S\rangle$} & Shape & 0.6072 & 0.6377 & 0.949  & 0.9851 & 0.9971\\ \cline{2-7}
 & Amplitude & 5.7650 & 4.1559 & 1.4993 & 1.1968 & 1.0696\\ \hline
\multirow{2}{*}{$\langle\Theta^V\Theta^V\rangle$} & Shape & 0.6587& 0.7381   & 0.9335    & 0.9795  & 0.9962\\ \cline{2-7}
 & Amplitude & 2.4435 & 1.8765 & 1.3244 & 1.1524 & 1.0593 \\ \hline
\multirow{2}{*}{$\langle\Theta^T\Theta^T\rangle$} & Shape &  0.5632 & 0.6011 & 0.9180 & 0.9772 & 0.9961  \\ \cline{2-7}
 & Amplitude & 3.5414 & 2.8519 & 1.5209 & 1.2278 & 1.0833\\ \hline
\end{tabular}
\label{table5}
\end{table}

As the grid resolution is increased to $1280^3$, Fig. \ref{shape_amp} and Table \ref{table5} show very good convergence in both the shape and the amplitude for all the UETCs. The convergence at approximately 5\% is limited by numerical constraints. However, from Fig. \ref{resolution}, one can see that, although we are approaching convergence with the correlators, this has not been yet achieved. Between  $70<k\tau<80$, the energy density UETC decays by 2 to 3 orders of magnitude compared to $k\tau=O(1)$ and hence would make a comparatively small contribution to the power spectrum. For the region $k\tau<70$, there is a definite sign that the graphs are approaching convergence, though it is not completely achieved.

\section{Analytic UETC model}
An analytic model for the calculation of UETCs based on the USM model for Nambu-Goto strings has been developed in Ref. \cite{copeland}. The analytic model is based on the phenomenological USM model. The correlation length can be expressed in terms of a new parameter $\xi$ defined as $\xi=\frac{L}{a\tau}$. 

Using Eqs. (\ref{th00usm}) and (\ref{thijusm}), as well as the SVT decomposition, the relevant UETCs are obtained analytically by integrating over the string network, separately for each stress-energy component of interest:
\begin{eqnarray}
&& \langle \Theta(k, \tau_1) \Theta(k,\tau_2) \rangle = \nonumber \\
&& = \frac{2 f(\tau_1,\tau_2,\xi,L_f)}{16\pi^3} \int_0^{2 \pi} d \phi \int_0^{\pi} \sin \theta \, d \theta \times \nonumber \\
&& \times \int_0^{2 \pi} d \psi \int_0^{2 \pi} d \chi \, \Theta(k,\tau_1) \Theta(k,\tau_2) \
\label{integrala}
\end{eqnarray}
where the function $f$ quantifies the decrease in the number of segments by string decay. The anisotropic scalar, vector and tensor components are given in this case by Eqs. (\ref{thSusm})-(\ref{thTusm}).
The UETCs that are computed are compared with simulations produced with the CMBACT code for different values of the parameters.

The final results have only three free parameters: \textit{v}, $\alpha$, and $\xi$. They can be obtained by integrating Eq. (\ref{integrala}) and depend on integral expressions $A_i$:
\begin{eqnarray}
\langle \Theta(k, \tau_1) \Theta(k,\tau_2) \rangle = \frac{  f(\tau_1,\tau_2,\xi,L_f) \mu ^2}{k^2 \left(1-v^2\right)} \times  \nonumber \\ \times \sum_{i=1}^6 A_i \left[ I_i (x_-, \rho) - I_i (x_+, \rho) \right] \,
\end{eqnarray} 
where $\rho = k |\tau_1 - \tau_2| v$, $x_{\pm} = k \xi (\tau_1 \pm \tau_2)/2$ and the expressions $A_i$ depend again on the three parameters and can be found in the Appendix of Ref. \cite{copeland}.

\subsection*{Fit to the analytic model}
To be able to compare the simulated UETCs with the analytical ones from Ref. \cite{copeland}, we have added the two vector and two tensor components and we have obtained the five functions used in Ref. \cite{copeland}. The analytical model depends on three parameters, $v$, $\alpha$, and $\xi$. The parameters have the following ranges: $v$ varies between 0 and 1, $\alpha$ is in the interval [1, 2] and $\xi$ is positive. We use again the \textit{shape} [see Eq. (\ref{shapec})] and \textit{amplitude correlators} [Eq. (\ref{amplitudec})], this time with \textit{A} representing the analytical UETC and \textit{B} representing the simulated one. The $s$'s and $r$'s have been tabled for parameters in the permitted ranges and the values of the shape correlators have been maximised. The amplitude correlators have been chosen to be as close to one another as possible (due to different normalisation factors).

The best-fit parameters are as follows: $\xi=0.2$ for the first two simulations and $\xi=0.3$ for the third; while $v$ is 0.5, 0.1, and 0.6, respectively; and $\alpha$ is 1.5, 1.3, and 1.3. The best results obtained for the three simulations for the shape and amplitude correlators are presented in Table \ref{table6}. We will show a comparison between these ``best fit'' power spectra and the ones that we have obtained using the eigenvectors in Sec. \ref{eigendecomp}. 

\begin{table*}[!htb]
\setlength{\tabcolsep}{4pt}
\centering
\caption{Shape and amplitude correlators for UETCs in the three simulations}
\begin{tabular}{|c|c|c|c|c|c|c|}
\hline
Simulation & Correlator &  $\langle \Theta_{00}\Theta_{00} \rangle$ & $\langle\Theta^S\Theta^S\rangle$ & $\langle\Theta_{00}\Theta^S\rangle$ & $\langle\Theta^V\Theta^V\rangle$ & $\langle\Theta^T\Theta^T\rangle$ \\ \hline
\multirow{2}{*}{Radiation era} & Shape & 0.710 & 0.841  & 0.188 & 0.815& 0.738       \\ \cline{2-7}
 & Amplitude & 1.009 & 0.985 & 0.338 & 0.656 & 0.932 \\ \hline
\multirow{2}{*}{Matter era} & Shape &0.667 & 0.801 & 0.132 & 0.744 & 0.663\\ \cline{2-7}
 & Amplitude & 1.000 & 0.978 & 0.343 & 0.693 & 1.086  \\ \hline
\multirow{2}{*}{Matter + $\Lambda$ eras } & Shape & 0.751 & 0.820 & 0.212 & 0.803& 0.718\\ \cline{2-7}
 & Amplitude & 0.998 & 1.094 & 0.365 & 0.746 & 0.928\\ \hline
\end{tabular}
\label{table6}
\end{table*}

\begin{figure}[!htb]
\begin{center}$
\begin{array}{c}
\includegraphics[width=3in]{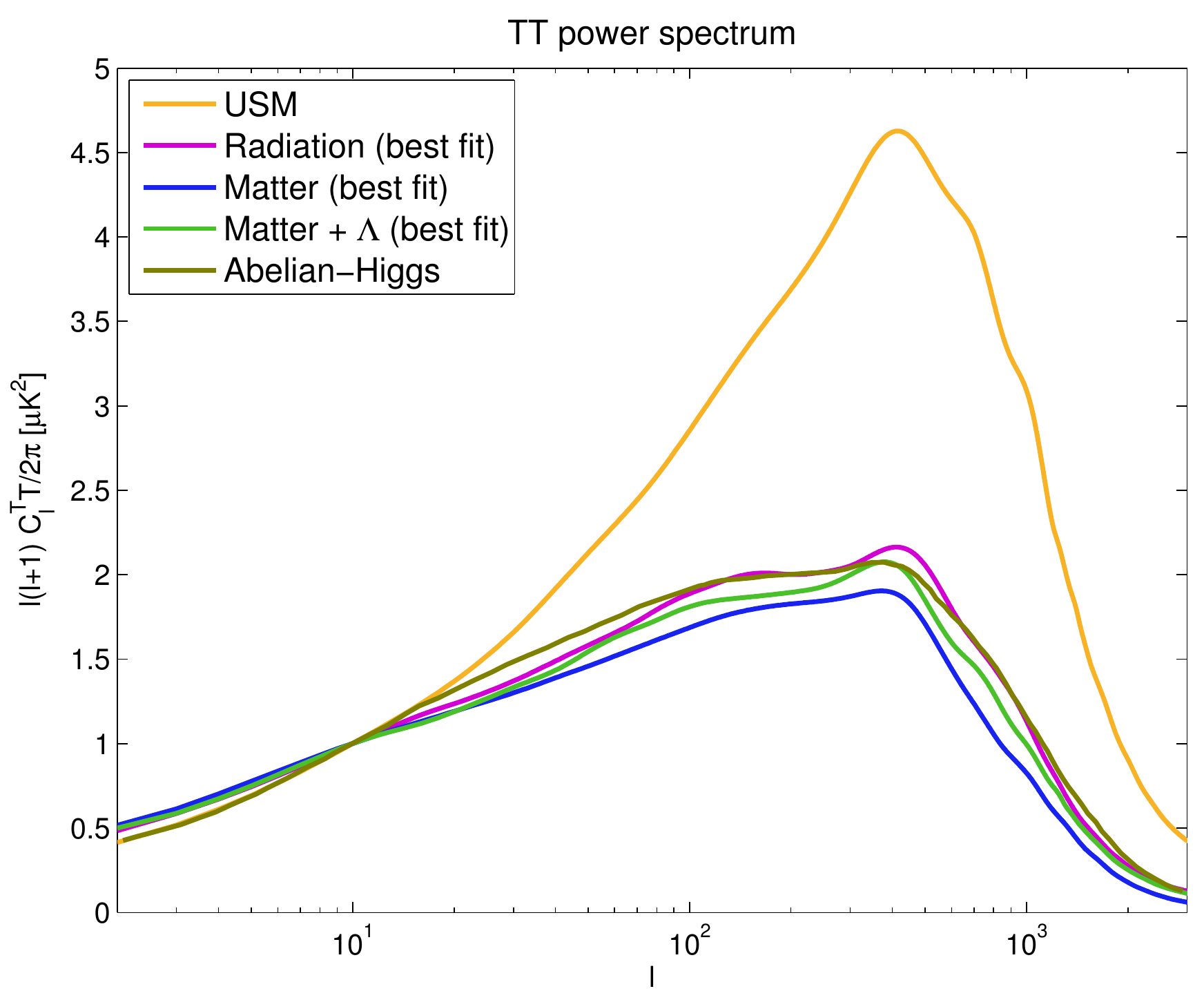}
\end{array}$
\caption{Comparison between the TT power spectrum for USM , Abelian-Higgs and Nambu-Goto simulated strings. Simulation 1 covers the radiation era, Simulation 2 the matter era, and Simulation 3 matter and cosmological constant eras. The USM and Abelian-Higgs power spectra are the standard results used in the Planck cosmic defects paper \cite{planckstr}}
\label{TTp}
\end{center}
\end{figure}

We have updated CMBACT with the latest published Planck parameters \cite{planckparams}, and we have taken $G\mu=2.07\times 10^{-6}$, as in Ref. \cite{copeland}. We then ran the code with 500 string segments and 400 realisations with the parameters found for the best fit. We have obtained these values of the parameters by fixing the values of $v$, $\alpha$, and $\xi$ on all scales. Otherwise, the parameters are just initial conditions for the differential equations in the VOS model, and hence the results vary only weakly with them.  The cosmological parameters chosen were the \textit{Planck+WP+high L+BAO} parameters from the 2013 Planck results \cite{planckres}. For comparison, we have also run the default CMBACT \cite{cmbact} with default initial parameters ($v=0.65$, $\alpha=1.9$, $\xi=0.13$) with Planck cosmology. We have taken the Abelian-Higgs power spectrum data from Ref. \cite{PhysRevD.82.065004}, and we have plotted in Fig. \ref{TTp} all three power spectra from simulations on the same graph in terms of the multipole $l$, in logarithmic scales, together with the USM and Abelian-Higgs ones.

The power spectrum for the Nambu-Goto strings, obtained from simulations, is situated between the power spectra of the USM and the Abelian-Higgs models. This was expected, as the USM model is unable to capture very accurately the entire small scale behaviour of the cosmic strings, while the Abelian-Higgs model does not have enough dynamic range. It can be seen that the position of the peak corresponds to approximately the same $l$ in all cases and that the power spectrum in the three cases is very similar for $l<30$. The matter era spectrum has a smaller peak amplitude and is straighter for large $l$. The CMB power spectra obtained from the simulations are very similar because of the fact that we are only using CMBACT with different parameters to obtain them. In the next section we will describe the power spectrum obtained using UETCs directly. 

\section{Power spectrum obtained from eigendecomposition of UETCs}
\label{eigendecomp}
\begin{figure*}[!htb]
\begin{center}$
\begin{array}{ccc}
\includegraphics[width=2.07in]{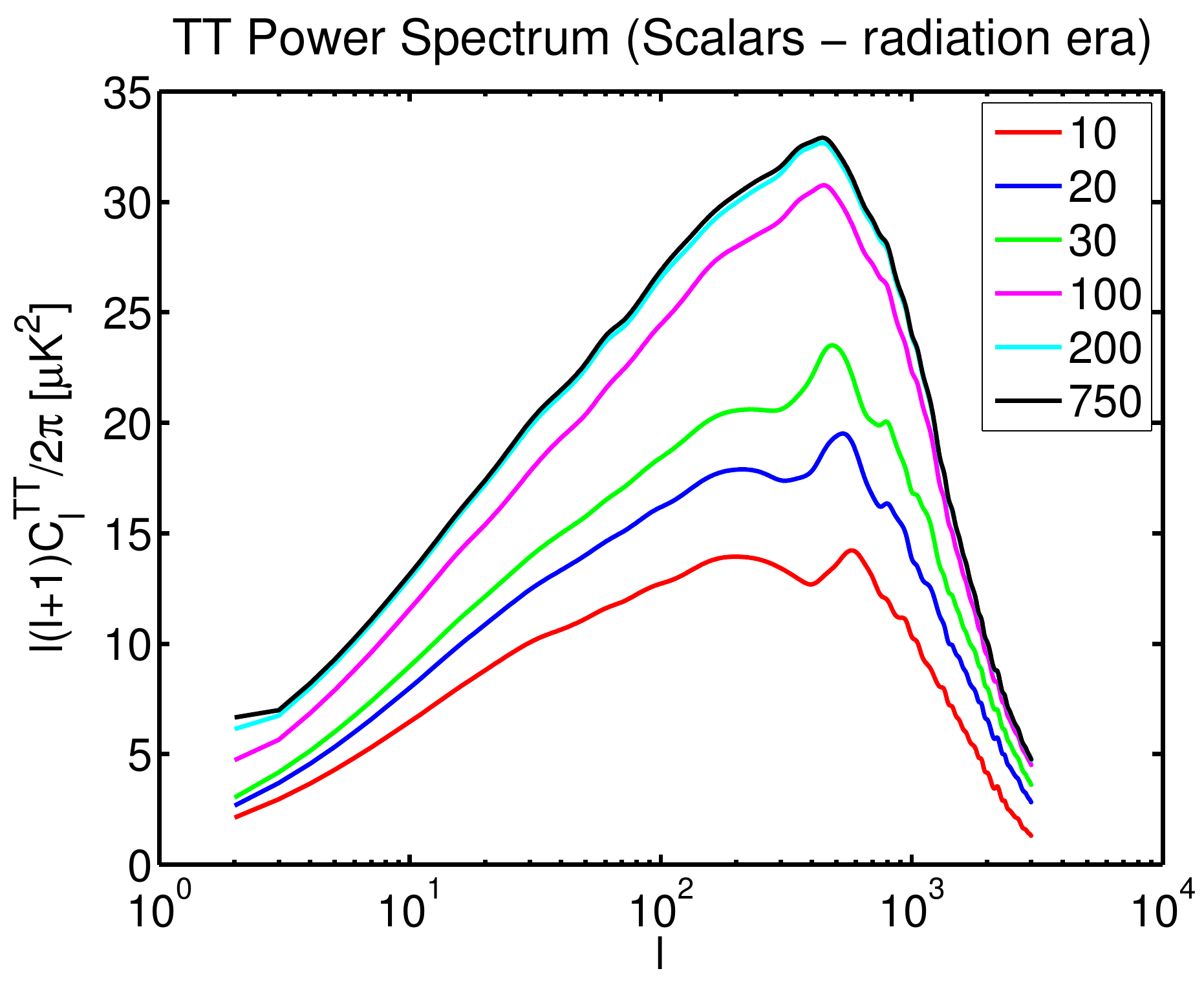} &
\includegraphics[width=2.07in]{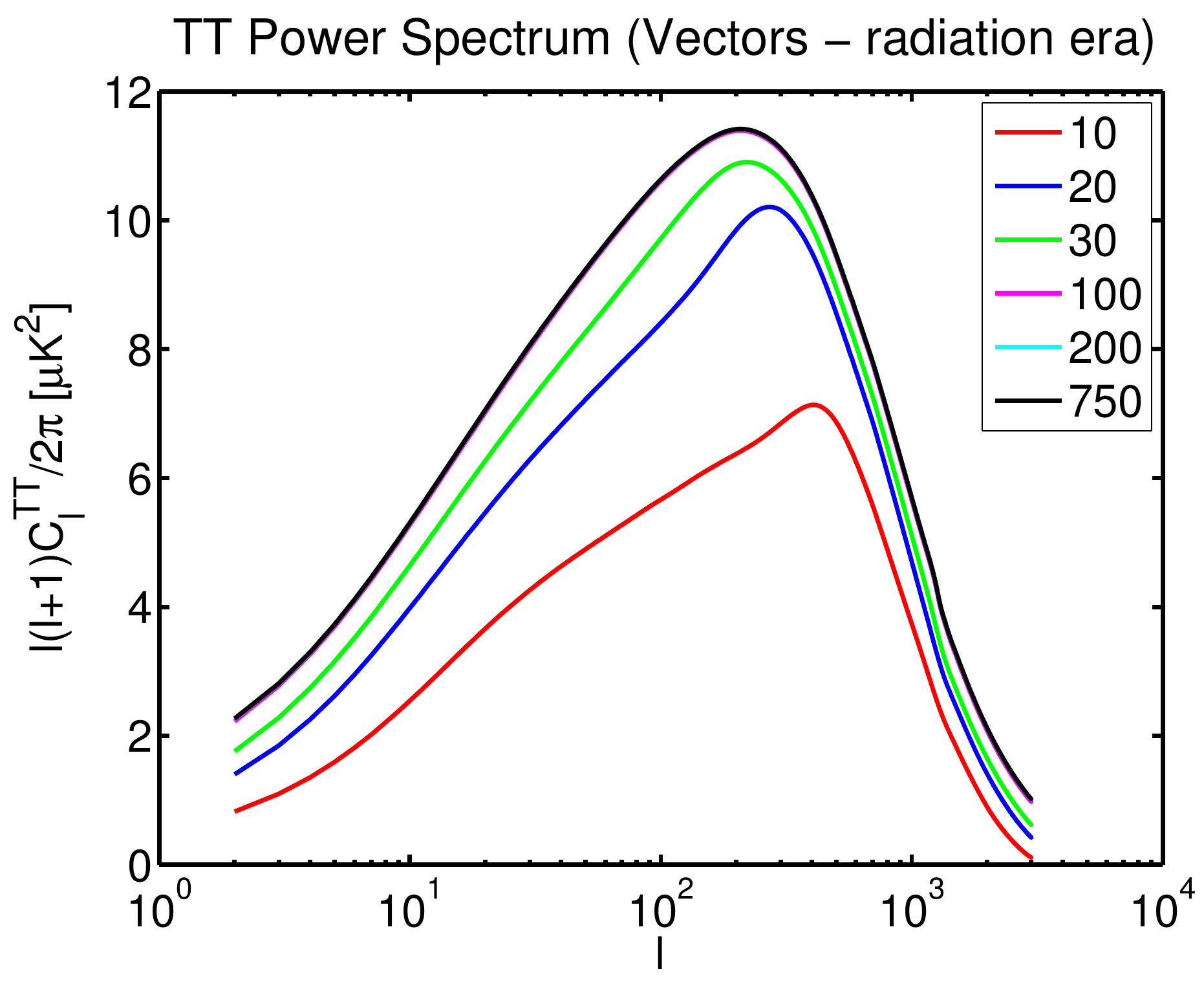} &
\includegraphics[width=2.07in]{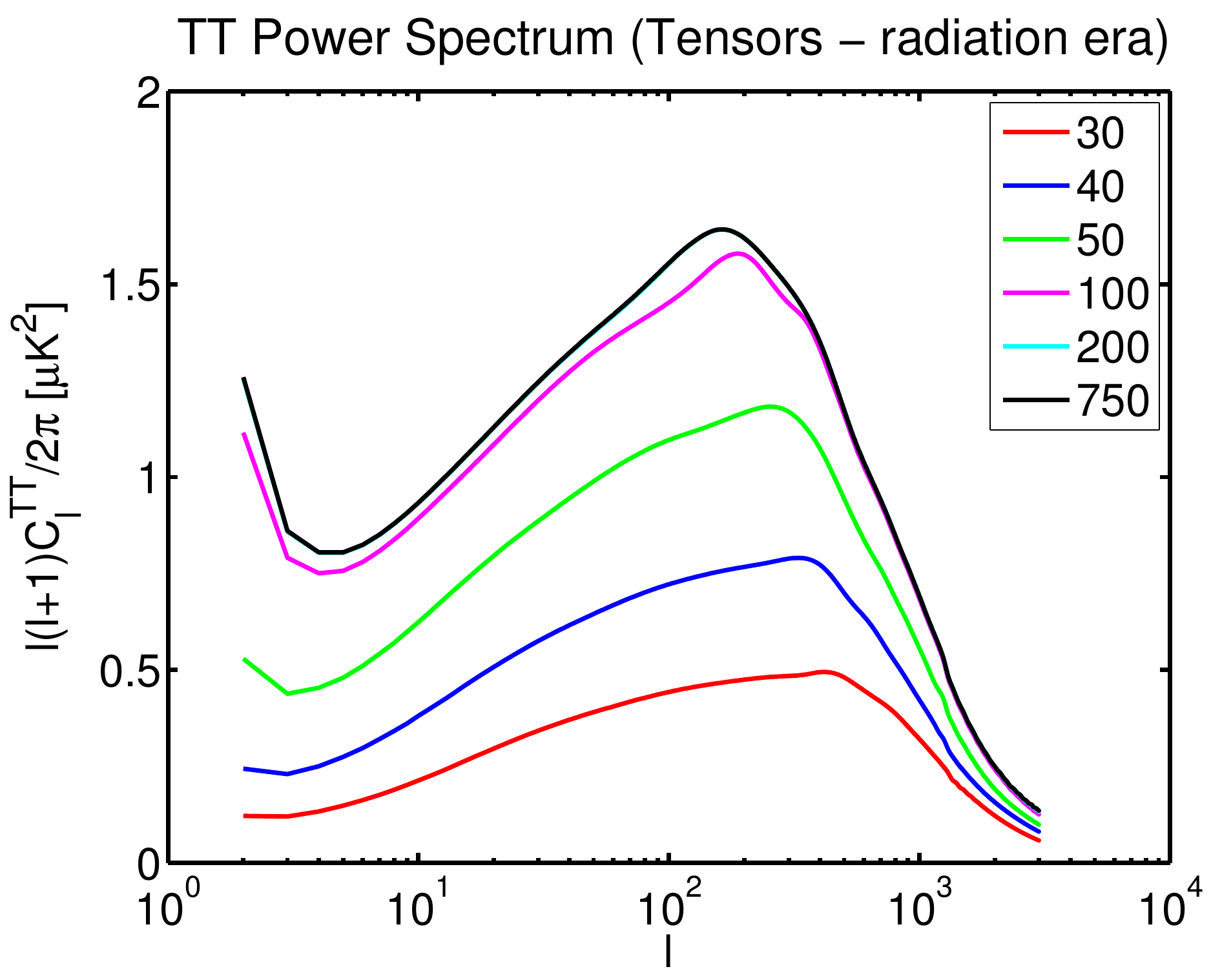} \\
\includegraphics[width=2.07in]{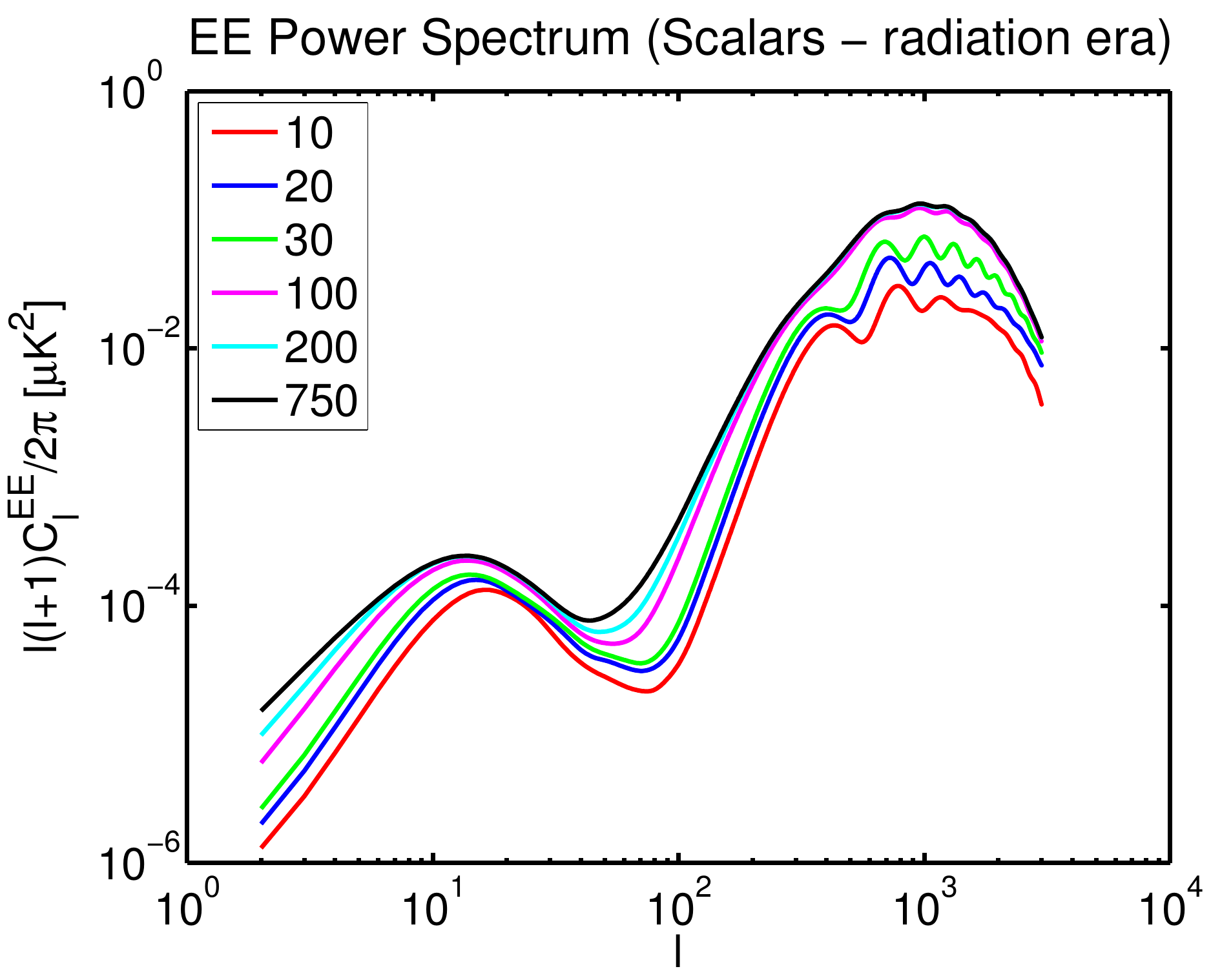} &
\includegraphics[width=2.07in]{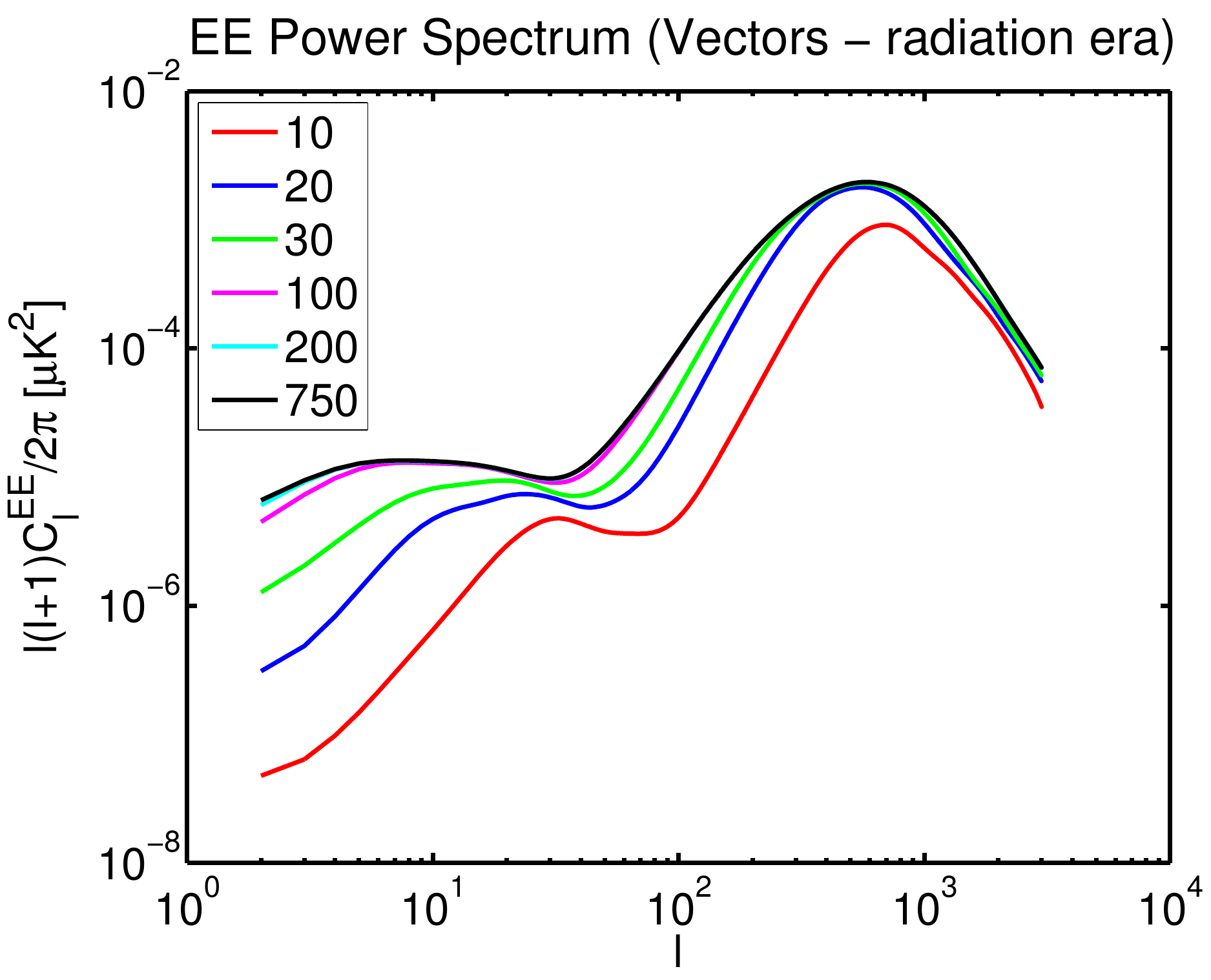} &
\includegraphics[width=2.07in]{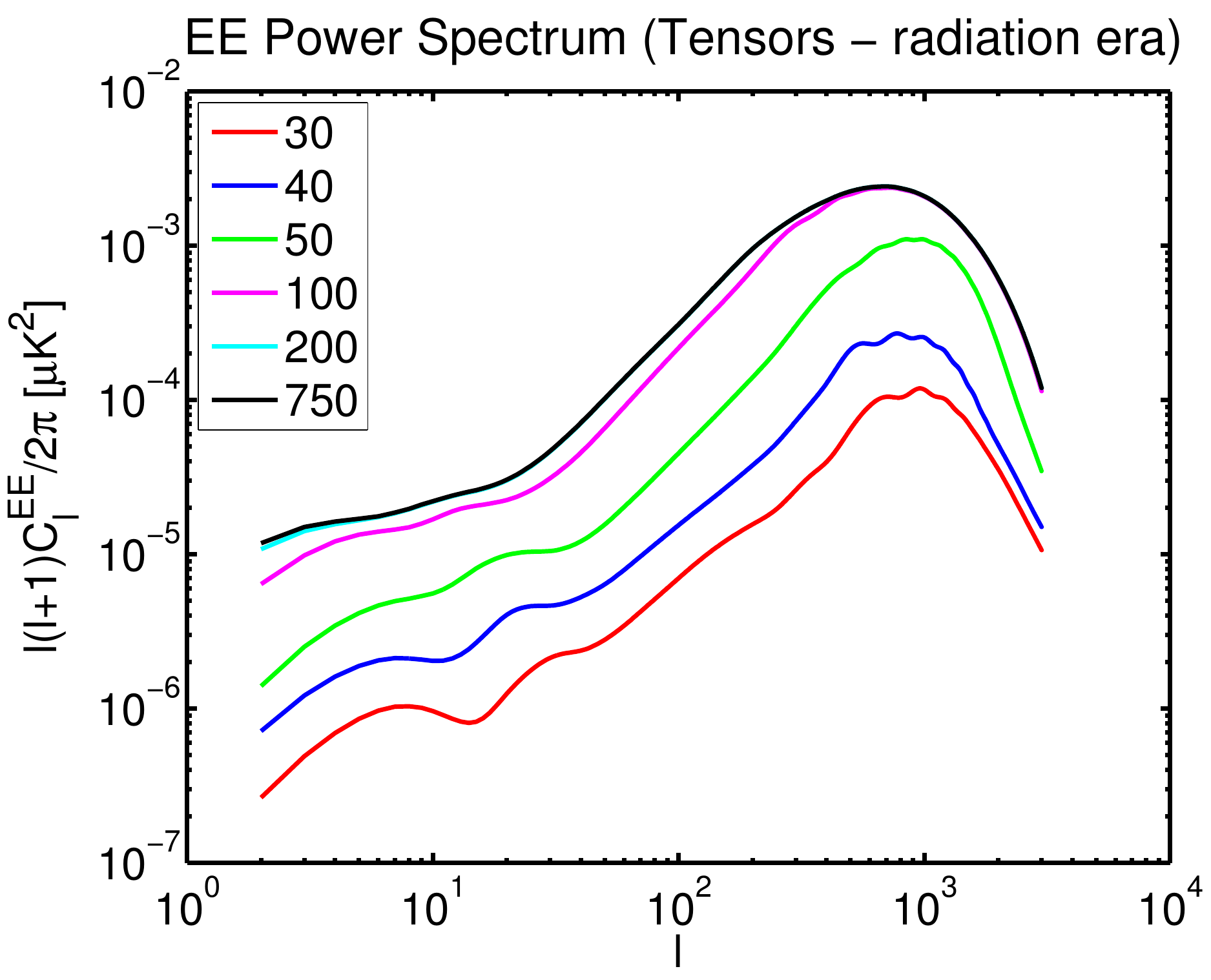} \\
\includegraphics[width=2.07in]{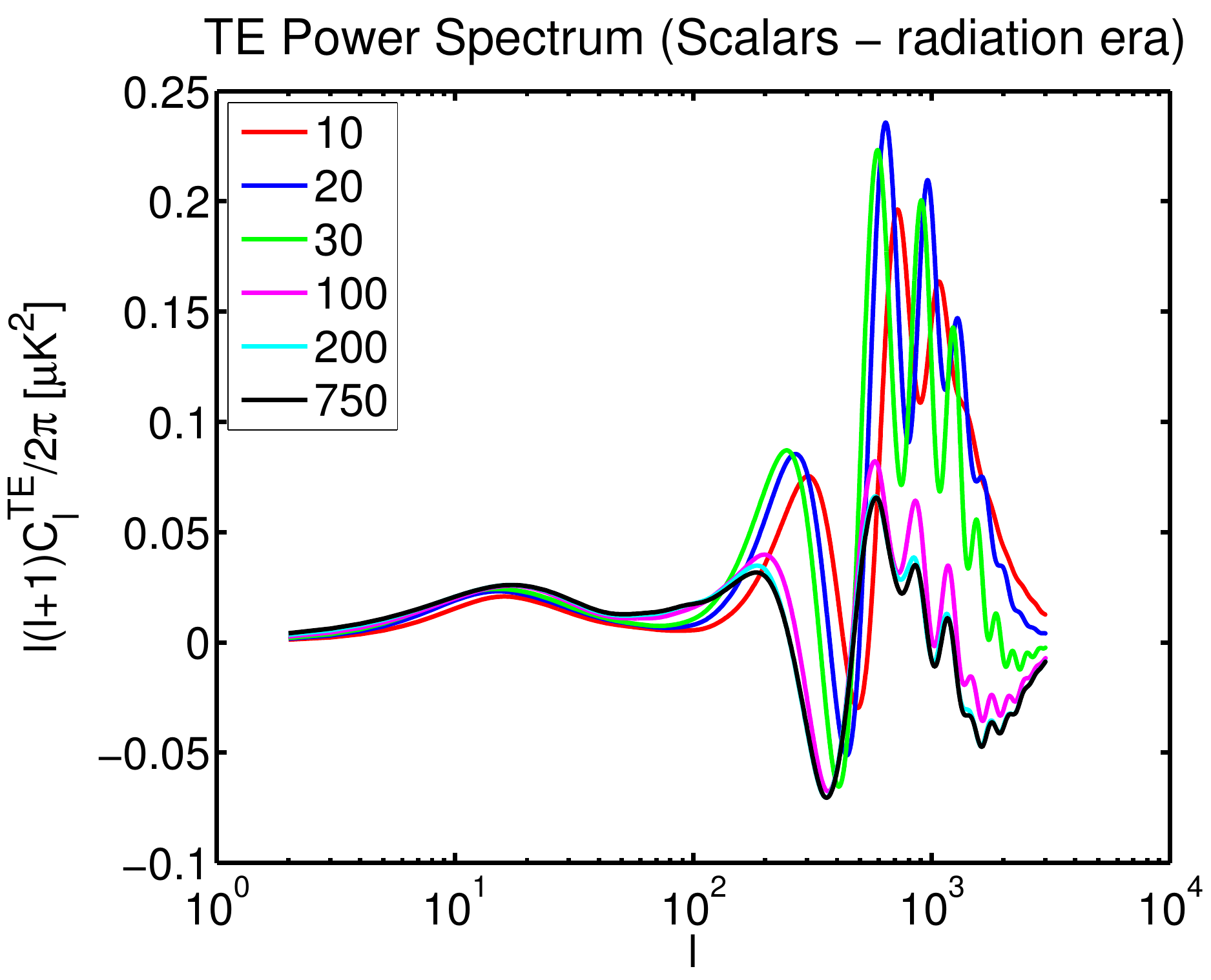} &
\includegraphics[width=2.07in]{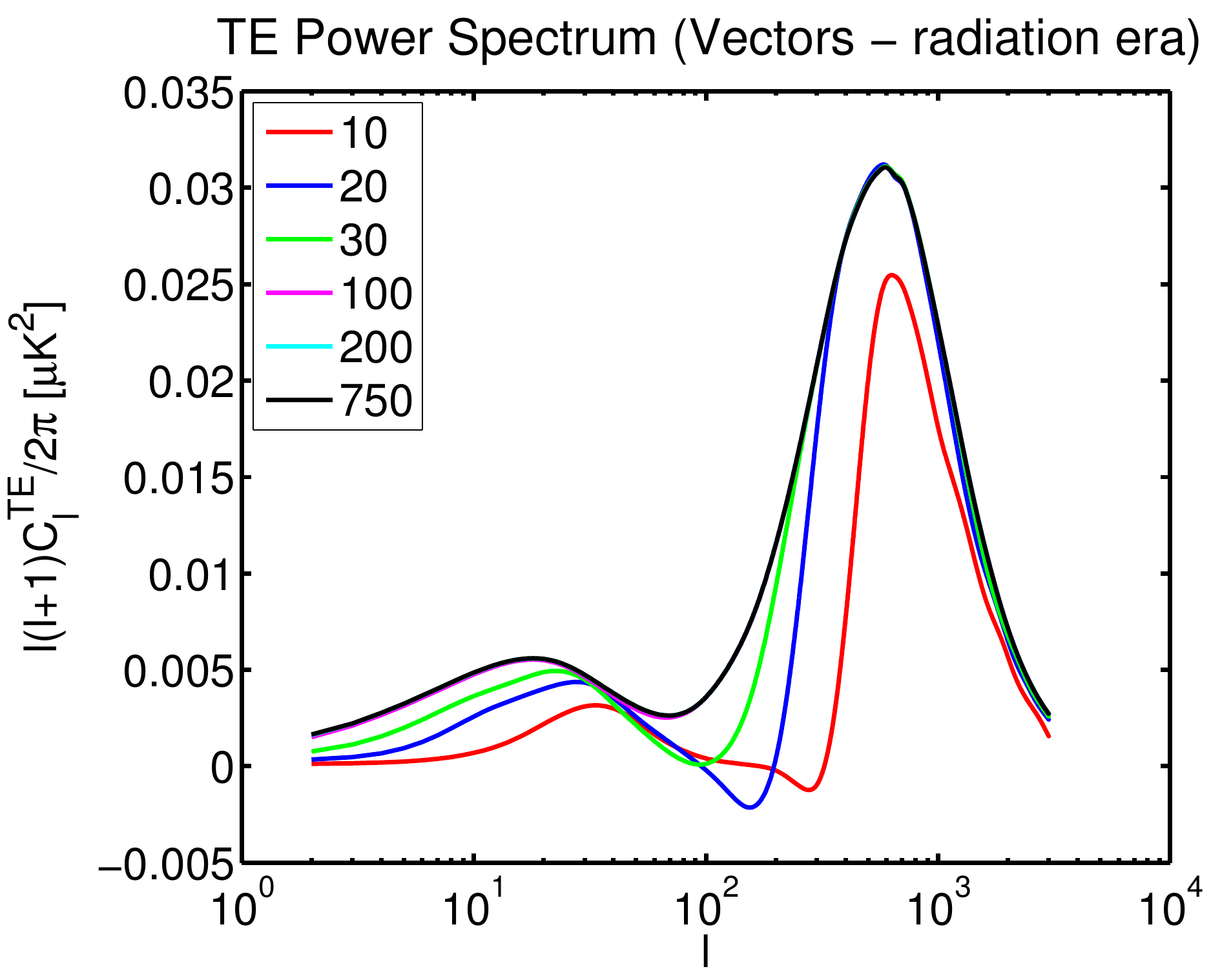} &
\includegraphics[width=2.07in]{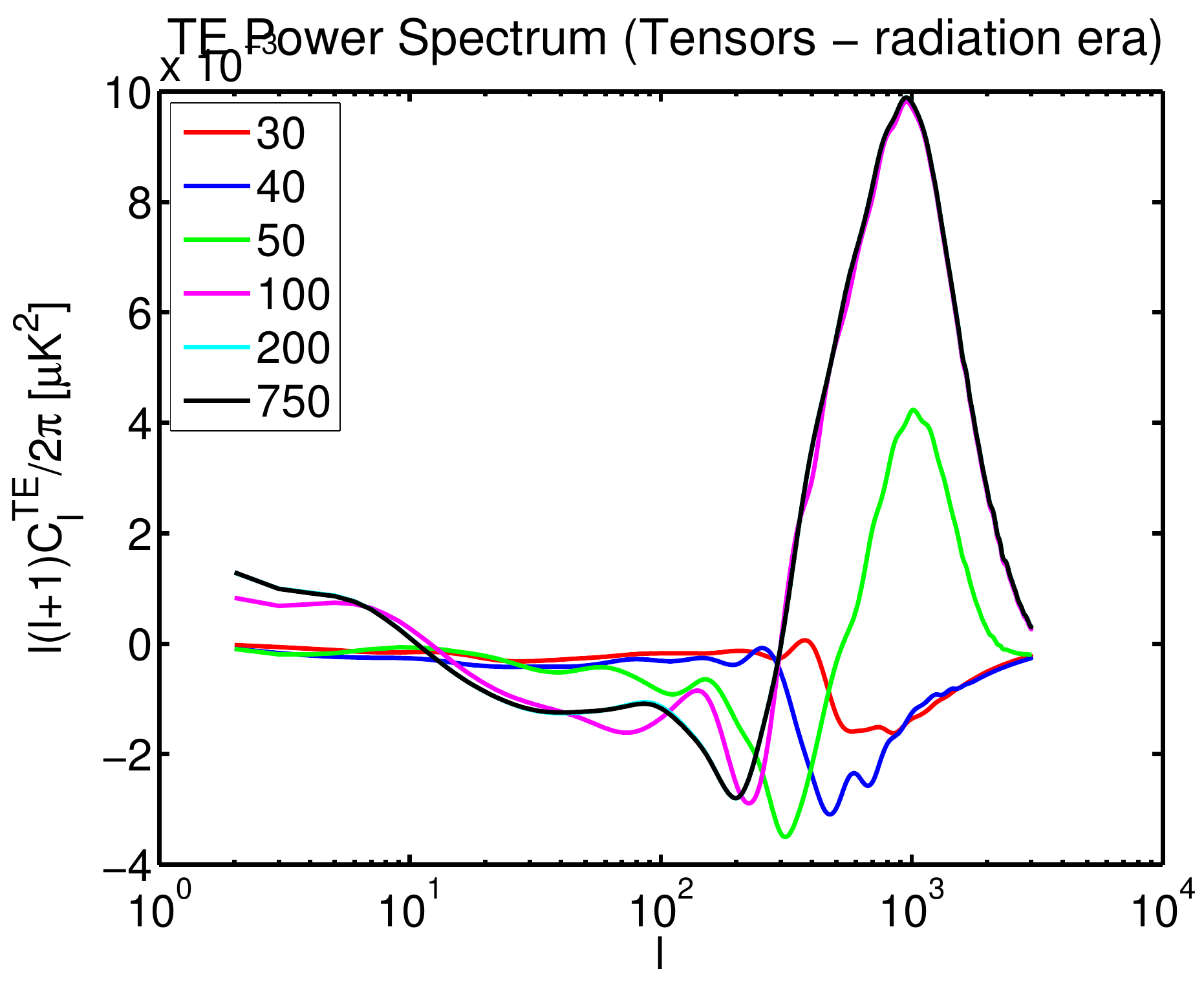} \\
 &
\includegraphics[width=2.07in]{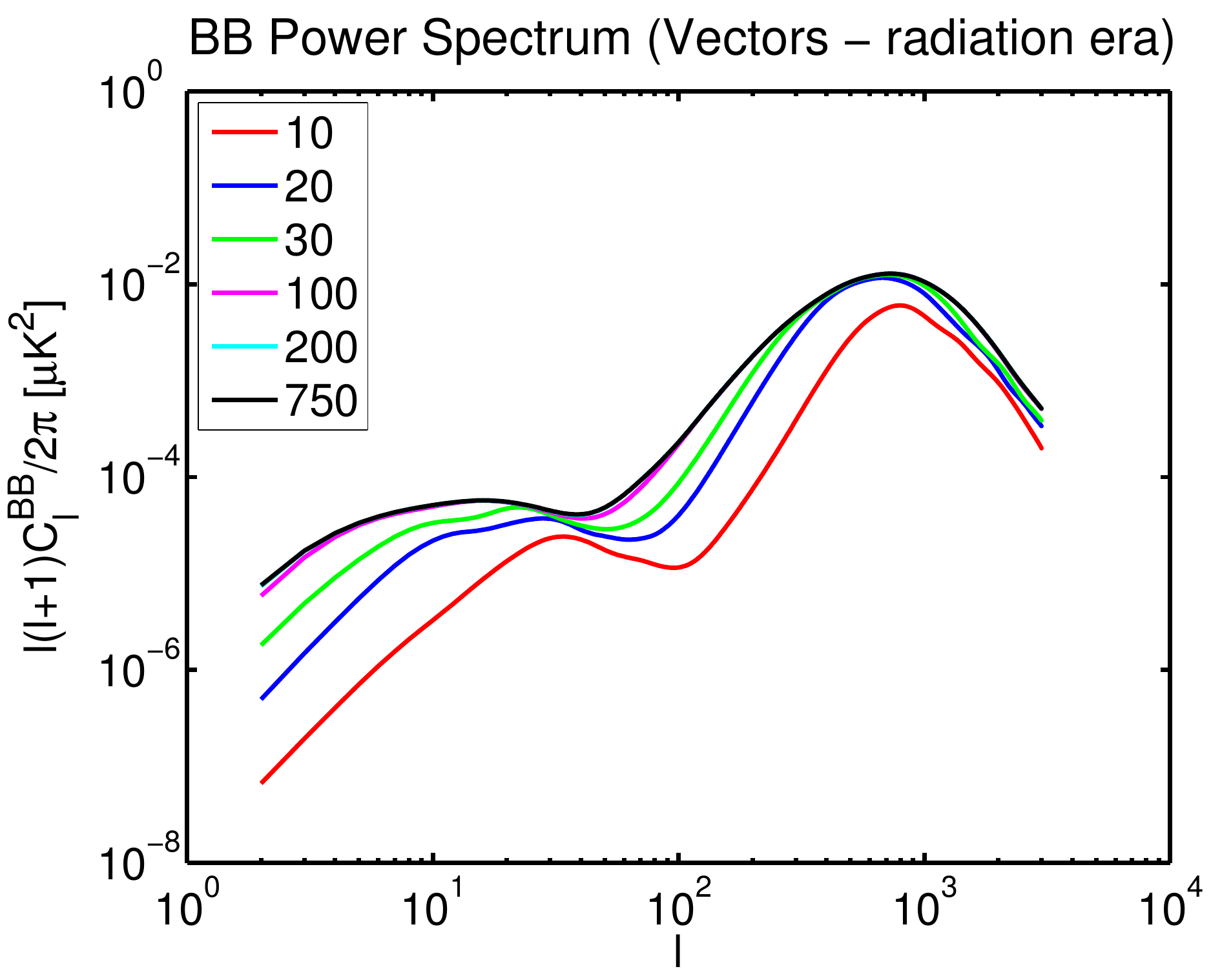} &
\includegraphics[width=2.07in]{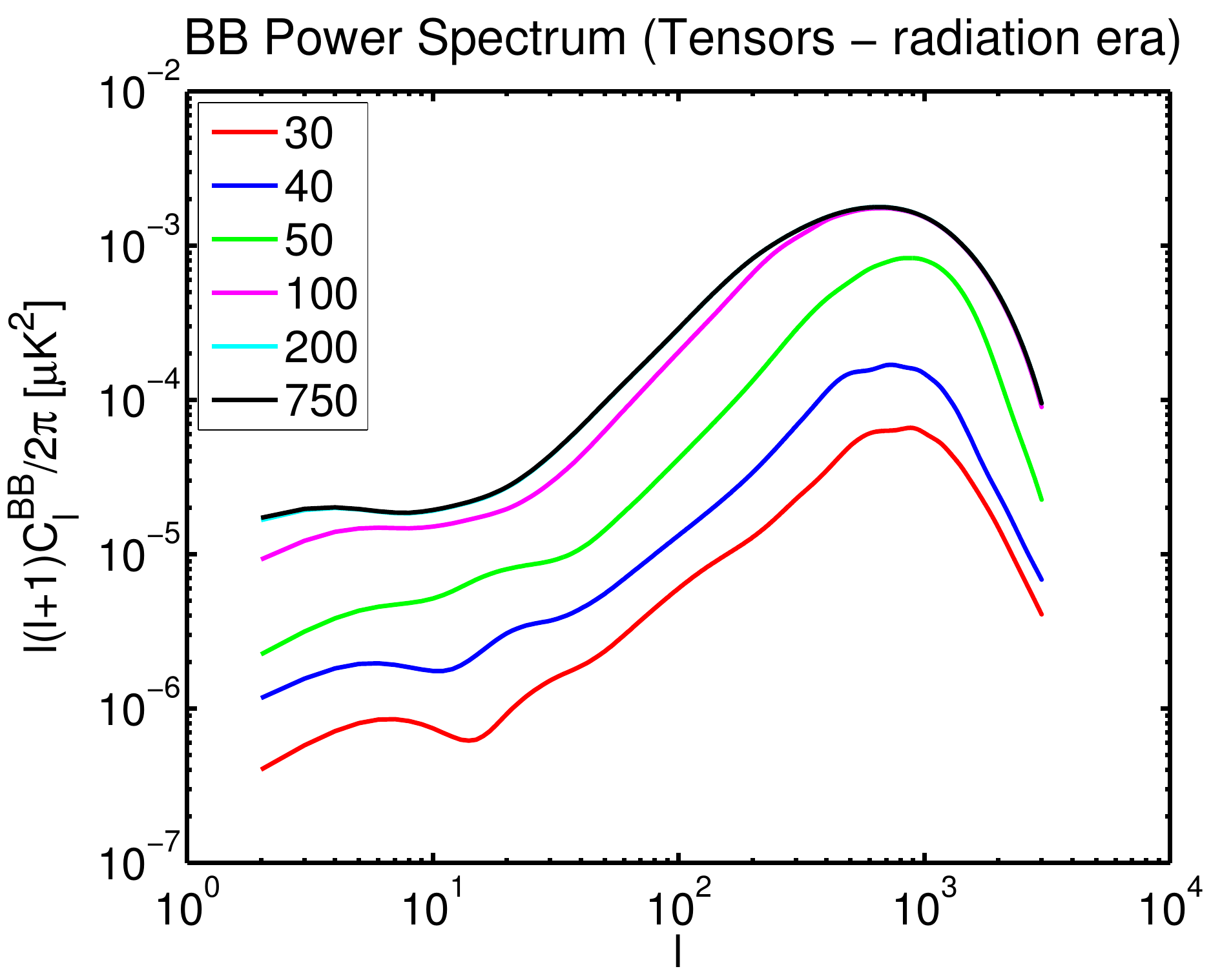} 
\end{array}$
\caption{Power spectra of the cosmic strings obtained from the simulations in the radiation era assuming scale invariance. From left to right:  scalar, vector, and tensor power spectra; from top to bottom: TT, EE, TE, and BB power spectra ($G\mu=1.5 \times 10^{-7}$). The numbers in the legend represent the number of eigenvectors used. The colours in the tensor spectra plots represent different numbers of eigenvectors used compared to the scalar and vector spectra.}
\label{radiation_ps_uetc}
\end{center}
\end{figure*}

Using the formulae in Sec. \ref{UETCapp}, we have run our code and we have computed the power spectra from the three simulations that span the whole cosmological time. The power spectra have been calculated first by using each of the individual simulations and extending their validity to the whole cosmological time by assuming scaling. For example, even though we have determined the UETCs using just cosmic strings that have evolved in the radiation era, we assume that the UETCs would be valid for all times. The matrices corresponding to them have been diagonalised, and their corresponding eigenvectors have been sorted in terms of the magnitude of their eigenvalues (from largest to lowest). We determined the power spectra from each of the eigenvectors and then we summed up the results. Although in principle all the eigenvectors have to be used in order to obtain an exact result, in practice using Eq. (\ref{sumcls}) it can be noticed that for very small eigenvalues the contribution to the overall angular power spectra becomes insignificant. We have analysed this problem in detail and it turns out that for all four power spectra considered using roughly 200 eigenvectors gives a very good convergence for the power spectra. We have checked this in all our results. This is illustrated in Fig. \ref{radiation_ps_uetc} with the power spectra that we have obtained in the radiation era. The power spectra obtained from the scalar, vector and tensor components have been plotted on separate figures, and the convergence in terms of the number of eigenvectors used has been  shown. In the case of the vectors, for the TT and TE power spectra we get excellent convergence using just 100 eigenvectors. For the EE and BB vector power spectra as for all tensor components we need 200 eigenvectors to get a very good convergence.

Later (see Fig. \ref{mixed_ps_uetc}), we will show a comparison between the results that we have obtained by assuming scale invariance throughout the history of the Universe vs scale invariance in each of the cosmological eras (radiation, matter, and $\Lambda$ domination). 

\begin{figure}[!htb]
\begin{center}$
\begin{array}{c}
\includegraphics[width=3in]{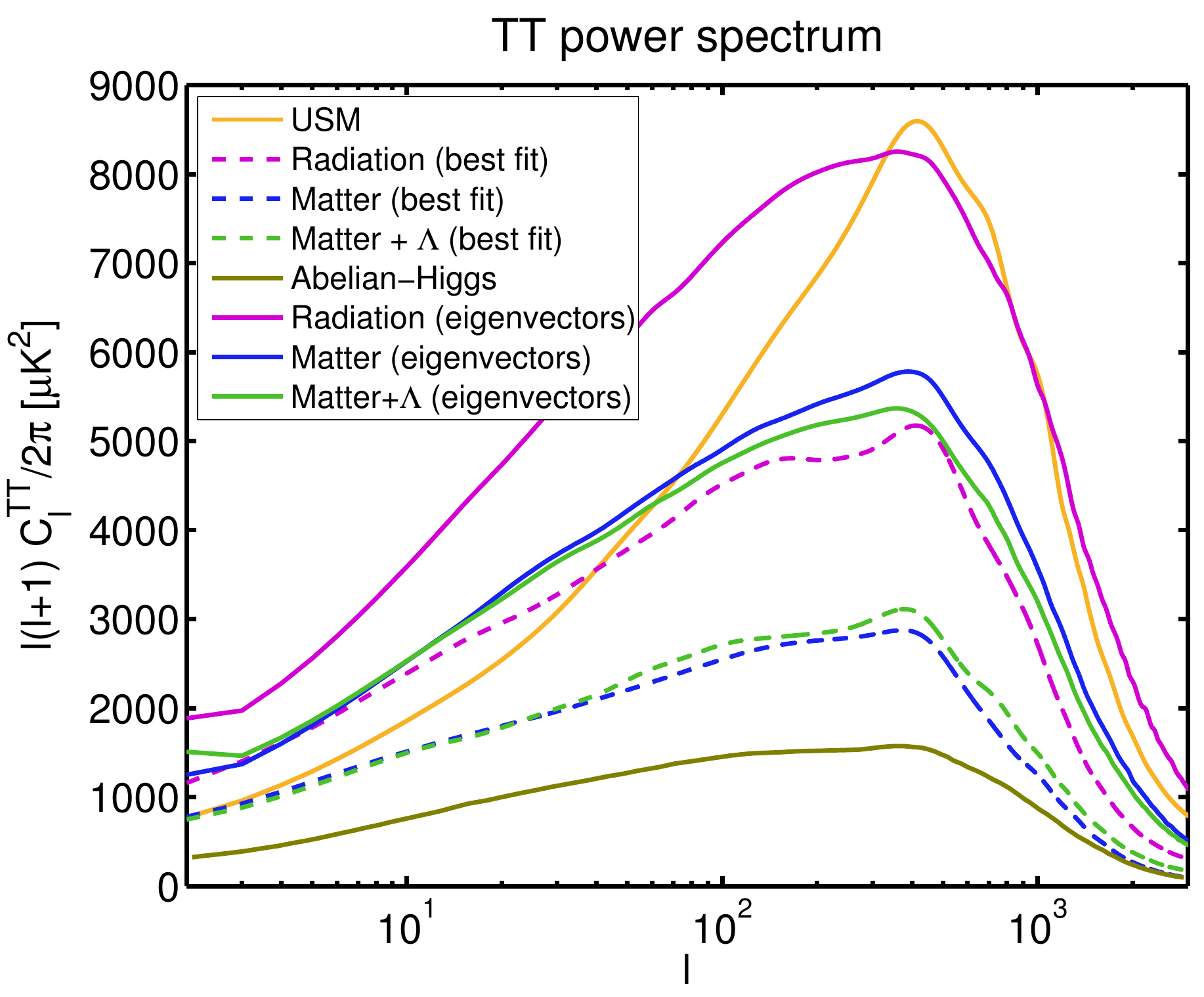}
\end{array}$
\caption{Comparison between the TT power spectra obtained through the best fit method and using eigenvectors}
\label{TTall}
\end{center}
\end{figure}

\begin{figure*}[!htb]
\begin{center}$
\begin{array}{ccc}
\includegraphics[width=2.07in]{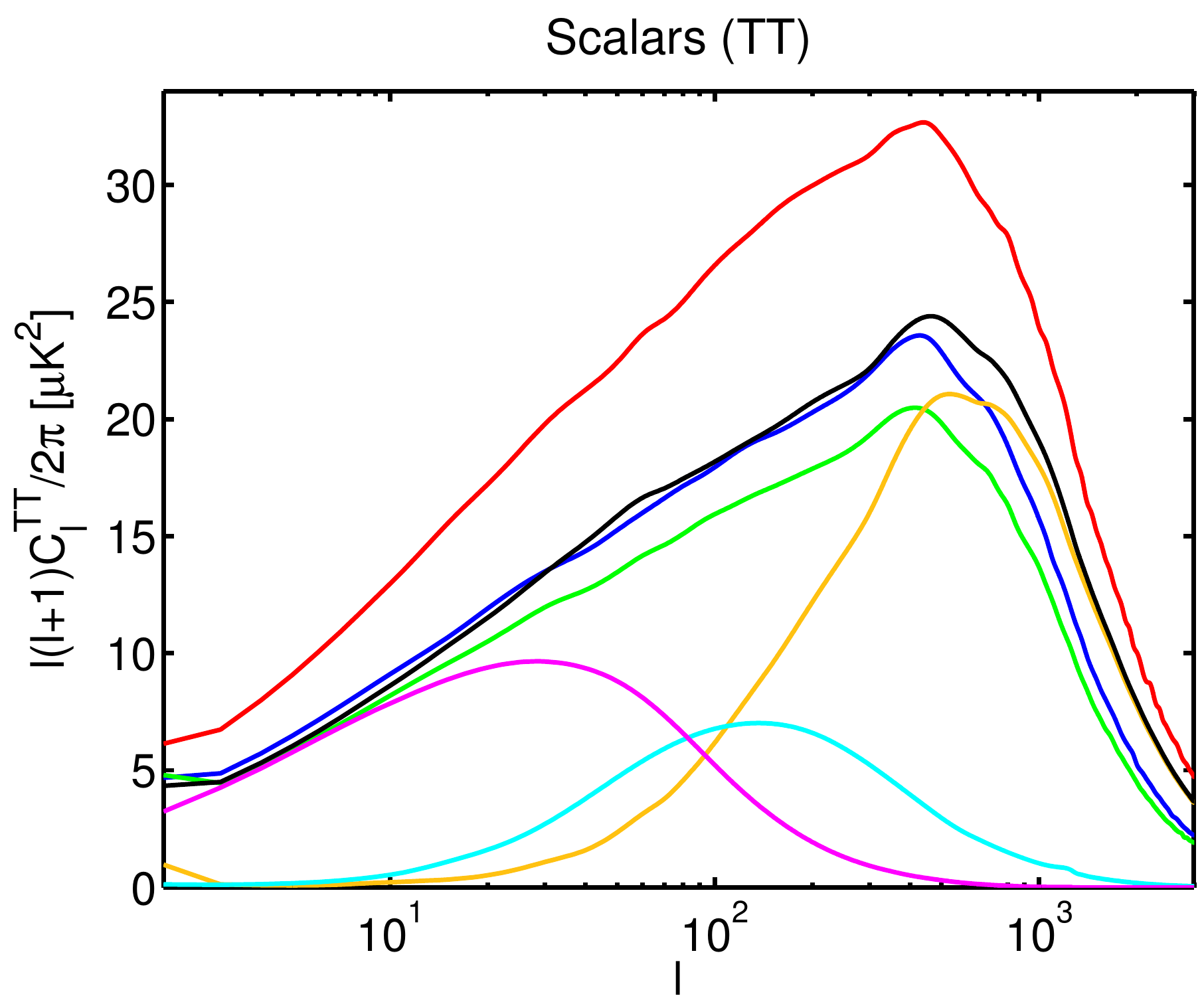} &
\includegraphics[width=2.07in]{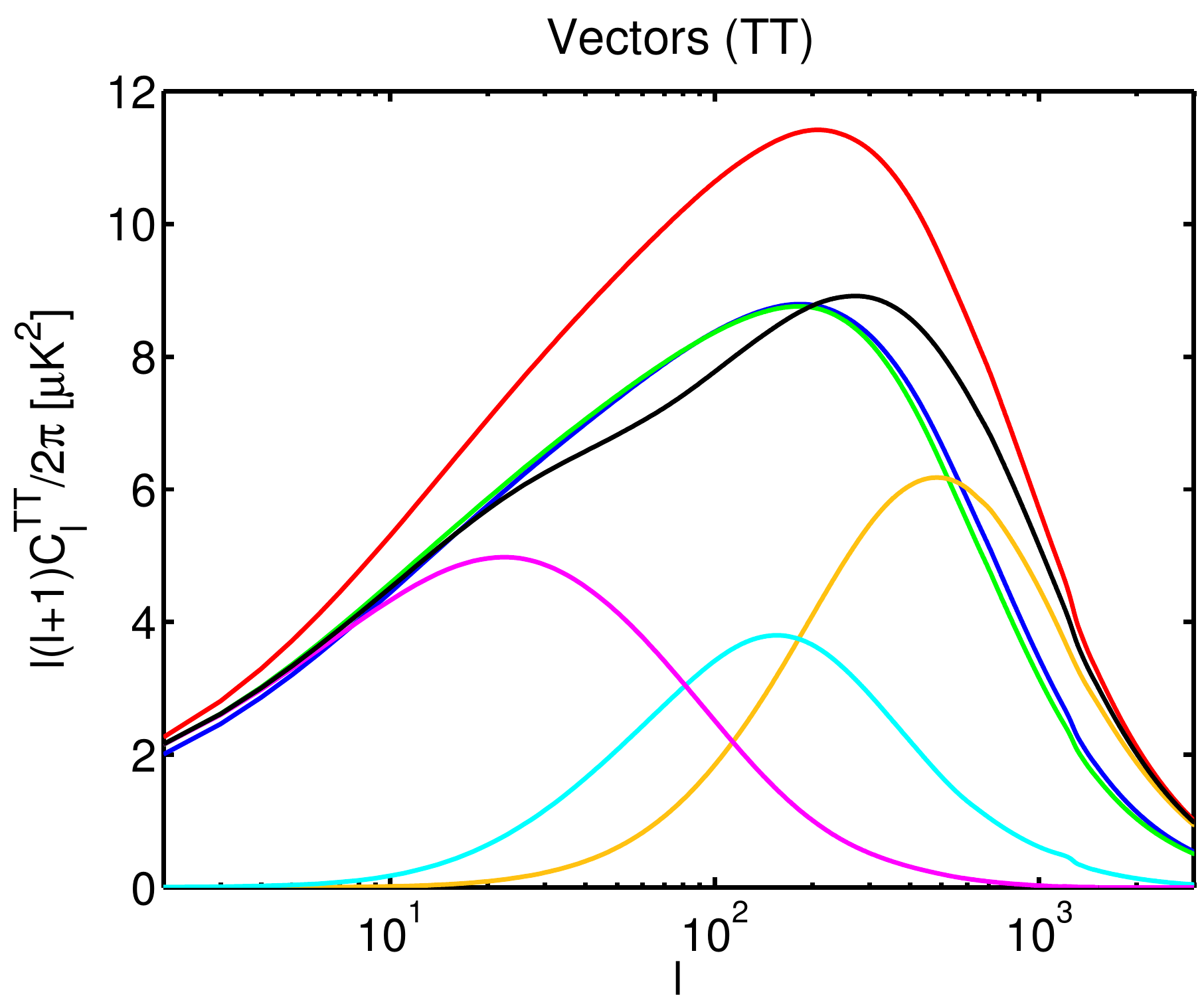} &
\includegraphics[width=2.07in]{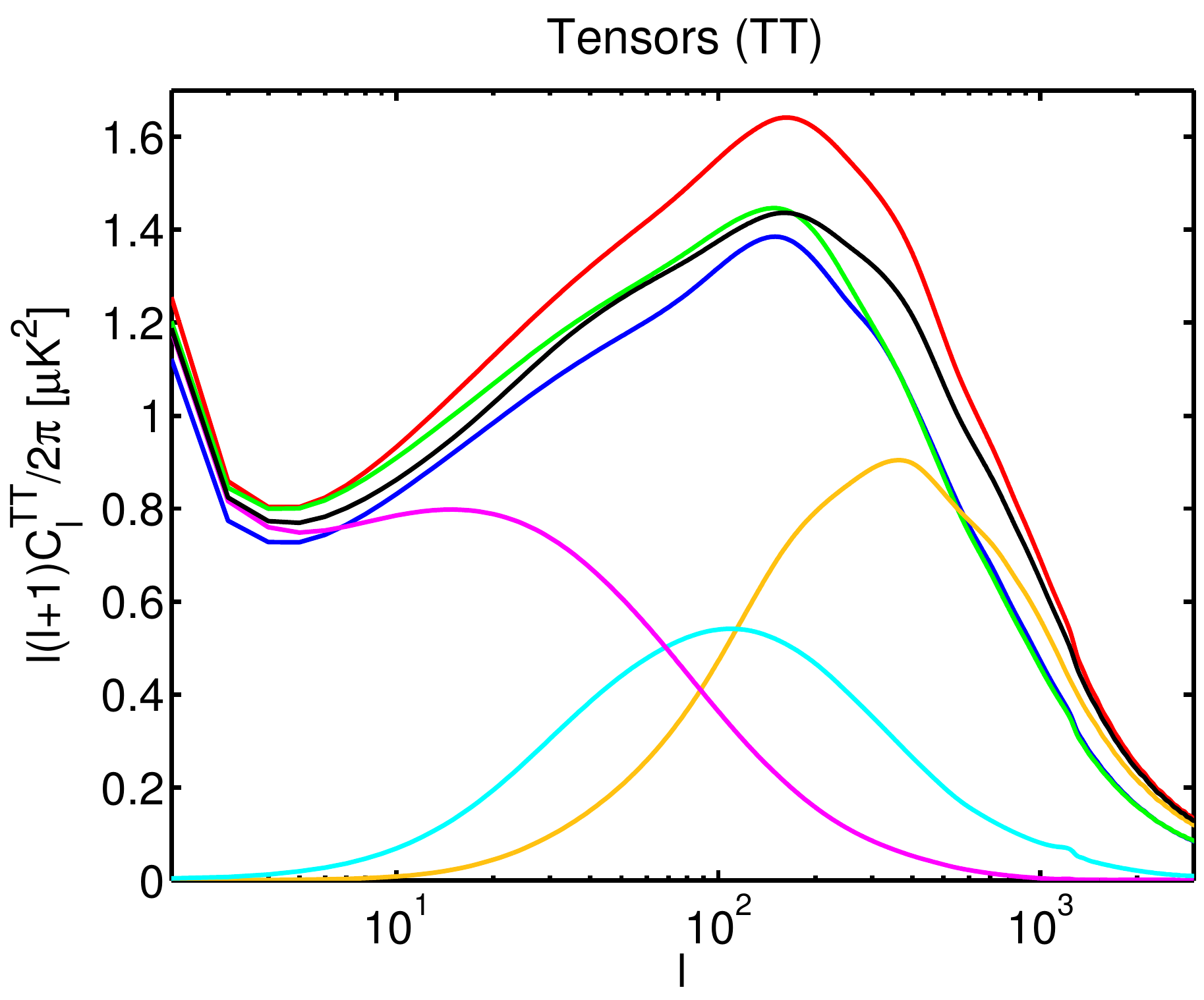} \\
\includegraphics[width=2.07in]{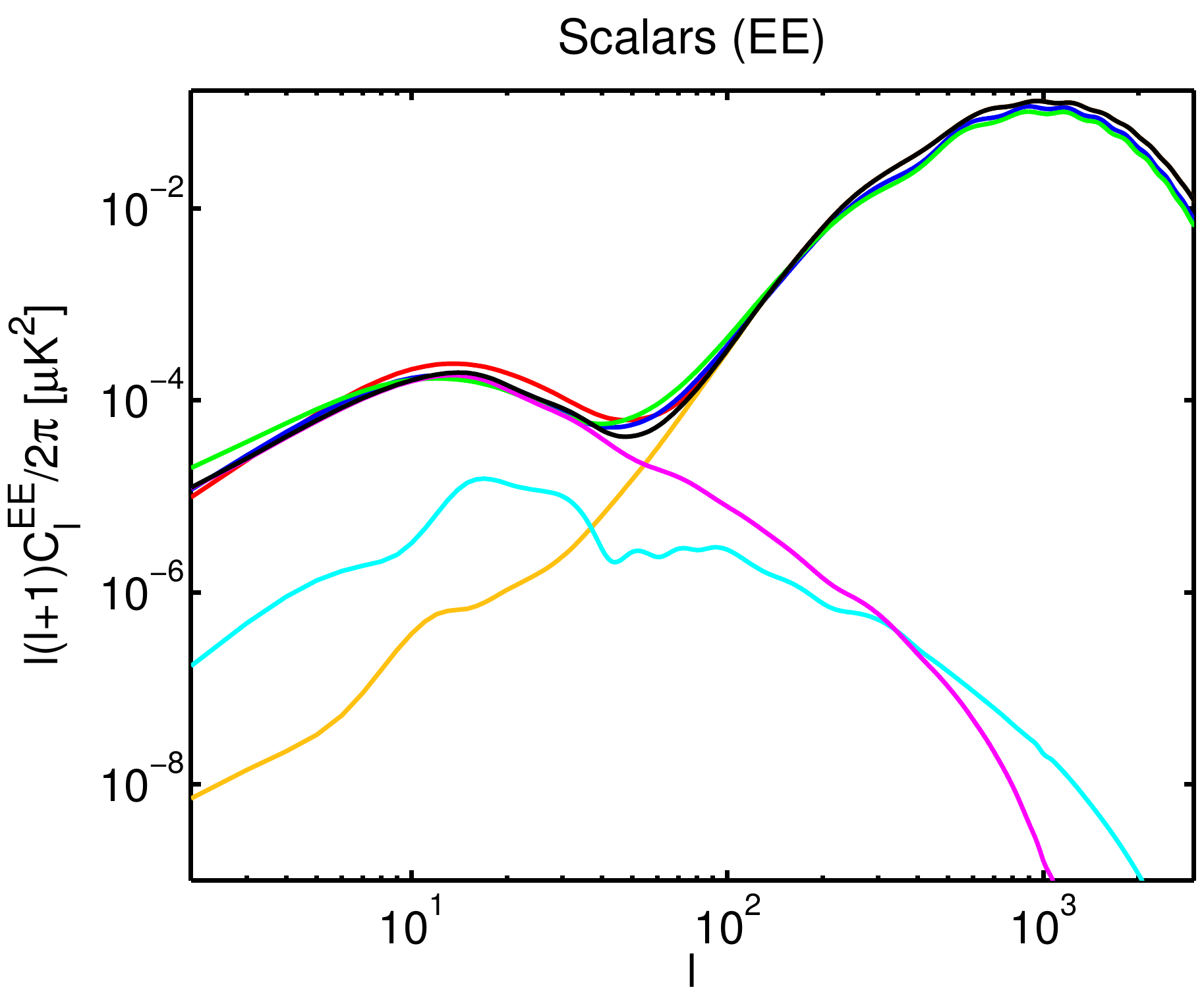} &
\includegraphics[width=2.07in]{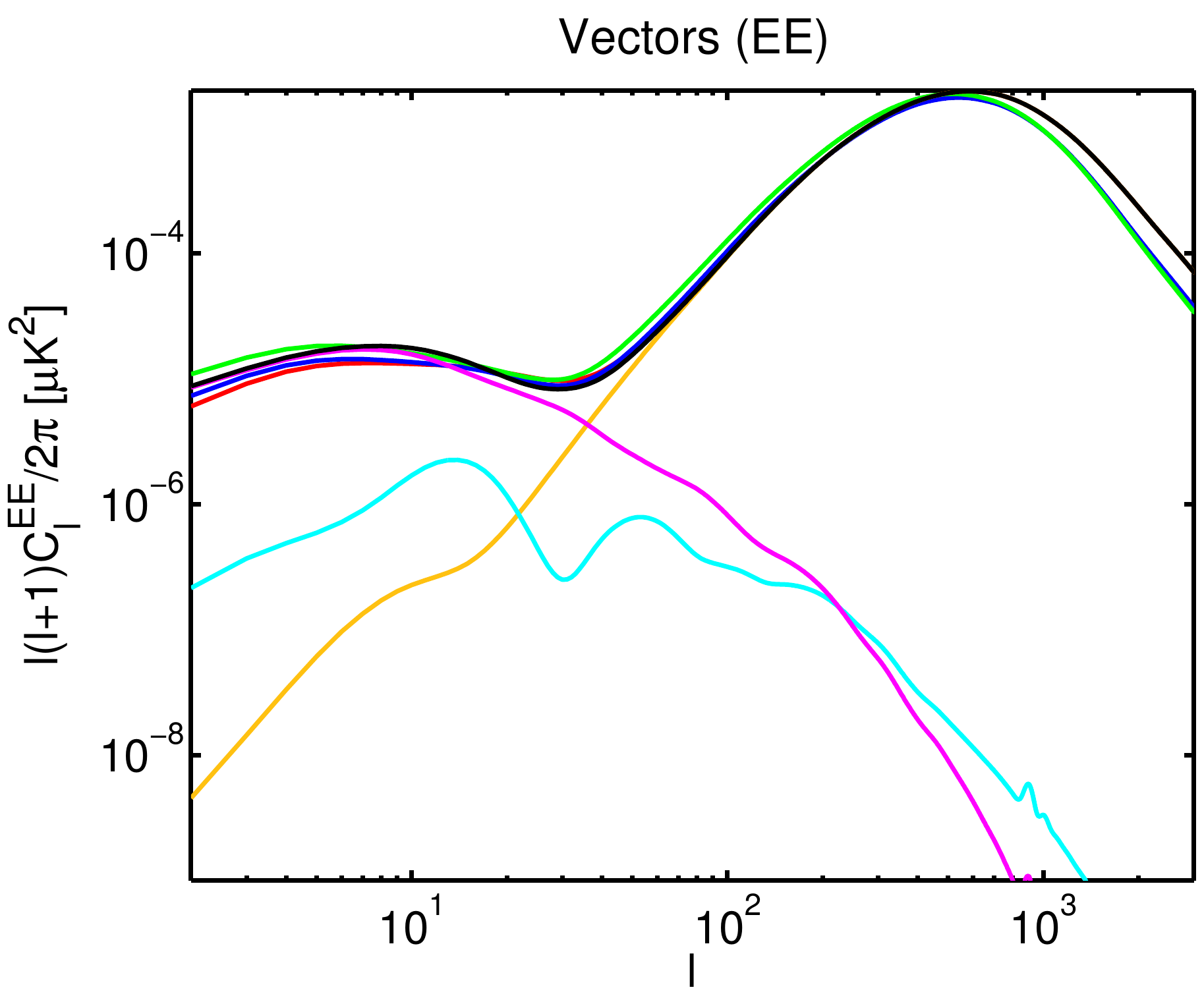} &
\includegraphics[width=2.07in]{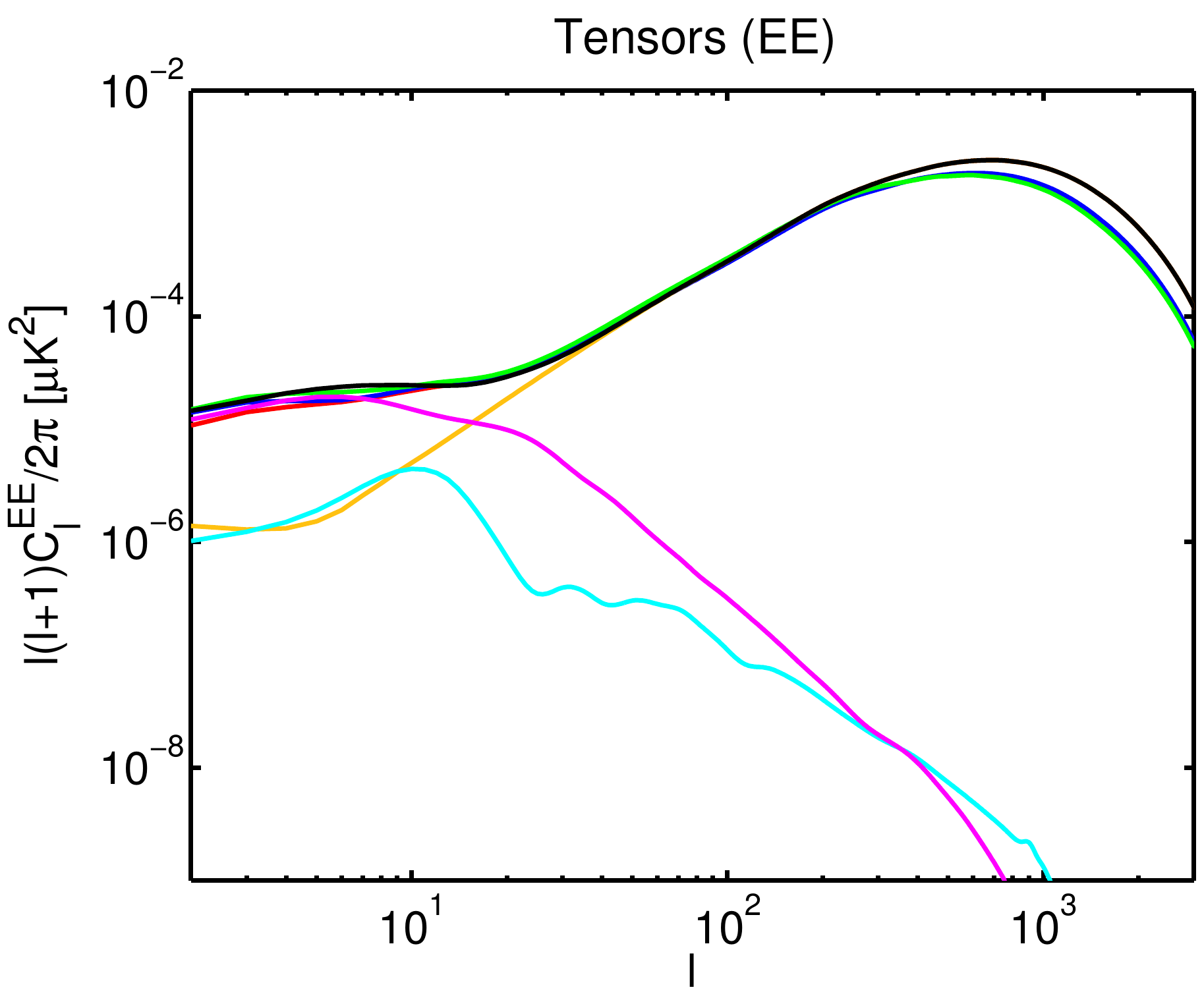} \\
\includegraphics[width=2.07in]{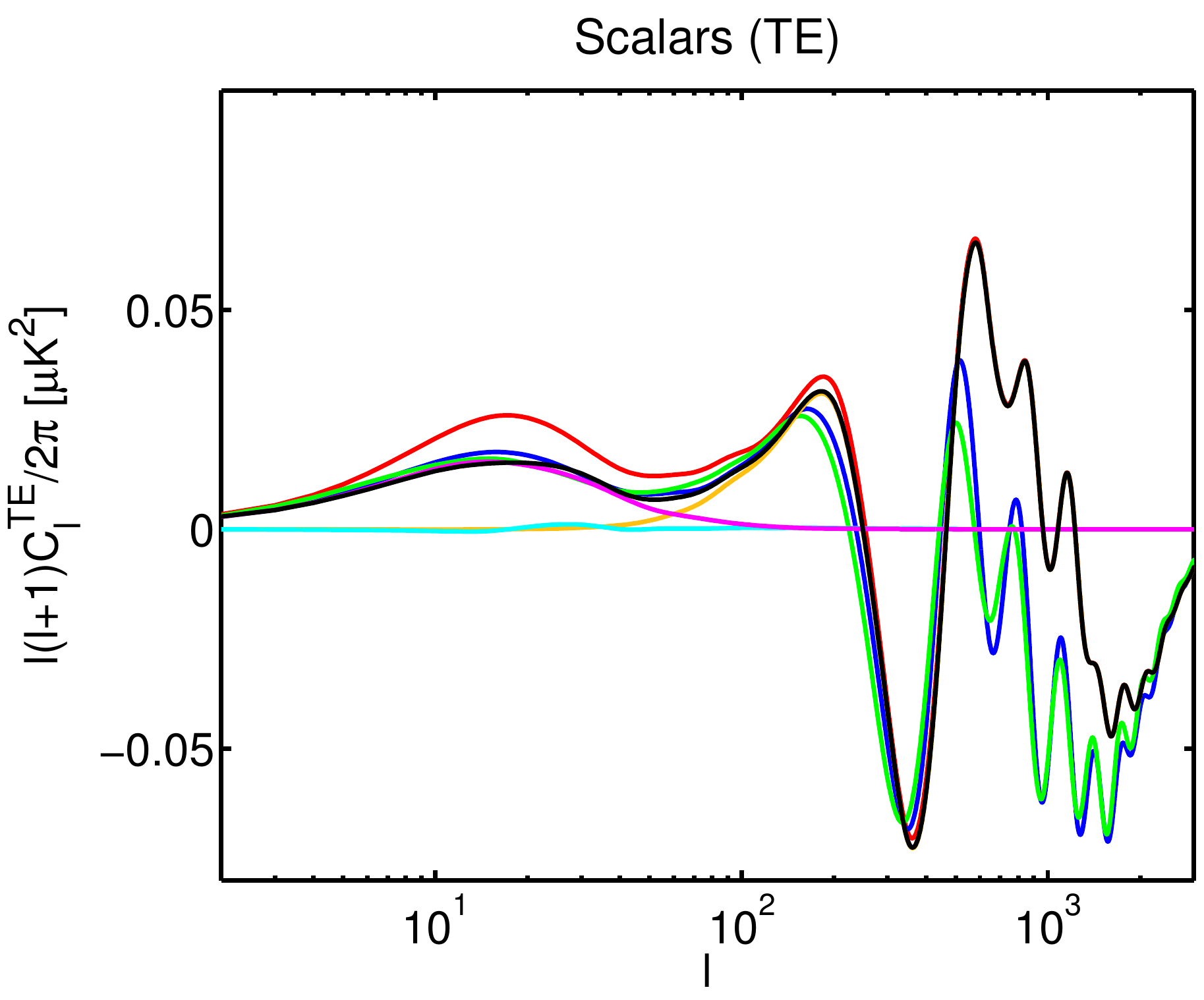} &
\includegraphics[width=2.07in]{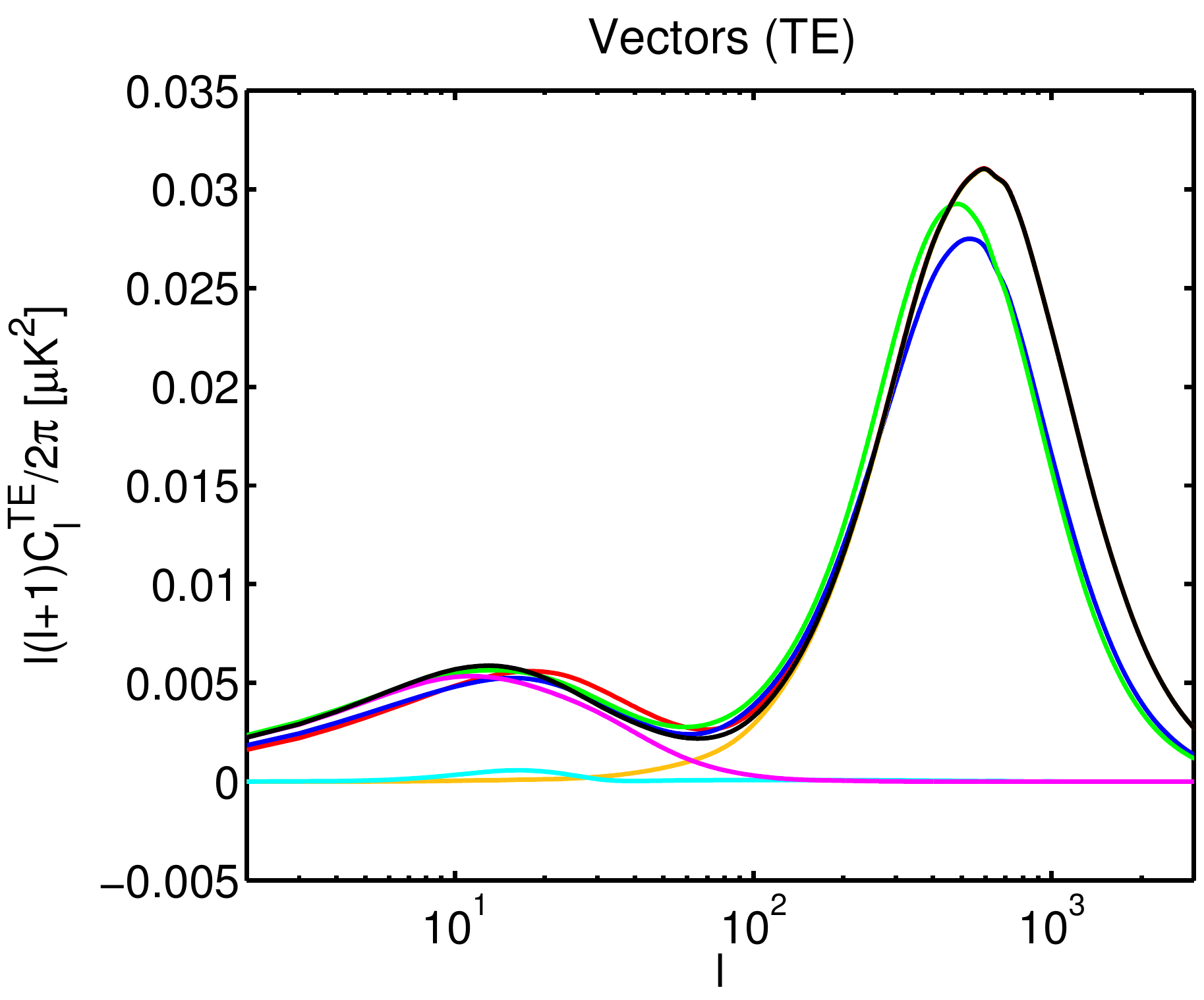} &
\includegraphics[width=2.07in]{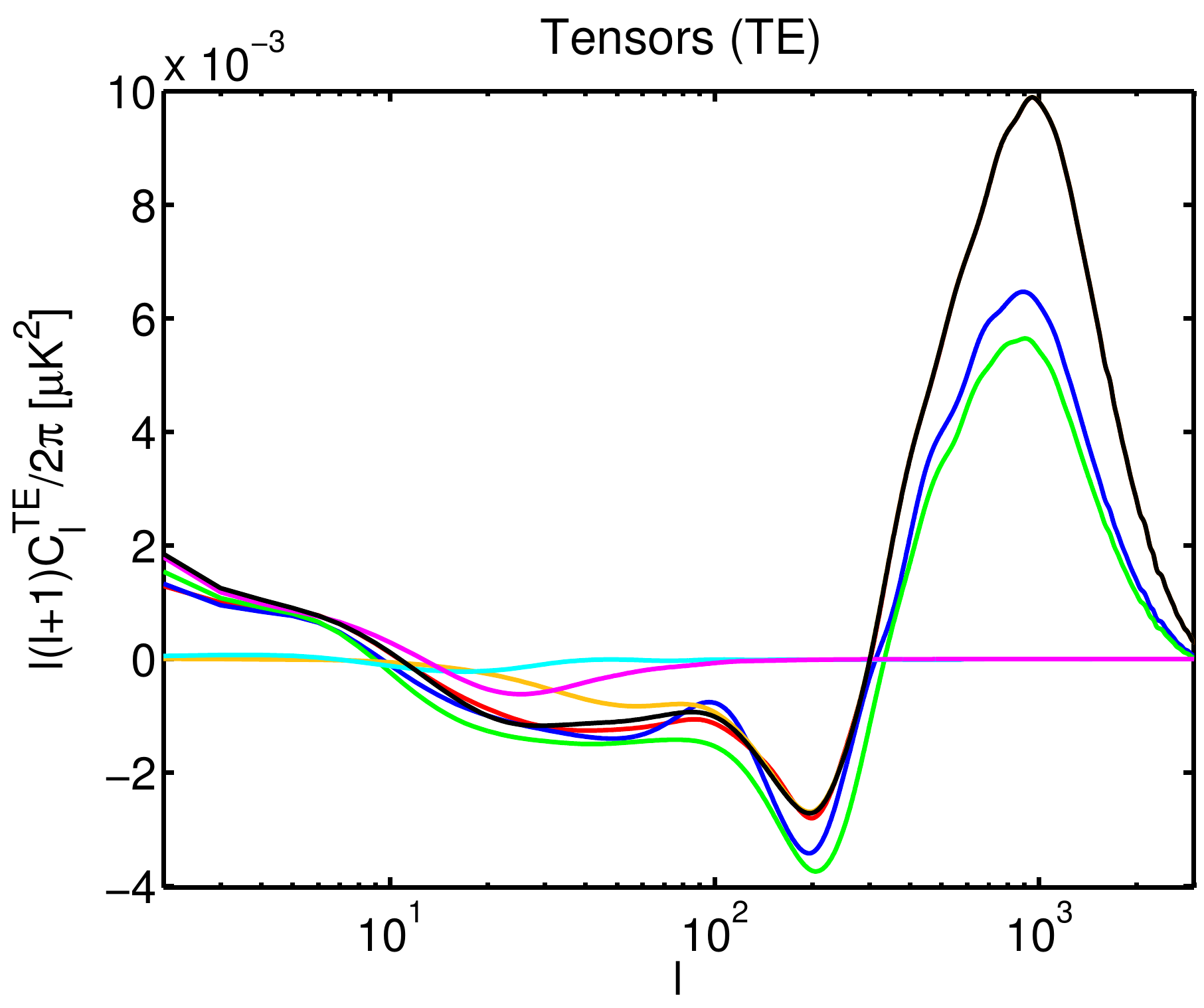} \\
\includegraphics[width=2.07in]{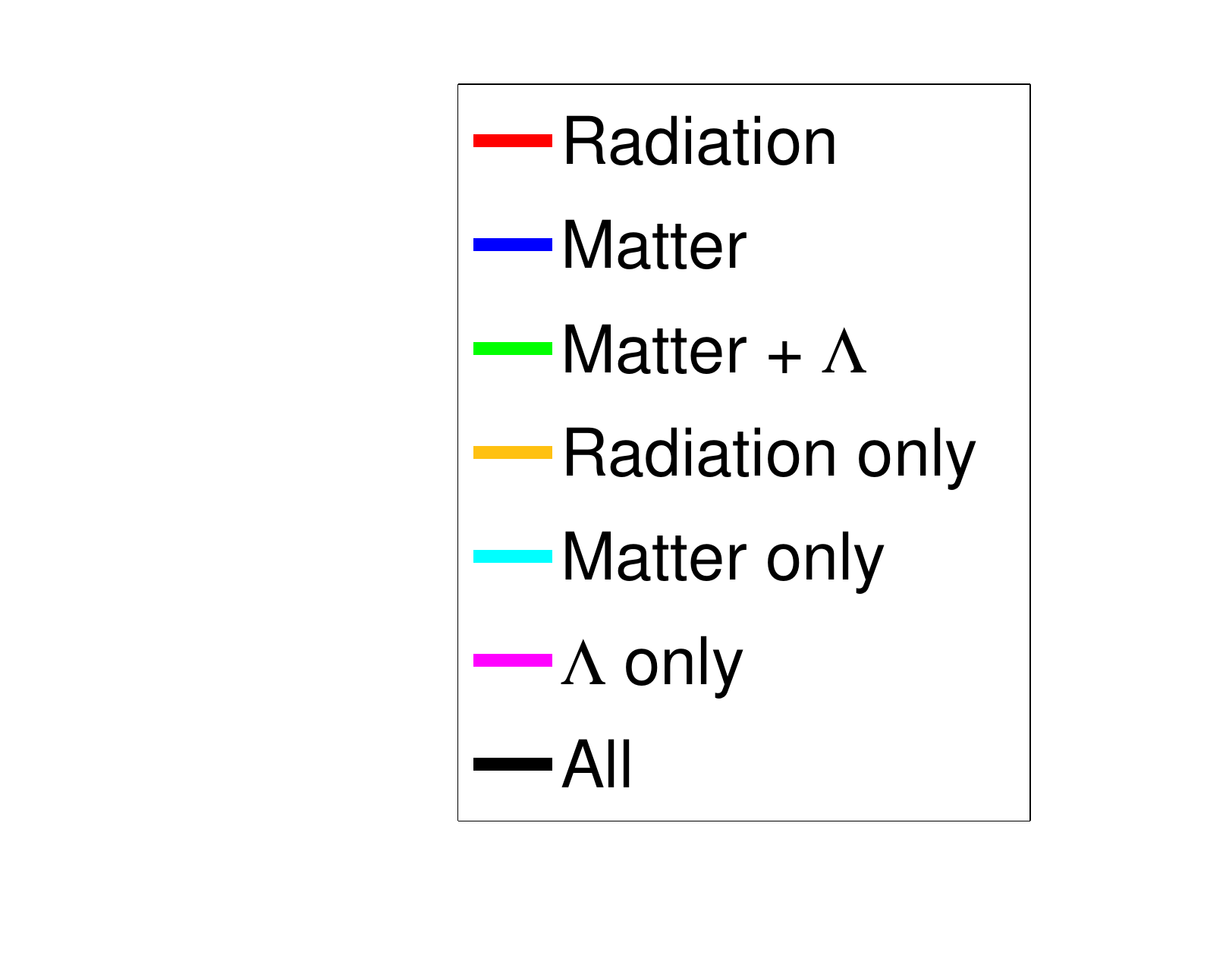} &
\includegraphics[width=2.07in]{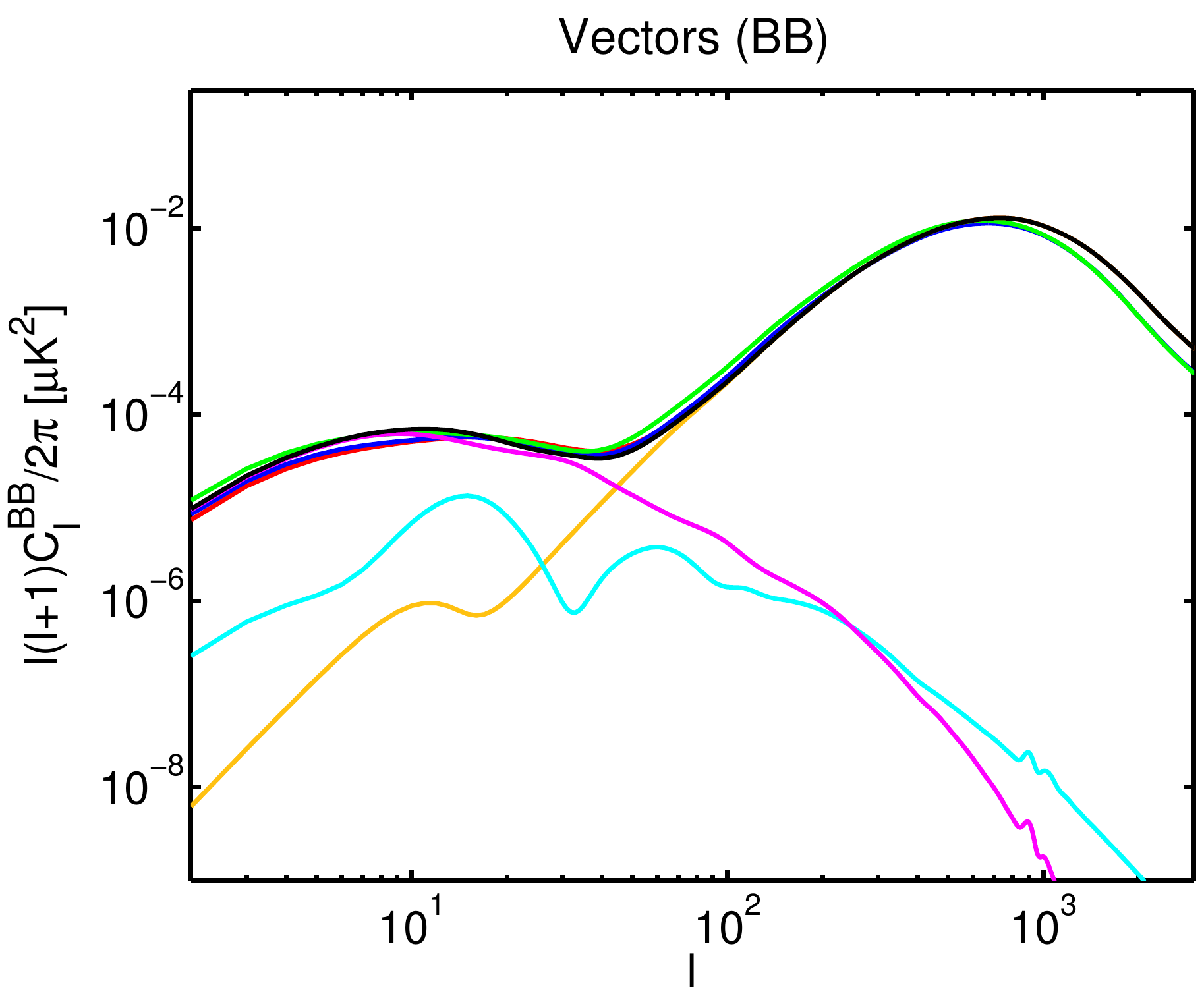} &
\includegraphics[width=2.07in]{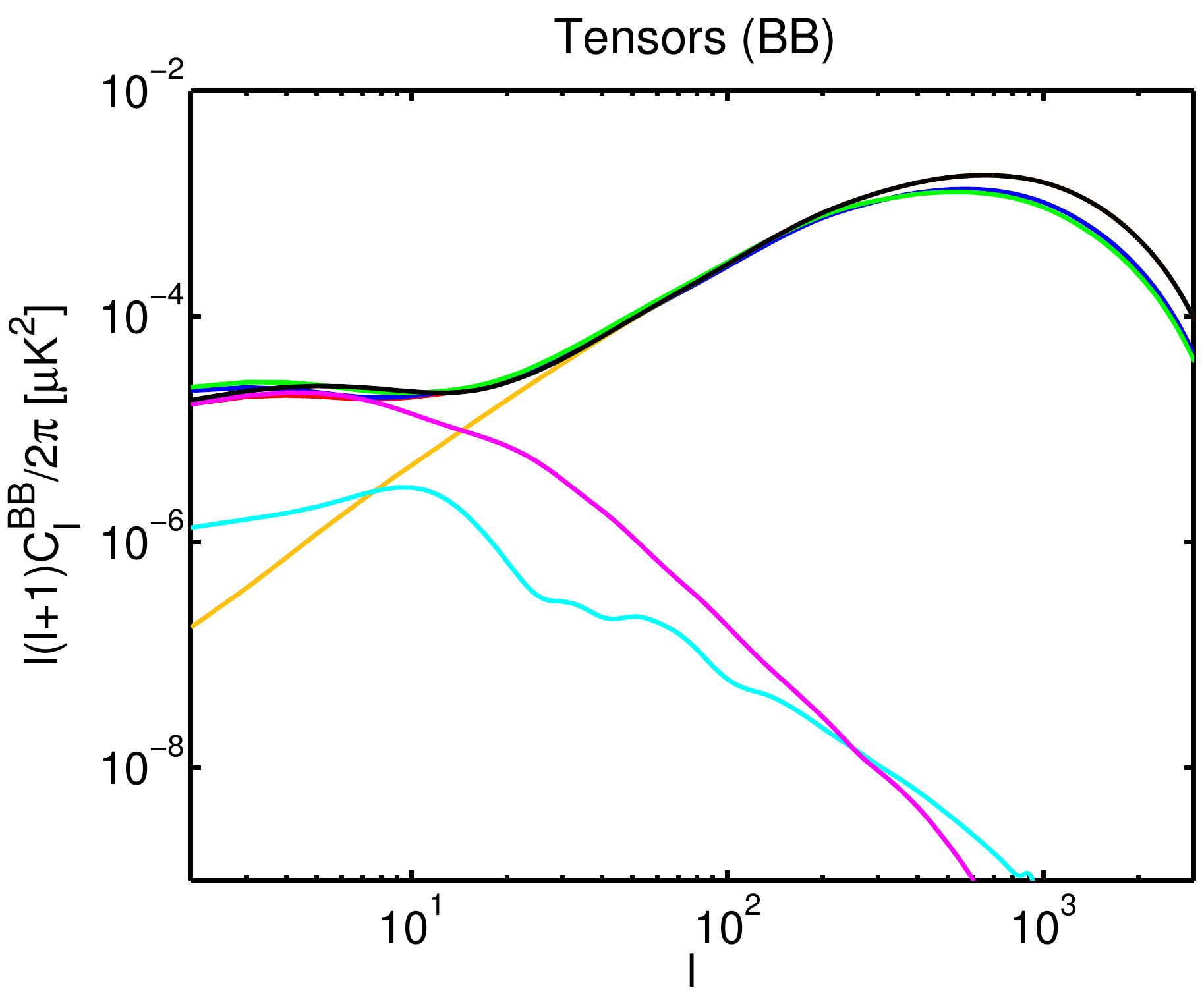} \\
\end{array}$
\caption{Power spectra of the cosmic strings obtained by using each of the three sets of UETCs and assuming scaling for the whole history of the Universe. The red, green, and blue show the power spectra considering the extrapolation of the results obtained in the radiation, matter, and matter + $\Lambda$ epochs. The contributions from the UETCs from just the time interval where they are valid are plotted in the yellow, cyan, and and magenta curves, and their sum is in black. The black curve represents the final overall power spectrum obtained. From left to right: The scalar, vector, and tensor power spectra; from top to bottom the TT, EE, TE, and BB power spectra ($G\mu=1.5 \times 10^{-7}$).}
\label{mixed_ps_uetc}
\end{center}
\end{figure*}

From the comparison of the results obtained from the simulations with the ones found by fitting the three parameters in CMBACT we notice that, unfortunately, the fits do not match the results from the simulations very well. The comparison between the simulations and the fits in the case of the temperature power spectrum is illustrated in Fig. \ref{TTall}. The standard USM and Abelian-Higgs power spectra are also plotted for comparison.

We used the three simulations separately, assuming their validity in the redshift range in which they were run, and we calculated the relevant $C_l$'s in each case, and then we added the results up. We will show the methodology used for combining the results from the three simulations for calculating the total combined angular power spectrum in the following paragraph.

Let $\hat{n}$ be the direction of the photon propagation, $\mu=\cos \left (\vec{k},\vec{p} \right )$ the angle between the wave vector and the momentum of the photon, and $\tau$ the conformal time, $\tau_0$ the conformal time today. We define the brightness function in terms of the relative variation of the temperature by:
\begin{equation}
\Delta(k,\mu,\tau)=4\frac{\Delta T}{T}
\end{equation}

By using the collisional Boltzmann equation, and using the perturbations from Eq. (\ref{pert}), it can be shown that the brightness function $\Delta$ satisfies the differential equation:
\begin{equation}
\Delta'+ik\mu\Delta=-2h_{ij}'\hat{n}^i\hat{n}^j+\dot{\tau}(\delta_\gamma+4\hat{n} \cdot v-\Delta)
\label{delta}
\end{equation}
where $\dot{\tau}$ is the differential Thomson cross section and $\delta_\gamma$ is the photon perturbation.

We assume that the cosmic string energy-momentum tensor (in this case the corresponding eigenvector) is nonzero only in a conformal time interval $(\tau^{(A)},\tau^{(B)})$. We will show that the time derivative of $h^{\alpha}$ tends to zero outside this interval. Equations. (\ref{modif2})-(\ref{modif3}) are linear and their initial conditions are $h^{\alpha}=\dot{h}^{\alpha}=0$ at $\tau=0$, with $\alpha$ corresponding to the scalars, vectors, and tensors. Hence, $h^{\alpha}(\tau)=\dot{h}^{\alpha}(\tau)=0$ for $\tau<\tau^{(A)}$. For $\tau>\tau^{(B)}$, there is no longer any source present, and hence $h^{(\alpha)}$ would at most remain constant while its time derivative would quickly decay. Hence, $\Delta=0$ in the absence of cosmic strings (due to the suitable initial conditions).

Equation (\ref{delta}) can be decomposed into eigenmodes using Legendre polynomials:
\begin{equation}
\Delta(k,\mu,\tau_0)=\sum_0^{\infty} \frac{2l+1}{i^l}\Delta_l(k,\tau_0)P_l(\mu)
\end{equation}

The integral identities involving the Legendre polynomials and the spherical Bessel functions:
\begin{eqnarray}
\int_{-1}^{1}P_m(x)P_n(x)dx=\frac{2}{2n+1}\delta_{mn} \\
\frac{i^l}{2}\int_{-1}^{1}P_l(\mu)e^{ik\mu(\tau-\tau_0)}d\mu=j_l((k(\tau_0-\tau)) 
\end{eqnarray}
can be used.
After splitting Eq. (\ref{delta}) into scalar, vector, and tensor modes, let $S=S(\dot{h}^{\alpha})$ be the source function due to strings in each of the cases above. 
\begin{equation}
\Delta_l(k,\tau_0)=\int_0^{\tau_0}d\tau S(k,\tau)j_l(k(\tau_0-\tau))
\label{delta_source}
\end{equation}

The corresponding angular power spectrum is expressed as:
\begin{equation}
C_l=\frac{2}{\pi}\int_0^{\infty}dk k^2 \Delta_l(k,\tau_0)^2
\label{cl}
\end{equation}

Each of the simulations considered is valid in a different time range. In the previous section we have extended the validity of the simulations by assuming scaling. However, scaling is not perfect throughout the history of Universe, as can be seen from the power spectra that we have obtained by making this assumption (Fig. \ref{mixed_ps_uetc}). If scaling were perfect, the power spectra from the three simulations would have to be identical. We consider the energy-momentum tensor as follows:
\begin{equation}
\Theta(k,\tau) \to \left\{
\begin{array}{cl}
\frac{v_{\text{radiation}}(k\tau)}{\sqrt{\tau}} & \text{if } \tau \in \text{radiation era} \\
\frac{v_{\text{matter}}(k\tau)}{\sqrt{\tau}} & \text{if } \tau \in \text{matter era}\\
\frac{v_{\text{matter+}\Lambda}(k\tau)}{\sqrt{\tau}} & \text{if } \tau \in \Lambda \text{ era} 
\end{array} \right.
\label{impartire}
\end{equation}

Equation (\ref{delta}) is a differential equation which is linear in the cosmic string sources and hence Eq. (\ref{delta_source}) has the same property for all values of $l$.  This shows that splitting the sources into three parts, computing the $\Delta_l$ functions separately, and then summing up the results would not change the integral. We will now consider that the string sources only act in the time interval where they are defined and we will split the calculation into three parts, corresponding to each of the epochs. For example, for the radiation era, we shall take the energy momentum-tensor from Eq. (\ref{impartire}) as:
\begin{equation}
\Theta(k,\tau) \to \left\{
\begin{array}{cc}
\frac{v_{\text{radiation}}(k\tau)}{\sqrt{\tau}} & \text{if } \tau \in \text{radiation era} \\
0 & \text{if } \tau \not\in \text{radiation era}\\
\end{array} \right.
\label{impartire_rad}
\end{equation}

More generally, we will assume that the sources $S$ from Eq. (\ref{delta_source}) can be written as a sum as: 
\begin{equation}
S(k,\tau)=\sum S_i(k,\tau)
\end{equation}
where each of the $S_i's$ is defined on an interval $(\tau_i^{(A)},\tau_i^{(B)})$. These intervals are disjoint. This is possible because the differential involved for $\Delta_l$ and $h$ are linear. However, in the expression for $C_l$ there is a square of $\Delta_l$. So we can reexpress Eq. (\ref{cl}) as:
\begin{eqnarray}
C_l=\sum_i C_l^{i}+\frac{4}{\pi}\sum_{i<j}\int_0^{\infty}dk k^2 \int_0^{\tau_0} d\tau_1 \int_0^{\tau_0} d\tau_2 \nonumber \\
S_i(k,\tau_1)S_j(k,\tau_2)j_l(k(\tau_0-\tau_1))j_l(k(\tau_0-\tau_2))
\label{cltot}
\end{eqnarray}
where $C_l^{i}$ represents the contribution to the angular power spectrum obtained only from source $i$ (e.g. only radiation era). We will now show that the last sum of integrals from Eq. (\ref{cltot}) is negligible compared to each of the terms in the first sum. We note that the sources $S$ oscillate much less in $k$ compared to the Bessel functions and hence, after changing the order of integration, a typical integral term from this sum can be reexpressed as:
\begin{gather}
\int_0^{\tau_0} d\tau_1 \int_0^{\tau_0} d\tau_2 S_i(\tau_1)S_j(\tau_2) \times \nonumber \\
\times \int_0^{\infty}dk k^2 j_l(k(\tau_0-\tau_1))j_l(k(\tau_0-\tau_2)) \sim \nonumber \\
\int_0^{\tau_0} d\tau_1 \int_0^{\tau_0} d\tau_2 S_i(\tau_1)S_j(\tau_2) \delta(\tau_2-\tau_1) = \nonumber \\
\int_0^{\tau_0} d\tau_1 \ S_i(\tau_1)S_j(\tau_1)
\label{cl_2}
\end{gather}
using the properties of the spherical Bessel functions. We now assume $i<j$ and we take into account that the cosmic strings only source the perturbation equations in the intervals $(\tau_i^{(A)},\tau_i^{(B)})$ and $(\tau_j^{(A)},\tau_j^{(B)})$. The contribution from the first source will only start at $\tau_i^{(A)}$ and end at $\tau_i^{(B)}$. Hence, $S_1$ will be zero before $\tau_i^{(A)}$ and start decaying after $\tau_i^{(B)}$. The decay of the sources after there are no strings is exponential in time. A similar behaviour is expected from the second cosmic string region. Hence, the integral (\ref{cl_2}) will only have a nonzero contribution in the region where the contribution of the first source has not completely decayed and the second source has an increasing contribution. As this contribution is suppressed due to the time decay of the sources $S_i$, this last integral will give a very small contribution and we will neglect it.

The results that we obtained show that, in the TT spectrum, the cosmological constant era contributes at  $l<100$ with a peak at $l=30$, the matter era contributes in the range $50<l<400$, and the radiation simulation for $l>200$, as expected. The total power spectrum converges to the matter and $\Lambda$ era result for low $l$ and the radiation era one at high $l$. The final results resemble most the extrapolated matter era simulation, in agreement with the results reported in Ref. \cite{durrer}. The other three spectra (TE, EE, and BB) exhibit a similar behaviour but the signal is dominated by the one from the radiation era. The individual results are shown in Fig. \ref{mixed_ps_uetc}. We have used 200 eigenvectors for each of the lines in the plots.

\begin{figure*}[!htb]
\begin{center}$
\begin{array}{cc}
\includegraphics[width=2.5in]{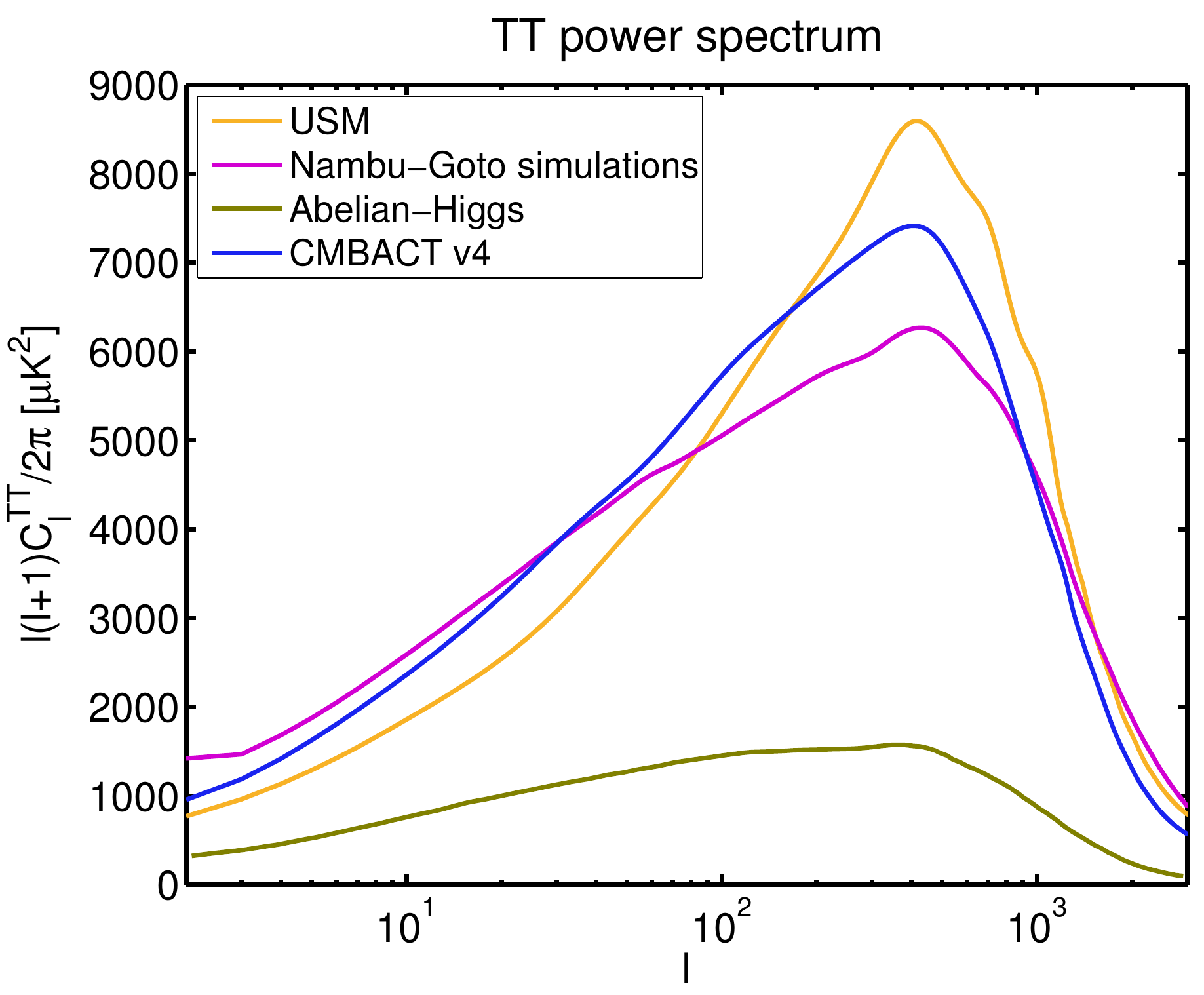} &
\includegraphics[width=2.5in]{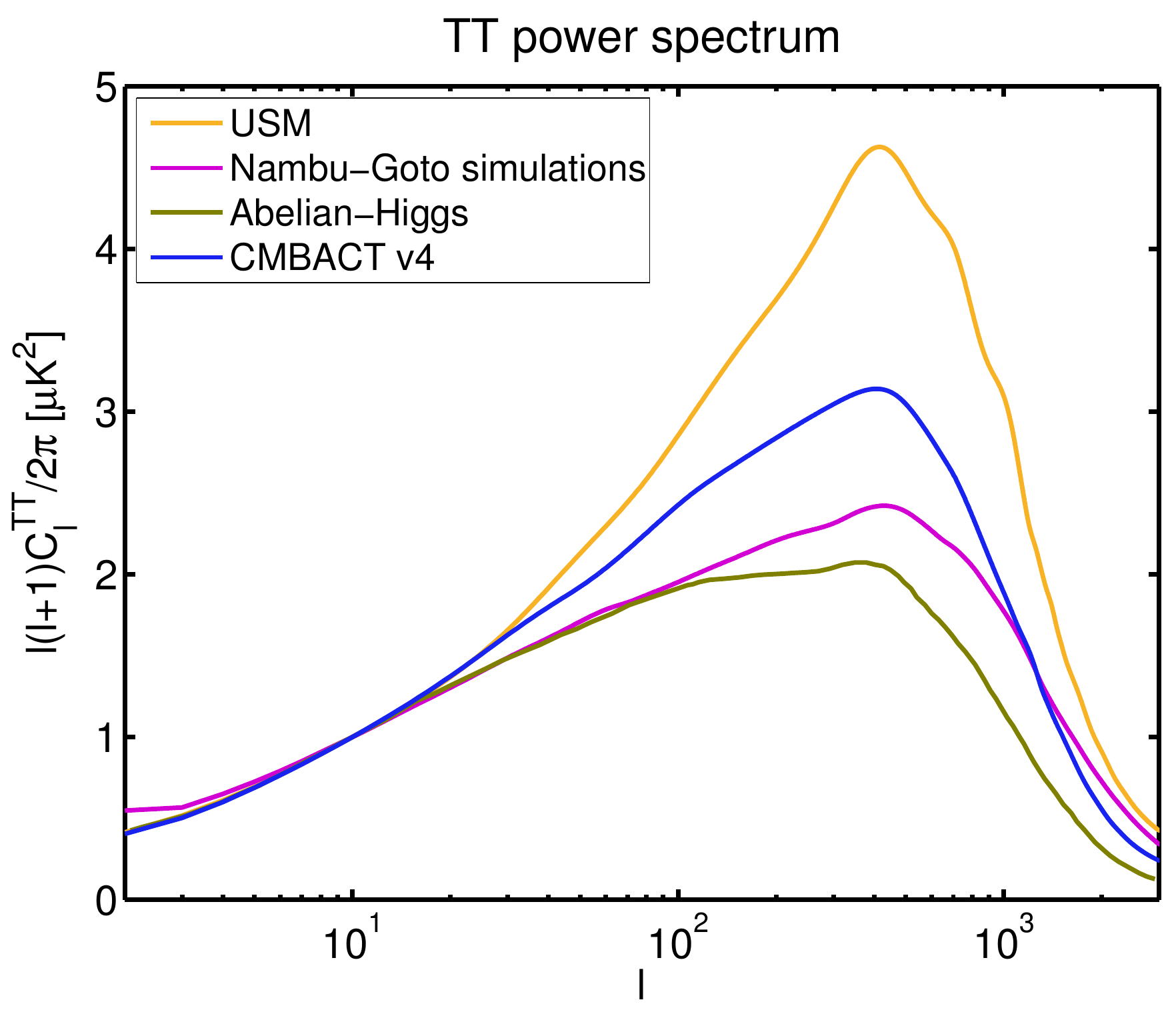}
\end{array}$
\caption{Comparison between the TT power spectra obtained using the three simulations and the USM, Abelian-Higgs (standard results) and CMBACT version 4. The string tension is taken $G\mu=2.07 \times 10^{-6}$. On the left the results are represented showing that USM and Nambu-Goto have different amplitudes to Abelian-Higgs cosmic strings, while the right shows similar shapes (normalised at $l=10$).}
\label{TTfinal}
\end{center}
\end{figure*}

In Fig. \ref{TTfinal} we show the final TT power spectrum obtained from the three Nambu-Goto simulations (combined), together with the USM and Abelian-Higgs ones. In addition, we also plot the results obtained with the fourth version of the code CMBACT \cite{cmbact}, in which the author has corrected various bugs but also updated the VOS model. This new version gives a lower amplitude for the temperature power spectrum and its overall shape resembles more the Abelian-Higgs one. Using our simulations, we obtain an even lower amplitude for the power spectrum. The peak remains at roughly the same position as in the USM case. The shape of our TT power spectrum is more similar in terms of amplitude to the USM result, but its shape resembles more the Abelian-Higgs spectrum.

\section{String tension constraints}

To constrain the power spectrum contribution from the Nambu-Goto string simulations, we have used a Markov chain Monte Carlo method, using a  modified version of the COSMOMC code \cite{cosmomc, cosmomc2}. This method involves evaluating the power spectrum each time the parameters are modified, by calling an instance of the code CAMB \cite{CAMB}. The total power spectrum is obtained from the sum between the inflationary spectrum and the one obtained from cosmic strings because the cosmic string sources, which are active sources, are uncorrelated with the primordial perturbations \cite{wyman}. This would in principle require the calculation of the cosmic string power spectrum many thousands of times, for each choice of cosmological parameters, which is not feasible because calculating the cosmic string power spectrum by itself requires several hours of computational work. Fortunately it has been suggested \cite{Pogosian, albrecht} that it evolves much slower as a function of the parameters compared to its inflationary counterpart. In Ref. \cite{PhysRevD.69.023003} it has been explicitly shown that the cosmological constant varies less than 10\% with $G\mu/c^2$. The cosmic strings are expected to contribute less than 5\% in the total power spectrum, so as the cosmological parameters are varied in the allowed regions, the string power spectrum does not vary more than 20\% \cite{kunz-cosmomc, battye-cosmomc}. This gives overall better than 1\% accuracy for the contribution of cosmic strings, which is greater than the accuracy of CAMB. Hence we have calculated the cosmic string power spectrum for a particular set of cosmological parameters and we only allow the overall string contribution to vary, through the parameter $f_{10}$, which represents the fractional power of the cosmic strings compared to the inflationary power at the tenth multipole \cite{battye2-cosmomc, planckstr}:
\begin{equation}
f_{10}=\frac{C_{10}^{\text{string}}}{C_{10}^{\text{total}}} 
\end{equation}
We also use the relation $f_{10}\propto (G\mu)^2$ to relate the new parameter to the string tension \cite{urrestilla-cosmomc}. We have used the latest version of the COSMOMC code, together with the BICEP data \cite{bicep}, in order to obtain accurate constraints on the maximum allowed string tension. We have used the standard cosmological parameters (the baryon density $\Omega_bh^2$, the cold dark matter density $\Omega_hh^2$, the optical depth to reionisation $\tau$, the expansion rate today $H_0$, the acoustic scale $\theta$, the amplitude $A_s$, and the spectral index of density fluctuations $n_s$), together with the Planck nuisance parameters in order to obtain a full likelihood calculation. In this paper, we present the result that we have obtained from using cosmic strings together with the $\Lambda$CDM parameters only in the Planck + WP case. In the situation where we have included BICEP2 likelihoods, we also need to include tensor modes. A more detailed description of the procedures involved, where we analyse degeneracies between cosmic strings and different nonstandard cosmological parameters (running of the spectral index, increasing the number of degrees of freedom, and  adding tensor modes and neutrinos in sterile states) is the object of another paper \cite{1475-7516-2015-02-024}. The two results that we have obtained at 95\% confidence level are $G\mu/c^2<1.49 \times 10^{-7}$ in the Planck only case and $G\mu/c^2<1.44 \times 10^{-7}$ for the BICEP2 and \textit{r} case. Our constraint in the Planck case is slightly stronger than the Planck Collaboration result \cite{planckstr}. This is due to the slightly different shape of our power spectrum to the USM one. We have validated our formalism by obtaining the Planck constraint with the USM data. In Table \ref{table-gmu} we present the results that we have obtained together with the results without cosmic strings, but otherwise using the same parameters. We observe that the standard cosmological parameters do not shift significantly. 
 
\begin{table*}[!htb]
\centering
\caption{Constraints on the fitted cosmological parameters, together with $1\sigma$ error bars in a full likelihood analysis (with all relevant nuisance parameters) with and without cosmic strings in the case of Planck and WMAP polarisation (left) and Planck, WMAP polarisation, BICEP2 likelihoods and tensor modes (right).}
\begin{tabular}{|c|c|c|c|c|}
\hline
  & \multicolumn{2}{ |c| }{Planck + WP} & \multicolumn{2}{ |c| }{Planck + WP + BICEP2 + $r$}\\
\hline
Parameter                           & No strings         & Strings                & No strings          & Strings    \\ \hline
$G\mu/c^2 < (2\sigma)$              & -                  & $1.49 \times 10^{-7}$  & -                   & $1.44 \times 10^{-7}$  \\ \hline
$G\mu/c^2$ (best fit)               & -                  & $4.99 \times 10^{-8}$  & -                   & $8.30 \times 10^{-8}$  \\ \hline
$r$                                 & -                  & -                      & $0.15 \pm 0.04$     & $0.15 \pm 0.04$        \\ \hline
$H_0$                               & $67.20\pm 1.16$    & $67.42 \pm 1.20$       & $67.72 \pm 1.10$    & $67.95 \pm 1.20$      \\ \hline 
$100\Omega_b h^2$                   & $2.202\pm 0.027$   & $2.209 \pm 0.029$      & $2.203 \pm 0.028$   & $2.210 \pm 0.029$        \\ \hline
$\Omega_c h^2$                      & $0.120 \pm 0.003$  & $0.119 \pm 0.003$      & $0.119 \pm 0.003$   & $0.118 \pm 0.003$       \\ \hline
$\tau$                              & $0.089 \pm 0.013$  & $0.087 \pm 0.013$      & $0.089 \pm 0.013$   & $0.088 \pm 0.013$       \\ \hline
$100\theta_{MC}$                    & $1.0412 \pm 0.0006$& $1.0412 \pm 0.0006$    & $1.0413 \pm 0.0006$ & $1.0414 \pm 0.0007$        \\ \hline
$ln(10^{10}A_s)$                    & $3.088 \pm 0.025$  & $3.078 \pm 0.026$      & $3.085 \pm 0.025$   & $3.075 \pm 0.025$    \\ \hline
$n_s$                               & $0.959 \pm 0.007$  & $0.958\pm 0.007$       & $0.964 \pm 0.007$   & $0.964 \pm 0.0007$                   \\ \hline
\end{tabular}
\label{table-gmu}
\end{table*}

While the Planck results are very robust and consistent with results obtained by other authors with and without cosmic strings \cite{planckres, planckstr}, the BICEP2 observations still require confirmation by independent experiments \cite{2013arXiv1303.5082P} until definitive results can be claimed \cite{2014arXiv1409.5738P}.

\section{Summary and outlook}

In this paper we have used high precision numerical simulations of the time evolution of Nambu-Goto strings to determine the UETCs of the energy-momentum tensor components. We have established the resolution required to obtain robust and accurate results for the cosmic string power spectrum. The resulting Nambu-Goto spectrum is situated between that expected from the Abelian-Higgs and USM models (Fig. \ref{TTp}). In the case of the Planck data, this gives a slightly weaker constraint on $G\mu/c^2$ compared to the one expected from the USM model but tighter compared to that from the Abelian-Higgs model. The string tension is constrained to be $G\mu/c^2<1.49\times 10^{-7}$ and the fractional power to  $f_{10}<0.019$ when using the the Planck data.

In a companion publication \cite{1475-7516-2015-02-024}, we have used the temperature power spectrum likelihoods from the Planck mission \cite{planckres} and the polarisation data from WMAP and BICEP2 \cite{bicep} to obtain strong and robust constraints on the string tension. In the future we will also use polarisation information to obtain stronger constraints from the next Planck data release. 

\section{Acknowledgements}
The authors are very grateful to Carlos Martins and Levon Pogosian for many enlightening discussions. AL is supported by STFC. This work was supported by an STFC Consolidated Grant No. ST/L000636/1. This work was undertaken on the COSMOS Shared Memory system at DAMTP, University of Cambridge operated on behalf of the STFC DiRAC HPC Facility. This equipment is funded by BIS National E-Infrastructure Capital Grant No. ST/J005673/1 and STFC Grants No. ST/H008586/1 and No. ST/K00333X/1.

\bibliography{Bibliografie}{}

\begin{thebibliography}{70}%
\makeatletter
\providecommand \@ifxundefined [1]{%
 \@ifx{#1\undefined}
}%
\providecommand \@ifnum [1]{%
 \ifnum #1\expandafter \@firstoftwo
 \else \expandafter \@secondoftwo
 \fi
}%
\providecommand \@ifx [1]{%
 \ifx #1\expandafter \@firstoftwo
 \else \expandafter \@secondoftwo
 \fi
}%
\providecommand \natexlab [1]{#1}%
\providecommand \enquote  [1]{``#1''}%
\providecommand \bibnamefont  [1]{#1}%
\providecommand \bibfnamefont [1]{#1}%
\providecommand \citenamefont [1]{#1}%
\providecommand \href@noop [0]{\@secondoftwo}%
\providecommand \href [0]{\begingroup \@sanitize@url \@href}%
\providecommand \@href[1]{\@@startlink{#1}\@@href}%
\providecommand \@@href[1]{\endgroup#1\@@endlink}%
\providecommand \@sanitize@url [0]{\catcode `\\12\catcode `\$12\catcode
  `\&12\catcode `\#12\catcode `\^12\catcode `\_12\catcode `\%12\relax}%
\providecommand \@@startlink[1]{}%
\providecommand \@@endlink[0]{}%
\providecommand \url  [0]{\begingroup\@sanitize@url \@url }%
\providecommand \@url [1]{\endgroup\@href {#1}{\urlprefix }}%
\providecommand \urlprefix  [0]{URL }%
\providecommand \Eprint [0]{\href }%
\providecommand \doibase [0]{http://dx.doi.org/}%
\providecommand \selectlanguage [0]{\@gobble}%
\providecommand \bibinfo  [0]{\@secondoftwo}%
\providecommand \bibfield  [0]{\@secondoftwo}%
\providecommand \translation [1]{[#1]}%
\providecommand \BibitemOpen [0]{}%
\providecommand \bibitemStop [0]{}%
\providecommand \bibitemNoStop [0]{.\EOS\space}%
\providecommand \EOS [0]{\spacefactor3000\relax}%
\providecommand \BibitemShut  [1]{\csname bibitem#1\endcsname}%
\let\auto@bib@innerbib\@empty
\bibitem [{\citenamefont {Vilenkin}\ and\ \citenamefont
  {Shellard}(1994)}]{shbook}%
  \BibitemOpen
  \bibfield  {author} {\bibinfo {author} {\bibfnamefont {A.}~\bibnamefont
  {Vilenkin}}\ and\ \bibinfo {author} {\bibfnamefont {E.~P.~S.}\ \bibnamefont
  {Shellard}},\ }\href@noop {} {\emph {\bibinfo {title} {{Cosmic Strings and
  Other Topological Defects}}}}\ (\bibinfo  {publisher} {Cambridge University
  Press},\ \bibinfo {year} {1994})\BibitemShut {NoStop}%
\bibitem [{\citenamefont {Kaiser}\ and\ \citenamefont
  {Stebbins}(1984)}]{kaiser}%
  \BibitemOpen
  \bibfield  {author} {\bibinfo {author} {\bibfnamefont {N.}~\bibnamefont
  {Kaiser}}\ and\ \bibinfo {author} {\bibfnamefont {A.}~\bibnamefont
  {Stebbins}},\ }\href {http://dx.doi.org/10.1038/310391a0} {\bibfield
  {journal} {\bibinfo  {journal} {Nature}\ }\textbf {\bibinfo {volume} {310}},\
  \bibinfo {pages} {391} (\bibinfo {year} {1984})}\BibitemShut {NoStop}%
\bibitem [{\citenamefont {Gott}(1985)}]{Gott:1984ef}%
  \BibitemOpen
  \bibfield  {author} {\bibinfo {author} {\bibfnamefont {I.}~\bibnamefont
  {Gott}, \bibfnamefont {J.~Richard}},\ }\href {\doibase 10.1086/162808}
  {\bibfield  {journal} {\bibinfo  {journal} {Astrophys.J.}\ }\textbf {\bibinfo
  {volume} {288}},\ \bibinfo {pages} {422} (\bibinfo {year}
  {1985})}\BibitemShut {NoStop}%
\bibitem [{\citenamefont {{Planck
  Collaboration}}(2014{\natexlab{a}})}]{planckres}%
  \BibitemOpen
  \bibfield  {author} {\bibinfo {author} {\bibnamefont {{Planck
  Collaboration}}},\ }\href {\doibase 10.1051/0004-6361/201321529} {\bibfield
  {journal} {\bibinfo  {journal} {A\&A}\ }\textbf {\bibinfo {volume} {571}},\
  \bibinfo {pages} {A1} (\bibinfo {year} {2014}{\natexlab{a}})}\BibitemShut
  {NoStop}%
\bibitem [{\citenamefont {Pogosian}\ and\ \citenamefont
  {Vachaspati}(1999)}]{Pogosian}%
  \BibitemOpen
  \bibfield  {author} {\bibinfo {author} {\bibfnamefont {L.}~\bibnamefont
  {Pogosian}}\ and\ \bibinfo {author} {\bibfnamefont {T.}~\bibnamefont
  {Vachaspati}},\ }\href {\doibase 10.1103/PhysRevD.60.083504} {\bibfield
  {journal} {\bibinfo  {journal} {Phys. Rev. D}\ }\textbf {\bibinfo {volume}
  {60}},\ \bibinfo {pages} {083504} (\bibinfo {year} {1999})}\BibitemShut
  {NoStop}%
\bibitem [{\citenamefont {Bevis}\ \emph {et~al.}(2007)\citenamefont {Bevis},
  \citenamefont {Hindmarsh}, \citenamefont {Kunz},\ and\ \citenamefont
  {Urrestilla}}]{hindmarsh2006}%
  \BibitemOpen
  \bibfield  {author} {\bibinfo {author} {\bibfnamefont {N.}~\bibnamefont
  {Bevis}}, \bibinfo {author} {\bibfnamefont {M.}~\bibnamefont {Hindmarsh}},
  \bibinfo {author} {\bibfnamefont {M.}~\bibnamefont {Kunz}}, \ and\ \bibinfo
  {author} {\bibfnamefont {J.}~\bibnamefont {Urrestilla}},\ }\href {\doibase
  10.1103/PhysRevD.75.065015} {\bibfield  {journal} {\bibinfo  {journal} {Phys.
  Rev. D}\ }\textbf {\bibinfo {volume} {75}},\ \bibinfo {pages} {065015}
  (\bibinfo {year} {2007})}\BibitemShut {NoStop}%
\bibitem [{\citenamefont {Fraisse}\ \emph {et~al.}(2008)\citenamefont
  {Fraisse}, \citenamefont {Ringeval}, \citenamefont {Spergel},\ and\
  \citenamefont {Bouchet}}]{PhysRevD.78.043535}%
  \BibitemOpen
  \bibfield  {author} {\bibinfo {author} {\bibfnamefont {A.}~\bibnamefont
  {Fraisse}}, \bibinfo {author} {\bibfnamefont {C.}~\bibnamefont {Ringeval}},
  \bibinfo {author} {\bibfnamefont {D.}~\bibnamefont {Spergel}}, \ and\
  \bibinfo {author} {\bibfnamefont {F.}~\bibnamefont {Bouchet}},\ }\href
  {\doibase 10.1103/PhysRevD.78.043535} {\bibfield  {journal} {\bibinfo
  {journal} {Phys. Rev. D}\ }\textbf {\bibinfo {volume} {78}},\ \bibinfo
  {pages} {043535} (\bibinfo {year} {2008})}\BibitemShut {NoStop}%
\bibitem [{\citenamefont {Ringeval}\ and\ \citenamefont
  {Bouchet}(2012)}]{PhysRevD.86.023513}%
  \BibitemOpen
  \bibfield  {author} {\bibinfo {author} {\bibfnamefont {C.}~\bibnamefont
  {Ringeval}}\ and\ \bibinfo {author} {\bibfnamefont {F.}~\bibnamefont
  {Bouchet}},\ }\href {\doibase 10.1103/PhysRevD.86.023513} {\bibfield
  {journal} {\bibinfo  {journal} {Phys. Rev. D}\ }\textbf {\bibinfo {volume}
  {86}},\ \bibinfo {pages} {023513} (\bibinfo {year} {2012})}\BibitemShut
  {NoStop}%
\bibitem [{\citenamefont {Regan}\ and\ \citenamefont
  {Shellard}(2010)}]{PhysRevD.82.063527}%
  \BibitemOpen
  \bibfield  {author} {\bibinfo {author} {\bibfnamefont {D.}~\bibnamefont
  {Regan}}\ and\ \bibinfo {author} {\bibfnamefont {E.}~\bibnamefont
  {Shellard}},\ }\href {\doibase 10.1103/PhysRevD.82.063527} {\bibfield
  {journal} {\bibinfo  {journal} {Phys. Rev. D}\ }\textbf {\bibinfo {volume}
  {82}},\ \bibinfo {pages} {063527} (\bibinfo {year} {2010})}\BibitemShut
  {NoStop}%
\bibitem [{\citenamefont {Sanidas}\ \emph {et~al.}(2012)\citenamefont
  {Sanidas}, \citenamefont {Battye},\ and\ \citenamefont
  {Stappers}}]{PhysRevD.85.122003}%
  \BibitemOpen
  \bibfield  {author} {\bibinfo {author} {\bibfnamefont {S.~A.}\ \bibnamefont
  {Sanidas}}, \bibinfo {author} {\bibfnamefont {R.~A.}\ \bibnamefont {Battye}},
  \ and\ \bibinfo {author} {\bibfnamefont {B.~W.}\ \bibnamefont {Stappers}},\
  }\href {\doibase 10.1103/PhysRevD.85.122003} {\bibfield  {journal} {\bibinfo
  {journal} {Phys. Rev. D}\ }\textbf {\bibinfo {volume} {85}},\ \bibinfo
  {pages} {122003} (\bibinfo {year} {2012})}\BibitemShut {NoStop}%
\bibitem [{\citenamefont {Blanco-Pillado}\ \emph {et~al.}(2014)\citenamefont
  {Blanco-Pillado}, \citenamefont {Olum},\ and\ \citenamefont
  {Shlaer}}]{PhysRevD.89.023512}%
  \BibitemOpen
  \bibfield  {author} {\bibinfo {author} {\bibfnamefont {J.~J.}\ \bibnamefont
  {Blanco-Pillado}}, \bibinfo {author} {\bibfnamefont {K.~D.}\ \bibnamefont
  {Olum}}, \ and\ \bibinfo {author} {\bibfnamefont {B.}~\bibnamefont
  {Shlaer}},\ }\href {\doibase 10.1103/PhysRevD.89.023512} {\bibfield
  {journal} {\bibinfo  {journal} {Phys. Rev. D}\ }\textbf {\bibinfo {volume}
  {89}},\ \bibinfo {pages} {023512} (\bibinfo {year} {2014})}\BibitemShut
  {NoStop}%
\bibitem [{\citenamefont {Avgoustidis}\ \emph {et~al.}(2012)\citenamefont
  {Avgoustidis}, \citenamefont {Copeland}, \citenamefont {Moss},\ and\
  \citenamefont {Skliros}}]{copeland}%
  \BibitemOpen
  \bibfield  {author} {\bibinfo {author} {\bibfnamefont {A.}~\bibnamefont
  {Avgoustidis}}, \bibinfo {author} {\bibfnamefont {E.}~\bibnamefont
  {Copeland}}, \bibinfo {author} {\bibfnamefont {A.}~\bibnamefont {Moss}}, \
  and\ \bibinfo {author} {\bibfnamefont {D.}~\bibnamefont {Skliros}},\ }\href
  {\doibase 10.1103/PhysRevD.86.123513} {\bibfield  {journal} {\bibinfo
  {journal} {Phys. Rev. D}\ }\textbf {\bibinfo {volume} {86}},\ \bibinfo
  {pages} {123513} (\bibinfo {year} {2012})}\BibitemShut {NoStop}%
\bibitem [{\citenamefont {Pogosian}(2014)}]{cmbact}%
  \BibitemOpen
  \bibfield  {author} {\bibinfo {author} {\bibfnamefont {L.}~\bibnamefont
  {Pogosian}},\ }\href@noop {} {\enquote {\bibinfo {title} {{CMBACT code -
  \url{http://www.sfu.ca/~levon/cmbact.html}}},}\ } (\bibinfo {year}
  {2009-2014})\BibitemShut {NoStop}%
\bibitem [{\citenamefont {Kirzhnits}\ and\ \citenamefont
  {Linde}(1972)}]{Kirzhnits}%
  \BibitemOpen
  \bibfield  {author} {\bibinfo {author} {\bibfnamefont {D.}~\bibnamefont
  {Kirzhnits}}\ and\ \bibinfo {author} {\bibfnamefont {A.}~\bibnamefont
  {Linde}},\ }\href {\doibase http://dx.doi.org/10.1016/0370-2693(72)90109-8}
  {\bibfield  {journal} {\bibinfo  {journal} {Phys. Lett. B}\ }\textbf
  {\bibinfo {volume} {42}},\ \bibinfo {pages} {471 } (\bibinfo {year}
  {1972})}\BibitemShut {NoStop}%
\bibitem [{\citenamefont {Weinberg}(1979)}]{weinberg1979prd}%
  \BibitemOpen
  \bibfield  {author} {\bibinfo {author} {\bibfnamefont {E.~J.}\ \bibnamefont
  {Weinberg}},\ }\href {\doibase 10.1103/PhysRevD.19.3008} {\bibfield
  {journal} {\bibinfo  {journal} {Phys. Rev. D}\ }\textbf {\bibinfo {volume}
  {19}},\ \bibinfo {pages} {3008} (\bibinfo {year} {1979})}\BibitemShut
  {NoStop}%
\bibitem [{\citenamefont {{Contaldi}}(2000)}]{2000astro.ph..5115C}%
  \BibitemOpen
  \bibfield  {author} {\bibinfo {author} {\bibfnamefont {C.~R.}\ \bibnamefont
  {{Contaldi}}},\ }\href@noop {} {\bibfield  {journal} {\bibinfo  {journal}
  {ArXiv Astrophysics e-prints}\ } (\bibinfo {year} {2000})},\ \Eprint
  {http://arxiv.org/abs/astro-ph/0005115} {astro-ph/0005115} \BibitemShut
  {NoStop}%
\bibitem [{\citenamefont {{Planck
  Collaboration}}(2014{\natexlab{b}})}]{planckstr}%
  \BibitemOpen
  \bibfield  {author} {\bibinfo {author} {\bibnamefont {{Planck
  Collaboration}}},\ }\href {\doibase 10.1051/0004-6361/201321621} {\bibfield
  {journal} {\bibinfo  {journal} {A\&A}\ }\textbf {\bibinfo {volume} {571}},\
  \bibinfo {pages} {A25} (\bibinfo {year} {2014}{\natexlab{b}})}\BibitemShut
  {NoStop}%
\bibitem [{\citenamefont {Battye}\ and\ \citenamefont
  {Moss}(2010{\natexlab{a}})}]{battye}%
  \BibitemOpen
  \bibfield  {author} {\bibinfo {author} {\bibfnamefont {R.}~\bibnamefont
  {Battye}}\ and\ \bibinfo {author} {\bibfnamefont {A.}~\bibnamefont {Moss}},\
  }\href {\doibase 10.1103/PhysRevD.82.023521} {\bibfield  {journal} {\bibinfo
  {journal} {Phys. Rev. D}\ }\textbf {\bibinfo {volume} {82}},\ \bibinfo
  {pages} {023521} (\bibinfo {year} {2010}{\natexlab{a}})}\BibitemShut
  {NoStop}%
\bibitem [{\citenamefont {Polchinski}(2005)}]{polchinski}%
  \BibitemOpen
  \bibfield  {author} {\bibinfo {author} {\bibfnamefont {J.}~\bibnamefont
  {Polchinski}},\ }\href {\doibase 10.1142/S0217751X05026686} {\bibfield
  {journal} {\bibinfo  {journal} {Int. J. Mod Phys A}\ }\textbf {\bibinfo
  {volume} {20}},\ \bibinfo {pages} {3413} (\bibinfo {year}
  {2005})}\BibitemShut {NoStop}%
\bibitem [{\citenamefont {Copeland}\ \emph {et~al.}(2004)\citenamefont
  {Copeland}, \citenamefont {Myers},\ and\ \citenamefont
  {Polchinski}}]{Copeland2004}%
  \BibitemOpen
  \bibfield  {author} {\bibinfo {author} {\bibfnamefont {E.~J.}\ \bibnamefont
  {Copeland}}, \bibinfo {author} {\bibfnamefont {R.~C.}\ \bibnamefont {Myers}},
  \ and\ \bibinfo {author} {\bibfnamefont {J.}~\bibnamefont {Polchinski}},\
  }\href {\doibase http://dx.doi.org/10.1016/j.crhy.2004.10.008} {\bibfield
  {journal} {\bibinfo  {journal} {Comptes Rendus Physique}\ }\textbf {\bibinfo
  {volume} {5}},\ \bibinfo {pages} {1021 } (\bibinfo {year}
  {2004})}\BibitemShut {NoStop}%
\bibitem [{\citenamefont {Arkani-Hamed}\ \emph {et~al.}(1998)\citenamefont
  {Arkani-Hamed}, \citenamefont {Dimopoulos},\ and\ \citenamefont
  {Dvali}}]{Arkani-Hamed}%
  \BibitemOpen
  \bibfield  {author} {\bibinfo {author} {\bibfnamefont {N.}~\bibnamefont
  {Arkani-Hamed}}, \bibinfo {author} {\bibfnamefont {S.}~\bibnamefont
  {Dimopoulos}}, \ and\ \bibinfo {author} {\bibfnamefont {G.}~\bibnamefont
  {Dvali}},\ }\href {\doibase http://dx.doi.org/10.1016/S0370-2693(98)00466-3}
  {\bibfield  {journal} {\bibinfo  {journal} {Physics Letters B}\ }\textbf
  {\bibinfo {volume} {429}},\ \bibinfo {pages} {263 } (\bibinfo {year}
  {1998})}\BibitemShut {NoStop}%
\bibitem [{\citenamefont {Randall}\ and\ \citenamefont
  {Sundrum}(1999)}]{Randall}%
  \BibitemOpen
  \bibfield  {author} {\bibinfo {author} {\bibfnamefont {L.}~\bibnamefont
  {Randall}}\ and\ \bibinfo {author} {\bibfnamefont {R.}~\bibnamefont
  {Sundrum}},\ }\href {\doibase 10.1103/PhysRevLett.83.4690} {\bibfield
  {journal} {\bibinfo  {journal} {Phys. Rev. Lett.}\ }\textbf {\bibinfo
  {volume} {83}},\ \bibinfo {pages} {4690} (\bibinfo {year}
  {1999})}\BibitemShut {NoStop}%
\bibitem [{\citenamefont {Jeannerot}\ \emph {et~al.}(2003)\citenamefont
  {Jeannerot}, \citenamefont {Rocher},\ and\ \citenamefont
  {Sakellariadou}}]{Jeannerot:2003qv}%
  \BibitemOpen
  \bibfield  {author} {\bibinfo {author} {\bibfnamefont {R.}~\bibnamefont
  {Jeannerot}}, \bibinfo {author} {\bibfnamefont {J.}~\bibnamefont {Rocher}}, \
  and\ \bibinfo {author} {\bibfnamefont {M.}~\bibnamefont {Sakellariadou}},\
  }\href {\doibase 10.1103/PhysRevD.68.103514} {\bibfield  {journal} {\bibinfo
  {journal} {Phys. Rev. D}\ }\textbf {\bibinfo {volume} {68}},\ \bibinfo
  {pages} {103514} (\bibinfo {year} {2003})}\BibitemShut {NoStop}%
\bibitem [{\citenamefont {Lizarraga}\ \emph {et~al.}(2014)\citenamefont
  {Lizarraga}, \citenamefont {Urrestilla}, \citenamefont {Daverio},
  \citenamefont {Hindmarsh}, \citenamefont {Kunz},\ and\ \citenamefont
  {Liddle}}]{2014arXiv1403.4924L}%
  \BibitemOpen
  \bibfield  {author} {\bibinfo {author} {\bibfnamefont {J.}~\bibnamefont
  {Lizarraga}}, \bibinfo {author} {\bibfnamefont {J.}~\bibnamefont
  {Urrestilla}}, \bibinfo {author} {\bibfnamefont {D.}~\bibnamefont {Daverio}},
  \bibinfo {author} {\bibfnamefont {M.}~\bibnamefont {Hindmarsh}}, \bibinfo
  {author} {\bibfnamefont {M.}~\bibnamefont {Kunz}}, \ and\ \bibinfo {author}
  {\bibfnamefont {A.~R.}\ \bibnamefont {Liddle}},\ }\href {\doibase
  10.1103/PhysRevLett.112.171301} {\bibfield  {journal} {\bibinfo  {journal}
  {Phys. Rev. Lett.}\ }\textbf {\bibinfo {volume} {112}},\ \bibinfo {pages}
  {171301} (\bibinfo {year} {2014})}\BibitemShut {NoStop}%
\bibitem [{\citenamefont {Moss}\ and\ \citenamefont
  {Pogosian}(2014)}]{2014arXiv1403.6105M}%
  \BibitemOpen
  \bibfield  {author} {\bibinfo {author} {\bibfnamefont {A.}~\bibnamefont
  {Moss}}\ and\ \bibinfo {author} {\bibfnamefont {L.}~\bibnamefont
  {Pogosian}},\ }\href {\doibase 10.1103/PhysRevLett.112.171302} {\bibfield
  {journal} {\bibinfo  {journal} {Phys. Rev. Lett.}\ }\textbf {\bibinfo
  {volume} {112}},\ \bibinfo {pages} {171302} (\bibinfo {year}
  {2014})}\BibitemShut {NoStop}%
\bibitem [{\citenamefont {Lazanu}\ and\ \citenamefont
  {Shellard}(2015)}]{1475-7516-2015-02-024}%
  \BibitemOpen
  \bibfield  {author} {\bibinfo {author} {\bibfnamefont {A.}~\bibnamefont
  {Lazanu}}\ and\ \bibinfo {author} {\bibfnamefont {P.}~\bibnamefont
  {Shellard}},\ }\href {http://stacks.iop.org/1475-7516/2015/i=02/a=024}
  {\bibfield  {journal} {\bibinfo  {journal} {Journal of Cosmology and
  Astroparticle Physics}\ }\textbf {\bibinfo {volume} {2015}},\ \bibinfo
  {pages} {024} (\bibinfo {year} {2015})}\BibitemShut {NoStop}%
\bibitem [{\citenamefont {Ade}\ \emph {et~al.}(2014)\citenamefont {Ade} \emph
  {et~al.}}]{bicep}%
  \BibitemOpen
  \bibfield  {author} {\bibinfo {author} {\bibfnamefont {P.}~\bibnamefont
  {Ade}} \emph {et~al.} (\bibinfo {collaboration} {(BICEP2 Collaboration)}),\
  }\href {\doibase 10.1103/PhysRevLett.112.241101} {\bibfield  {journal}
  {\bibinfo  {journal} {Phys. Rev. Lett.}\ }\textbf {\bibinfo {volume} {112}},\
  \bibinfo {pages} {241101} (\bibinfo {year} {2014})}\BibitemShut {NoStop}%
\bibitem [{\citenamefont {Jackson}\ \emph {et~al.}(2005)\citenamefont
  {Jackson}, \citenamefont {Jones},\ and\ \citenamefont
  {Polchinski}}]{1126-6708-2005-10-013}%
  \BibitemOpen
  \bibfield  {author} {\bibinfo {author} {\bibfnamefont {M.~G.}\ \bibnamefont
  {Jackson}}, \bibinfo {author} {\bibfnamefont {N.~T.}\ \bibnamefont {Jones}},
  \ and\ \bibinfo {author} {\bibfnamefont {J.}~\bibnamefont {Polchinski}},\
  }\href {http://stacks.iop.org/1126-6708/2005/i=10/a=013} {\bibfield
  {journal} {\bibinfo  {journal} {Journal of High Energy Physics}\ }\textbf
  {\bibinfo {volume} {2005}},\ \bibinfo {pages} {013} (\bibinfo {year}
  {2005})}\BibitemShut {NoStop}%
\bibitem [{\citenamefont {Abrikosov}(1957)}]{abrikosov}%
  \BibitemOpen
  \bibfield  {author} {\bibinfo {author} {\bibfnamefont {A.~A.}\ \bibnamefont
  {Abrikosov}},\ }\href@noop {} {\bibfield  {journal} {\bibinfo  {journal}
  {Soviet Physics JETP}\ }\textbf {\bibinfo {volume} {53}},\ \bibinfo {pages}
  {1174} (\bibinfo {year} {1957})}\BibitemShut {NoStop}%
\bibitem [{\citenamefont {Nielsen}\ and\ \citenamefont
  {Olesen}(1973)}]{nielsen}%
  \BibitemOpen
  \bibfield  {author} {\bibinfo {author} {\bibfnamefont {H.~B.}\ \bibnamefont
  {Nielsen}}\ and\ \bibinfo {author} {\bibfnamefont {P.}~\bibnamefont
  {Olesen}},\ }\href {\doibase 10.1016/0550-3213(73)90350-7} {\bibfield
  {journal} {\bibinfo  {journal} {Nuclear Physics B}\ }\textbf {\bibinfo
  {volume} {61}},\ \bibinfo {pages} {45} (\bibinfo {year} {1973})}\BibitemShut
  {NoStop}%
\bibitem [{\citenamefont {Hindmarsh}\ and\ \citenamefont
  {Kibble}(1995)}]{kibble}%
  \BibitemOpen
  \bibfield  {author} {\bibinfo {author} {\bibfnamefont {M.~B.}\ \bibnamefont
  {Hindmarsh}}\ and\ \bibinfo {author} {\bibfnamefont {T.~W.~B.}\ \bibnamefont
  {Kibble}},\ }\href {http://stacks.iop.org/0034-4885/58/i=5/a=001} {\bibfield
  {journal} {\bibinfo  {journal} {Reports on Progress in Physics}\ }\textbf
  {\bibinfo {volume} {58}},\ \bibinfo {pages} {477} (\bibinfo {year}
  {1995})}\BibitemShut {NoStop}%
\bibitem [{\citenamefont {Maeda}\ and\ \citenamefont
  {Turok}(1988)}]{turok1988}%
  \BibitemOpen
  \bibfield  {author} {\bibinfo {author} {\bibfnamefont {K.}~\bibnamefont
  {Maeda}}\ and\ \bibinfo {author} {\bibfnamefont {N.}~\bibnamefont {Turok}},\
  }\href {\doibase 10.1016/0370-2693(88)90488-1} {\bibfield  {journal}
  {\bibinfo  {journal} {Physics Letters B}\ }\textbf {\bibinfo {volume}
  {202}},\ \bibinfo {pages} {376} (\bibinfo {year} {1988})}\BibitemShut
  {NoStop}%
\bibitem [{\citenamefont {Vachaspati}\ and\ \citenamefont
  {Vilenkin}(1984)}]{vilenkin}%
  \BibitemOpen
  \bibfield  {author} {\bibinfo {author} {\bibfnamefont {T.}~\bibnamefont
  {Vachaspati}}\ and\ \bibinfo {author} {\bibfnamefont {A.}~\bibnamefont
  {Vilenkin}},\ }\href {\doibase 10.1103/PhysRevD.30.2036} {\bibfield
  {journal} {\bibinfo  {journal} {Phys. Rev. D}\ }\textbf {\bibinfo {volume}
  {30}},\ \bibinfo {pages} {2036} (\bibinfo {year} {1984})}\BibitemShut
  {NoStop}%
\bibitem [{\citenamefont {Bennett}\ and\ \citenamefont
  {Bouchet}(1990)}]{bouchet1990}%
  \BibitemOpen
  \bibfield  {author} {\bibinfo {author} {\bibfnamefont {D.~P.}\ \bibnamefont
  {Bennett}}\ and\ \bibinfo {author} {\bibfnamefont {F.~R.}\ \bibnamefont
  {Bouchet}},\ }\href {\doibase 10.1103/PhysRevD.41.2408} {\bibfield  {journal}
  {\bibinfo  {journal} {Phys. Rev. D}\ }\textbf {\bibinfo {volume} {41}},\
  \bibinfo {pages} {2408} (\bibinfo {year} {1990})}\BibitemShut {NoStop}%
\bibitem [{\citenamefont {Kibble}(1976)}]{kibble76}%
  \BibitemOpen
  \bibfield  {author} {\bibinfo {author} {\bibfnamefont {T.~W.~B.}\
  \bibnamefont {Kibble}},\ }\href {http://stacks.iop.org/0305-4470/9/i=8/a=029}
  {\bibfield  {journal} {\bibinfo  {journal} {Journal of Physics A:
  Mathematical and General}\ }\textbf {\bibinfo {volume} {9}},\ \bibinfo
  {pages} {1387} (\bibinfo {year} {1976})}\BibitemShut {NoStop}%
\bibitem [{\citenamefont {Martins}\ and\ \citenamefont
  {Shellard}(1996{\natexlab{a}})}]{shellard96}%
  \BibitemOpen
  \bibfield  {author} {\bibinfo {author} {\bibfnamefont {C.~J. A.~P.}\
  \bibnamefont {Martins}}\ and\ \bibinfo {author} {\bibfnamefont {E.~P.~S.}\
  \bibnamefont {Shellard}},\ }\href {\doibase 10.1103/PhysRevD.53.R575}
  {\bibfield  {journal} {\bibinfo  {journal} {Phys. Rev. D}\ }\textbf {\bibinfo
  {volume} {53}},\ \bibinfo {pages} {R575} (\bibinfo {year}
  {1996}{\natexlab{a}})}\BibitemShut {NoStop}%
\bibitem [{\citenamefont {Martins}\ and\ \citenamefont
  {Shellard}(1996{\natexlab{b}})}]{shellard96_2}%
  \BibitemOpen
  \bibfield  {author} {\bibinfo {author} {\bibfnamefont {C.~J. A.~P.}\
  \bibnamefont {Martins}}\ and\ \bibinfo {author} {\bibfnamefont {E.~P.~S.}\
  \bibnamefont {Shellard}},\ }\href {\doibase 10.1103/PhysRevD.54.2535}
  {\bibfield  {journal} {\bibinfo  {journal} {Phys. Rev. D}\ }\textbf {\bibinfo
  {volume} {54}},\ \bibinfo {pages} {2535} (\bibinfo {year}
  {1996}{\natexlab{b}})}\BibitemShut {NoStop}%
\bibitem [{\citenamefont {Martins}\ and\ \citenamefont
  {Shellard}(2002)}]{shellard2000}%
  \BibitemOpen
  \bibfield  {author} {\bibinfo {author} {\bibfnamefont {C.~J. A.~P.}\
  \bibnamefont {Martins}}\ and\ \bibinfo {author} {\bibfnamefont {E.~P.~S.}\
  \bibnamefont {Shellard}},\ }\href {\doibase 10.1103/PhysRevD.65.043514}
  {\bibfield  {journal} {\bibinfo  {journal} {Phys. Rev. D}\ }\textbf {\bibinfo
  {volume} {65}},\ \bibinfo {pages} {043514} (\bibinfo {year}
  {2002})}\BibitemShut {NoStop}%
\bibitem [{\citenamefont {Albrecht}\ \emph {et~al.}(1998)\citenamefont
  {Albrecht}, \citenamefont {Battye},\ and\ \citenamefont
  {Robinson}}]{albrecht}%
  \BibitemOpen
  \bibfield  {author} {\bibinfo {author} {\bibfnamefont {A.}~\bibnamefont
  {Albrecht}}, \bibinfo {author} {\bibfnamefont {R.~A.}\ \bibnamefont
  {Battye}}, \ and\ \bibinfo {author} {\bibfnamefont {J.}~\bibnamefont
  {Robinson}},\ }\href {\doibase 10.1103/PhysRevD.59.023508} {\bibfield
  {journal} {\bibinfo  {journal} {Phys. Rev. D}\ }\textbf {\bibinfo {volume}
  {59}},\ \bibinfo {pages} {023508} (\bibinfo {year} {1998})}\BibitemShut
  {NoStop}%
\bibitem [{\citenamefont {Battye}\ \emph {et~al.}(1998)\citenamefont {Battye},
  \citenamefont {Robinson},\ and\ \citenamefont {Albrecht}}]{robinson2}%
  \BibitemOpen
  \bibfield  {author} {\bibinfo {author} {\bibfnamefont {R.~A.}\ \bibnamefont
  {Battye}}, \bibinfo {author} {\bibfnamefont {J.}~\bibnamefont {Robinson}}, \
  and\ \bibinfo {author} {\bibfnamefont {A.}~\bibnamefont {Albrecht}},\ }\href
  {\doibase 10.1103/PhysRevLett.80.4847} {\bibfield  {journal} {\bibinfo
  {journal} {Phys. Rev. Lett.}\ }\textbf {\bibinfo {volume} {80}},\ \bibinfo
  {pages} {4847} (\bibinfo {year} {1998})}\BibitemShut {NoStop}%
\bibitem [{\citenamefont {Carter}(1990)}]{carter}%
  \BibitemOpen
  \bibfield  {author} {\bibinfo {author} {\bibfnamefont {B.}~\bibnamefont
  {Carter}},\ }\href {\doibase 10.1103/PhysRevD.41.3869} {\bibfield  {journal}
  {\bibinfo  {journal} {Phys. Rev. D}\ }\textbf {\bibinfo {volume} {41}},\
  \bibinfo {pages} {3869} (\bibinfo {year} {1990})}\BibitemShut {NoStop}%
\bibitem [{\citenamefont {{Seljak}}\ and\ \citenamefont
  {{Zaldarriaga}}(1996)}]{seljak}%
  \BibitemOpen
  \bibfield  {author} {\bibinfo {author} {\bibfnamefont {U.}~\bibnamefont
  {{Seljak}}}\ and\ \bibinfo {author} {\bibfnamefont {M.}~\bibnamefont
  {{Zaldarriaga}}},\ }\href {\doibase 10.1086/177793} {\bibfield  {journal}
  {\bibinfo  {journal} {\apj}\ }\textbf {\bibinfo {volume} {469}},\ \bibinfo
  {pages} {437} (\bibinfo {year} {1996})},\ \Eprint
  {http://arxiv.org/abs/astro-ph/9603033} {astro-ph/9603033} \BibitemShut
  {NoStop}%
\bibitem [{\citenamefont {M.Hindmarsh}(2011)}]{hindmarsh2011}%
  \BibitemOpen
  \bibfield  {author} {\bibinfo {author} {\bibnamefont {M.Hindmarsh}},\ }\href
  {\doibase 10.1143/PTPS.190.197} {\bibfield  {journal} {\bibinfo  {journal}
  {Progr. Theoret. Phys. Suppl.}\ }\textbf {\bibinfo {volume} {190}},\ \bibinfo
  {pages} {197} (\bibinfo {year} {2011})}\BibitemShut {NoStop}%
\bibitem [{\citenamefont {Allen}\ and\ \citenamefont {Shellard}(1990)}]{Allen}%
  \BibitemOpen
  \bibfield  {author} {\bibinfo {author} {\bibfnamefont {B.}~\bibnamefont
  {Allen}}\ and\ \bibinfo {author} {\bibfnamefont {E.~P.~S.}\ \bibnamefont
  {Shellard}},\ }\href {\doibase 10.1103/PhysRevLett.64.119} {\bibfield
  {journal} {\bibinfo  {journal} {Phys. Rev. Lett.}\ }\textbf {\bibinfo
  {volume} {64}},\ \bibinfo {pages} {119} (\bibinfo {year} {1990})}\BibitemShut
  {NoStop}%
\bibitem [{\citenamefont {Landriau}\ and\ \citenamefont
  {Shellard}(2003)}]{landriau2003}%
  \BibitemOpen
  \bibfield  {author} {\bibinfo {author} {\bibfnamefont {M.}~\bibnamefont
  {Landriau}}\ and\ \bibinfo {author} {\bibfnamefont {E.~P.~S.}\ \bibnamefont
  {Shellard}},\ }\href {\doibase 10.1103/PhysRevD.67.103512} {\bibfield
  {journal} {\bibinfo  {journal} {Phys. Rev. D}\ }\textbf {\bibinfo {volume}
  {67}},\ \bibinfo {pages} {103512} (\bibinfo {year} {2003})}\BibitemShut
  {NoStop}%
\bibitem [{\citenamefont {Landriau}\ and\ \citenamefont
  {Shellard}(2011)}]{PhysRevD.83.043516}%
  \BibitemOpen
  \bibfield  {author} {\bibinfo {author} {\bibfnamefont {M.}~\bibnamefont
  {Landriau}}\ and\ \bibinfo {author} {\bibfnamefont {E.~P.~S.}\ \bibnamefont
  {Shellard}},\ }\href {\doibase 10.1103/PhysRevD.83.043516} {\bibfield
  {journal} {\bibinfo  {journal} {Phys. Rev. D}\ }\textbf {\bibinfo {volume}
  {83}},\ \bibinfo {pages} {043516} (\bibinfo {year} {2011})}\BibitemShut
  {NoStop}%
\bibitem [{\citenamefont {Allen}\ \emph {et~al.}(1996)\citenamefont {Allen},
  \citenamefont {Caldwell}, \citenamefont {Shellard}, \citenamefont
  {Stebbins},\ and\ \citenamefont {Veeraraghavan}}]{PhysRevLett.77.3061}%
  \BibitemOpen
  \bibfield  {author} {\bibinfo {author} {\bibfnamefont {B.}~\bibnamefont
  {Allen}}, \bibinfo {author} {\bibfnamefont {R.~R.}\ \bibnamefont {Caldwell}},
  \bibinfo {author} {\bibfnamefont {E.~P.~S.}\ \bibnamefont {Shellard}},
  \bibinfo {author} {\bibfnamefont {A.}~\bibnamefont {Stebbins}}, \ and\
  \bibinfo {author} {\bibfnamefont {S.}~\bibnamefont {Veeraraghavan}},\ }\href
  {\doibase 10.1103/PhysRevLett.77.3061} {\bibfield  {journal} {\bibinfo
  {journal} {Phys. Rev. Lett.}\ }\textbf {\bibinfo {volume} {77}},\ \bibinfo
  {pages} {3061} (\bibinfo {year} {1996})}\BibitemShut {NoStop}%
\bibitem [{\citenamefont {Wu}\ \emph {et~al.}(2002)\citenamefont {Wu},
  \citenamefont {Avelino}, \citenamefont {Shellard},\ and\ \citenamefont
  {Allen}}]{doi:10.1142/S0218271802001299}%
  \BibitemOpen
  \bibfield  {author} {\bibinfo {author} {\bibfnamefont {J.-H.~P.}\
  \bibnamefont {Wu}}, \bibinfo {author} {\bibfnamefont {P.~P.}\ \bibnamefont
  {Avelino}}, \bibinfo {author} {\bibfnamefont {E.~P.~S.}\ \bibnamefont
  {Shellard}}, \ and\ \bibinfo {author} {\bibfnamefont {B.}~\bibnamefont
  {Allen}},\ }\href {\doibase 10.1142/S0218271802001299} {\bibfield  {journal}
  {\bibinfo  {journal} {International Journal of Modern Physics D}\ }\textbf
  {\bibinfo {volume} {11}},\ \bibinfo {pages} {61} (\bibinfo {year}
  {2002})}\BibitemShut {NoStop}%
\bibitem [{\citenamefont {Vanchurin}\ \emph {et~al.}(2006)\citenamefont
  {Vanchurin}, \citenamefont {Olum},\ and\ \citenamefont
  {Vilenkin}}]{PhysRevD.74.063527}%
  \BibitemOpen
  \bibfield  {author} {\bibinfo {author} {\bibfnamefont {V.}~\bibnamefont
  {Vanchurin}}, \bibinfo {author} {\bibfnamefont {K.~D.}\ \bibnamefont {Olum}},
  \ and\ \bibinfo {author} {\bibfnamefont {A.}~\bibnamefont {Vilenkin}},\
  }\href {\doibase 10.1103/PhysRevD.74.063527} {\bibfield  {journal} {\bibinfo
  {journal} {Phys. Rev. D}\ }\textbf {\bibinfo {volume} {74}},\ \bibinfo
  {pages} {063527} (\bibinfo {year} {2006})}\BibitemShut {NoStop}%
\bibitem [{\citenamefont {Olum}\ and\ \citenamefont
  {Vanchurin}(2007)}]{PhysRevD.75.063521}%
  \BibitemOpen
  \bibfield  {author} {\bibinfo {author} {\bibfnamefont {K.~D.}\ \bibnamefont
  {Olum}}\ and\ \bibinfo {author} {\bibfnamefont {V.}~\bibnamefont
  {Vanchurin}},\ }\href {\doibase 10.1103/PhysRevD.75.063521} {\bibfield
  {journal} {\bibinfo  {journal} {Phys. Rev. D}\ }\textbf {\bibinfo {volume}
  {75}},\ \bibinfo {pages} {063521} (\bibinfo {year} {2007})}\BibitemShut
  {NoStop}%
\bibitem [{\citenamefont {Turok}(1996)}]{turok-causality}%
  \BibitemOpen
  \bibfield  {author} {\bibinfo {author} {\bibfnamefont {N.}~\bibnamefont
  {Turok}},\ }\href {\doibase 10.1103/PhysRevD.54.R3686} {\bibfield  {journal}
  {\bibinfo  {journal} {Phys. Rev. D}\ }\textbf {\bibinfo {volume} {54}},\
  \bibinfo {pages} {R3686} (\bibinfo {year} {1996})}\BibitemShut {NoStop}%
\bibitem [{\citenamefont {Pen}\ \emph {et~al.}(1994)\citenamefont {Pen},
  \citenamefont {Spergel},\ and\ \citenamefont {Turok}}]{ue-li-pen2}%
  \BibitemOpen
  \bibfield  {author} {\bibinfo {author} {\bibfnamefont {U.-L.}\ \bibnamefont
  {Pen}}, \bibinfo {author} {\bibfnamefont {D.~N.}\ \bibnamefont {Spergel}}, \
  and\ \bibinfo {author} {\bibfnamefont {N.}~\bibnamefont {Turok}},\ }\href
  {\doibase 10.1103/PhysRevD.49.692} {\bibfield  {journal} {\bibinfo  {journal}
  {Phys. Rev. D}\ }\textbf {\bibinfo {volume} {49}},\ \bibinfo {pages} {692}
  (\bibinfo {year} {1994})}\BibitemShut {NoStop}%
\bibitem [{\citenamefont {Durrer}\ \emph {et~al.}(2002)\citenamefont {Durrer},
  \citenamefont {Kunz},\ and\ \citenamefont {Melchiorri}}]{durrer}%
  \BibitemOpen
  \bibfield  {author} {\bibinfo {author} {\bibfnamefont {R.}~\bibnamefont
  {Durrer}}, \bibinfo {author} {\bibfnamefont {M.}~\bibnamefont {Kunz}}, \ and\
  \bibinfo {author} {\bibfnamefont {A.}~\bibnamefont {Melchiorri}},\ }\href
  {\doibase http://dx.doi.org/10.1016/S0370-1573(02)00014-5} {\bibfield
  {journal} {\bibinfo  {journal} {Physics Reports}\ }\textbf {\bibinfo {volume}
  {364}},\ \bibinfo {pages} {1 } (\bibinfo {year} {2002})}\BibitemShut
  {NoStop}%
\bibitem [{\citenamefont {Contaldi}\ \emph {et~al.}(1999)\citenamefont
  {Contaldi}, \citenamefont {Hindmarsh},\ and\ \citenamefont
  {Magueijo}}]{contaldi}%
  \BibitemOpen
  \bibfield  {author} {\bibinfo {author} {\bibfnamefont {C.}~\bibnamefont
  {Contaldi}}, \bibinfo {author} {\bibfnamefont {M.}~\bibnamefont {Hindmarsh}},
  \ and\ \bibinfo {author} {\bibfnamefont {J.}~\bibnamefont {Magueijo}},\
  }\href {\doibase 10.1103/PhysRevLett.82.679} {\bibfield  {journal} {\bibinfo
  {journal} {Phys. Rev. Lett.}\ }\textbf {\bibinfo {volume} {82}},\ \bibinfo
  {pages} {679} (\bibinfo {year} {1999})}\BibitemShut {NoStop}%
\bibitem [{\citenamefont {Pen}\ \emph {et~al.}(1997)\citenamefont {Pen},
  \citenamefont {Seljak},\ and\ \citenamefont {Turok}}]{ue-li-pen}%
  \BibitemOpen
  \bibfield  {author} {\bibinfo {author} {\bibfnamefont {U.-L.}\ \bibnamefont
  {Pen}}, \bibinfo {author} {\bibfnamefont {U.}~\bibnamefont {Seljak}}, \ and\
  \bibinfo {author} {\bibfnamefont {N.}~\bibnamefont {Turok}},\ }\href
  {\doibase 10.1103/PhysRevLett.79.1611} {\bibfield  {journal} {\bibinfo
  {journal} {Phys. Rev. Lett.}\ }\textbf {\bibinfo {volume} {79}},\ \bibinfo
  {pages} {1611} (\bibinfo {year} {1997})}\BibitemShut {NoStop}%
\bibitem [{\citenamefont {{Ma}}\ and\ \citenamefont
  {{Bertschinger}}(1995)}]{ma}%
  \BibitemOpen
  \bibfield  {author} {\bibinfo {author} {\bibfnamefont {C.-P.}\ \bibnamefont
  {{Ma}}}\ and\ \bibinfo {author} {\bibfnamefont {E.}~\bibnamefont
  {{Bertschinger}}},\ }\href {\doibase 10.1086/176550} {\bibfield  {journal}
  {\bibinfo  {journal} {\apj}\ }\textbf {\bibinfo {volume} {455}},\ \bibinfo
  {pages} {7} (\bibinfo {year} {1995})},\ \Eprint
  {http://arxiv.org/abs/astro-ph/9506072} {astro-ph/9506072} \BibitemShut
  {NoStop}%
\bibitem [{\citenamefont {Landriau}\ and\ \citenamefont
  {Martins}()}]{landriau-later}%
  \BibitemOpen
  \bibfield  {author} {\bibinfo {author} {\bibfnamefont {M.}~\bibnamefont
  {Landriau}}\ and\ \bibinfo {author} {\bibfnamefont {C.~J. A.~P.}\
  \bibnamefont {Martins}},\ }\href@noop {} {\bibinfo  {journal} {In
  preparation}\ }\BibitemShut {NoStop}%
\bibitem [{\citenamefont {{Planck
  Collaboration}}(2014{\natexlab{c}})}]{planckparams}%
  \BibitemOpen
\bibfield  {journal} {  }\bibfield  {author} {\bibinfo {author} {\bibnamefont
  {{Planck Collaboration}}},\ }\href {\doibase 10.1051/0004-6361/201321591}
  {\bibfield  {journal} {\bibinfo  {journal} {A\&A}\ }\textbf {\bibinfo
  {volume} {571}},\ \bibinfo {pages} {A16} (\bibinfo {year}
  {2014}{\natexlab{c}})}\BibitemShut {NoStop}%
\bibitem [{\citenamefont {Bevis}\ \emph {et~al.}(2010)\citenamefont {Bevis},
  \citenamefont {Hindmarsh}, \citenamefont {Kunz},\ and\ \citenamefont
  {Urrestilla}}]{PhysRevD.82.065004}%
  \BibitemOpen
  \bibfield  {author} {\bibinfo {author} {\bibfnamefont {N.}~\bibnamefont
  {Bevis}}, \bibinfo {author} {\bibfnamefont {M.}~\bibnamefont {Hindmarsh}},
  \bibinfo {author} {\bibfnamefont {M.}~\bibnamefont {Kunz}}, \ and\ \bibinfo
  {author} {\bibfnamefont {J.}~\bibnamefont {Urrestilla}},\ }\href {\doibase
  10.1103/PhysRevD.82.065004} {\bibfield  {journal} {\bibinfo  {journal} {Phys.
  Rev. D}\ }\textbf {\bibinfo {volume} {82}},\ \bibinfo {pages} {065004}
  (\bibinfo {year} {2010})}\BibitemShut {NoStop}%
\bibitem [{\citenamefont {Lewis}(2013{\natexlab{a}})}]{cosmomc}%
  \BibitemOpen
  \bibfield  {author} {\bibinfo {author} {\bibfnamefont {A.}~\bibnamefont
  {Lewis}},\ }\href@noop {} {\enquote {\bibinfo {title} {{COSMOMC code -
  \url{http://cosmologist.info/cosmomc}}},}\ } (\bibinfo {year}
  {2013}{\natexlab{a}})\BibitemShut {NoStop}%
\bibitem [{\citenamefont {Lewis}\ and\ \citenamefont
  {Bridle}(2002)}]{cosmomc2}%
  \BibitemOpen
  \bibfield  {author} {\bibinfo {author} {\bibfnamefont {A.}~\bibnamefont
  {Lewis}}\ and\ \bibinfo {author} {\bibfnamefont {S.}~\bibnamefont {Bridle}},\
  }\href {\doibase 10.1103/PhysRevD.66.103511} {\bibfield  {journal} {\bibinfo
  {journal} {Phys. Rev. D}\ }\textbf {\bibinfo {volume} {66}},\ \bibinfo
  {pages} {103511} (\bibinfo {year} {2002})}\BibitemShut {NoStop}%
\bibitem [{\citenamefont {Lewis}(2013{\natexlab{b}})}]{CAMB}%
  \BibitemOpen
  \bibfield  {author} {\bibinfo {author} {\bibfnamefont {A.}~\bibnamefont
  {Lewis}},\ }\href@noop {} {\enquote {\bibinfo {title} {{CAMB code -
  http://camb.info}},}\ } (\bibinfo {year} {2013}{\natexlab{b}})\BibitemShut
  {NoStop}%
\bibitem [{\citenamefont {Wyman}\ \emph {et~al.}(2005)\citenamefont {Wyman},
  \citenamefont {Pogosian},\ and\ \citenamefont {Wasserman}}]{wyman}%
  \BibitemOpen
  \bibfield  {author} {\bibinfo {author} {\bibfnamefont {M.}~\bibnamefont
  {Wyman}}, \bibinfo {author} {\bibfnamefont {L.}~\bibnamefont {Pogosian}}, \
  and\ \bibinfo {author} {\bibfnamefont {I.}~\bibnamefont {Wasserman}},\ }\href
  {\doibase 10.1103/PhysRevD.72.023513} {\bibfield  {journal} {\bibinfo
  {journal} {Phys. Rev. D}\ }\textbf {\bibinfo {volume} {72}},\ \bibinfo
  {pages} {023513} (\bibinfo {year} {2005})}\BibitemShut {NoStop}%
\bibitem [{\citenamefont {Landriau}\ and\ \citenamefont
  {Shellard}(2004)}]{PhysRevD.69.023003}%
  \BibitemOpen
  \bibfield  {author} {\bibinfo {author} {\bibfnamefont {M.}~\bibnamefont
  {Landriau}}\ and\ \bibinfo {author} {\bibfnamefont {E.~P.~S.}\ \bibnamefont
  {Shellard}},\ }\href {\doibase 10.1103/PhysRevD.69.023003} {\bibfield
  {journal} {\bibinfo  {journal} {Phys. Rev. D}\ }\textbf {\bibinfo {volume}
  {69}},\ \bibinfo {pages} {023003} (\bibinfo {year} {2004})}\BibitemShut
  {NoStop}%
\bibitem [{\citenamefont {N.Bevis}\ \emph {et~al.}(2004)\citenamefont
  {N.Bevis}, \citenamefont {Hindmarsh},\ and\ \citenamefont
  {Kunz}}]{kunz-cosmomc}%
  \BibitemOpen
  \bibfield  {author} {\bibinfo {author} {\bibnamefont {N.Bevis}}, \bibinfo
  {author} {\bibfnamefont {M.}~\bibnamefont {Hindmarsh}}, \ and\ \bibinfo
  {author} {\bibfnamefont {M.}~\bibnamefont {Kunz}},\ }\href {\doibase
  10.1103/PhysRevD.70.043508} {\bibfield  {journal} {\bibinfo  {journal} {Phys.
  Rev. D}\ }\textbf {\bibinfo {volume} {70}},\ \bibinfo {pages} {043508}
  (\bibinfo {year} {2004})}\BibitemShut {NoStop}%
\bibitem [{\citenamefont {Battye}\ \emph {et~al.}(2006)\citenamefont {Battye},
  \citenamefont {Garbrecht},\ and\ \citenamefont {Moss}}]{battye-cosmomc}%
  \BibitemOpen
  \bibfield  {author} {\bibinfo {author} {\bibfnamefont {R.~A.}\ \bibnamefont
  {Battye}}, \bibinfo {author} {\bibfnamefont {B.}~\bibnamefont {Garbrecht}}, \
  and\ \bibinfo {author} {\bibfnamefont {A.}~\bibnamefont {Moss}},\ }\href
  {http://stacks.iop.org/1475-7516/2006/i=09/a=007} {\bibfield  {journal}
  {\bibinfo  {journal} {JCAP}\ }\textbf {\bibinfo {volume} {2006}},\ \bibinfo
  {pages} {007} (\bibinfo {year} {2006})}\BibitemShut {NoStop}%
\bibitem [{\citenamefont {Battye}\ and\ \citenamefont
  {Moss}(2010{\natexlab{b}})}]{battye2-cosmomc}%
  \BibitemOpen
  \bibfield  {author} {\bibinfo {author} {\bibfnamefont {R.}~\bibnamefont
  {Battye}}\ and\ \bibinfo {author} {\bibfnamefont {A.}~\bibnamefont {Moss}},\
  }\href {\doibase 10.1103/PhysRevD.82.023521} {\bibfield  {journal} {\bibinfo
  {journal} {Phys. Rev. D}\ }\textbf {\bibinfo {volume} {82}},\ \bibinfo
  {pages} {023521} (\bibinfo {year} {2010}{\natexlab{b}})}\BibitemShut
  {NoStop}%
\bibitem [{\citenamefont {J.Urrestilla}\ \emph {et~al.}(2011)\citenamefont
  {J.Urrestilla}, \citenamefont {Bevis}, \citenamefont {M.Hindmarsh},\ and\
  \citenamefont {Kunz}}]{urrestilla-cosmomc}%
  \BibitemOpen
  \bibfield  {author} {\bibinfo {author} {\bibnamefont {J.Urrestilla}},
  \bibinfo {author} {\bibfnamefont {N.}~\bibnamefont {Bevis}}, \bibinfo
  {author} {\bibnamefont {M.Hindmarsh}}, \ and\ \bibinfo {author}
  {\bibfnamefont {M.}~\bibnamefont {Kunz}},\ }\href
  {http://stacks.iop.org/1475-7516/2011/i=12/a=021} {\bibfield  {journal}
  {\bibinfo  {journal} {JCAP}\ }\textbf {\bibinfo {volume} {2011}},\ \bibinfo
  {pages} {021} (\bibinfo {year} {2011})}\BibitemShut {NoStop}%
\bibitem [{\citenamefont {{Planck Collaboration}}\ \emph
  {et~al.}(2014{\natexlab{a}})\citenamefont {{Planck Collaboration}},
  \citenamefont {{Ade, P. A. R.}} \emph {et~al.}}]{2013arXiv1303.5082P}%
  \BibitemOpen
  \bibfield  {author} {\bibinfo {author} {\bibnamefont {{Planck
  Collaboration}}}, \bibinfo {author} {\bibnamefont {{Ade, P. A. R.}}},  \emph
  {et~al.},\ }\href {\doibase 10.1051/0004-6361/201321569} {\bibfield
  {journal} {\bibinfo  {journal} {A\&A}\ }\textbf {\bibinfo {volume} {571}},\
  \bibinfo {pages} {A22} (\bibinfo {year} {2014}{\natexlab{a}})}\BibitemShut
  {NoStop}%
\bibitem [{\citenamefont {{Planck Collaboration}}\ \emph
  {et~al.}(2014{\natexlab{b}})\citenamefont {{Planck Collaboration}},
  \citenamefont {{Adam}} \emph {et~al.}}]{2014arXiv1409.5738P}%
  \BibitemOpen
  \bibfield  {author} {\bibinfo {author} {\bibnamefont {{Planck
  Collaboration}}}, \bibinfo {author} {\bibfnamefont {R.}~\bibnamefont
  {{Adam}}},  \emph {et~al.},\ }\href@noop {} {\bibfield  {journal} {\bibinfo
  {journal} {ArXiv e-prints}\ } (\bibinfo {year} {2014}{\natexlab{b}})},\
  \Eprint {http://arxiv.org/abs/1409.5738} {arXiv:1409.5738} \BibitemShut
  {NoStop}%
\end{thebibliography}%

\end{document}